\documentclass[aps,prb,twocolumn,amsfonts,showpacs,longbibliography,citeautoscript,superscriptaddress]{revtex4-2}
\usepackage{amsmath,amssymb}
\usepackage{graphicx}
\usepackage{color}
\usepackage{url}
\usepackage[utf8]{inputenc}
\usepackage{slashed}
\usepackage{subfigure}
\usepackage{slashed,bbm}
\usepackage{graphics,psfrag,epsfig}
\usepackage{dsfont}
\usepackage{setspace}
\usepackage[colorlinks=true,bookmarks=false,citecolor=blue,linkcolor=blue,hyperfootnotes=true,urlcolor=blue]{hyperref}

\newcommand{\abs}[1]{\left| #1\right|}

\newcommand{\be}{\begin{equation}}
\newcommand{\ee}{\end{equation}}
\newcommand{\bw}{\begin{widetext}}
\newcommand{\ew}{\end{widetext}}
\newcommand{\bea}{\begin{eqnarray}}
\newcommand{\eea}{\end{eqnarray}}

\newcommand{\dg}{\dagger}

\newcommand{\s}{\sigma}

\newcommand{\sgn}[1]{\text{\,sgn}\!\left(#1\right)}

\allowdisplaybreaks
\begin{document}

\title{Superconductivity of incoherent electrons in Yukawa-SYK model}

\author{Laura~Classen}
\affiliation{School of Physics and Astronomy, University of Minnesota, Minneapolis, MN 55455, USA}
\affiliation{Physics Department, Brookhaven National Laboratory, Bldg. 510A, Upton, NY 11973, USA}

\author{Andrey V. Chubukov}
\affiliation{School of Physics and Astronomy, University of Minnesota, Minneapolis, MN 55455, USA}

\begin{abstract}
We study a model of $N$ fermions in a quantum dot, coupled to $M$ bosons by a disorder-induced complex Yukawa coupling
(Yukawa-SYK model), in order to explore the interplay between non-Fermi liquid and superconductivity in a strongly coupled, (quantum-)critical environment.
We analyze the phase diagram of the model for an arbitrary complex interaction and arbitrary ratio of $N/M$, with special focus on the two regimes of non-Fermi-liquid behavior: an SYK-like behavior with a power-law frequency dependence of the fermionic  self-energy and an impurity-like behavior with frequency independent self-energy.  We show that the crossover between the two. can be reached by varying either the strength of the fermion-boson coupling or the ratio $M/N$.
We next argue that in both regimes the system is unstable to superconductivity if the strength of time-reversal-symmetry-breaking disorder is below a certain threshold.
 We show how the corresponding onset temperatures vary between the two regimes.
 We argue that the superconducting state is highly unconventional with an infinite set of minima of the condensation energy at $T=0$, corresponding to topologically different gap functions. We discuss in detail similarities and differences between this model and the model of dispersion-full fermions tuned to a metallic
 quantum-critical point, with an effective singular dynamical interaction $V(\Omega) \propto 1/|\Omega|^\gamma$ (the $\gamma-$model).
\end{abstract}

\maketitle

\section{Introduction}
The coupling of order-parameter fluctuations and electronic degrees of freedom in the vicinity of a quantum critical point (QCP) yields a variety of fascinating phenomena, which, as many believe, are responsible for the complex phase diagram of several strongly correlated electron systems \cite{RevModPhys.73.797,ReviewCuprates,doi:10.1146/annurev-conmatphys-031113-133921,doi:10.1146/annurev-conmatphys-070909-103925,Sachdev_2018,doi:10.1146/annurev-conmatphys-031016-025531,PhysRevLett.117.157002,PhysRevLett.119.157001,QCPheavyfermions,QCPheavyfermions2,PhysRevLett.124.076801,Boke1501329,Xu_2019}. At the heart of the complexity lies the interplay of two competing effects. On the one hand, critical order-parameter fluctuations make electronic excitations incoherent and give rise to non-Fermi-liquid (NFL) behavior, while incoherent electrons, in turn, modify the dynamics of the order-parameter excitations.
On the other hand, in most cases, order-parameter fluctuations mediate a strong attractive pairing interaction in at least one pairing channel.
The formation of Cooper pairs is hindered by the incoherence of electronic states, but if the pairing develops, it gaps out low-energy electronic modes and thereby reduces the tendency to NFL behavior.

The competition between NFL and pairing has been addressed within several models of fermions interacting with their collective spin and charge excitations\cite{PhysRevB.45.13047,PhysRevB.52.4607,Abanov_2001,PhysRevB.80.035117,PhysRevLett.83.3916,PhysRevLett.110.127001,Efetov2013,PhysRevB.89.195115,SCwophonons,PhysRevB.92.205104,PhysRevLett.114.097001,PhysRevB.91.115111,PhysRevB.54.16216,PhysRevB.92.125108,PhysRevLett.87.056401,PhysRevB.82.054501,PhysRevB.99.104501,PhysRevLett.116.096402,PhysRevB.88.024516,PhysRevB.89.165126,PhysRevLett.77.3009,PhysRevB.91.125136,PhysRevLett.87.257001,PhysRevB.63.140504,PhysRevLett.90.077002,Lederer4905,Klein2019,PhysRevB.98.220501,PhysRevLett.106.047004,PhysRevB.95.035124,PhysRevB.95.174520,doi:10.1146/annurev-conmatphys-031218-013339,PhysRevLett.125.247001,Berg1606,jiang2021pseudogap},
or generalized gapless bosonic modes \cite{PhysRevB.82.045121,PhysRevB.88.115116,PhysRevB.88.125116,PhysRevB.88.245106,PhysRevB.89.165114,PhysRevB.90.165144,PhysRevB.91.195135,PhysRevB.92.045118,PhysRevB.95.165137,gammamodel1,gammamodel2,gammamodel3,gammamodel4,gammamodel5,PhysRevB.99.144512,PhysRevB.99.180506,PhysRevB.99.014502,PhysRevLett.117.157001,PhysRevB.103.155161,PhysRevB.102.045147,PhysRevB.97.054502,PhysRevB.96.144508,PhysRevLett.123.096402},
for electron-phonon interaction in the limit of vanishing dressed Debye frequency \cite{PhysRevB.51.11625,Bergmann1973,Allen1991,PhysRevB.12.905,PhysRevB.43.5355,KARAKOZOV1991329,CHUBUKOV2020168190}, for fermions at a half-filled Landau level\cite{PhysRevLett.63.680,PhysRevB.47.3454,NAYAK1994359,PhysRevB.50.14048,PhysRevB.50.17917,PhysRevB.93.205401},
and even for quarks, interacting via a gluon exchange \cite{PhysRevD.59.094019,PhysRevB.72.174520}.

In this communication, we analyze this competition within a generalization of the Sachdev-Ye-Kitaev (SYK) model \cite{PhysRevLett.70.3339,Kitaevtalk,PhysRevLett.85.840,PhysRevLett.105.151602,Kitaev2018}. An appeal of SYK-type models is that they usually become exactly solvable in the limit of an infinite number of relevant degrees of freedom \cite{PhysRevLett.119.216601,PhysRevX.8.031024,PhysRevB.100.075101,PhysRevB.100.045124,PhysRevD.95.026009,PhysRevResearch.2.013301}. They also allow one to understand quantum critical, NFL behavior from a more generic perspective \cite{PhysRevX.5.041025,Polchinski2016,BAGRETS2016191,BAGRETS2017727} by. e.g., relating it to maximal quantum chaos\cite{Gu2017,PhysRevD.94.106002,PhysRevB.95.205105}. The interplay between NFL and pairing has recently been addressed for Yukawa-SYK (YSYK) models of $N$ dispersion-less fermions in a quantum dot, randomly coupled by a complex interaction to $M$ bosons that can represent either Einstein phonons or collective electronic excitations\cite{PhysRevB.100.115132,HAUCK2020168120,PhysRevLett.124.017002,2020arXiv200106586P,PhysRevResearch.2.033084,PhysRevB.103.195108,PhysRevResearch.3.013250} (see also Refs.~\onlinecite{PhysRevLett.121.187001,PhysRevB.99.024506}).
In these models, the random coupling between electrons and bosons is responsible for incoherent NFL behavior \emph{and} electronic pairing, like in more conventional models of dispersion-full electrons and non-random fermion-boson coupling.
However, in contrast to the conventional cases, when the system has to be tuned to a QCP by an external parameter (magnetic field, pressure, doping, etc), in the YSYK model this is not necessary. The reason is that in the limit of zero bandwidth the system tunes itself to criticality through the electron-boson interaction for any value of the bare bosonic mass $\omega_0$, i.e., no fine-tuning is required to reach the critical state. In this sense, the YSYK model can also be viewed as a toy model to study the interplay between NFL and pairing in a situation when the interaction is larger than the electronic bandwidth.

The outcome of the competition between NFL behavior and pairing in YSYK models depends on the ratio of the number of fermions and bosons, $N/M$, and the strength and symmetry of the Yukawa coupling.
In particular, by choosing real or complex Yukawa coupling, one can model disorder that either preserves or breaks time-reversal symmetry (TRS)\cite{HAUCK2020168120}.
Earlier works considered either a generic complex interaction, but with $N=M$ (Refs.  \onlinecite{PhysRevB.100.115132,HAUCK2020168120}) or an imaginary interaction with arbitrary $N/M$ (Ref. ~\onlinecite{PhysRevLett.124.017002,2020arXiv200106586P,PhysRevResearch.2.033084,PhysRevB.103.195108,PhysRevResearch.3.013250}).

In this article, we extend previous works and consider a generic complex interaction and arbitrary $N/M$.
We analyze in detail the emerging NFL behavior and pairing instability, and reveal special features of the pairing mediated by dispersionless critical fluctuations.
We report three sets of results, which are based on our analysis of coupled equations for the fermionic self-energy, the bosonic polarization, and the pairing vertex.
First, we obtain the normal state phase diagram as a function of temperature and fermion-boson coupling $g_0$ for various $N/M$.  Refs. \onlinecite{PhysRevB.100.115132,HAUCK2020168120} have found that for $N=M$ there are two types of NFL behavior -- an SYK-like one at smaller temperatures or
 smaller $g_0$, with characteristic power-law fermionic and bosonic self-energies, and an impurity-like one at higher temperatures and larger $g_0$, with a frequency-independent fermionic self-energy. We show that these two regimes exist for all values of $N/M$. 
 We also argue that there is a range of $T$ where fermions and bosons behave as almost free quasiparticles, and that at small 
 $N/M$ there are 
 additional intermediate regimes.
Second, we obtain the onset temperature for the pairing, $T^{(0)}_c$, as a function of three parameters:
the coupling $g_0$, the ratio $N/M$, and the degree of TRS-breaking disorder specified by a parameter $\alpha$.
We argue that the behavior of $T^{(0)}_c$ depends on out of which regime the pairing develops.
At small to intermediate $N/M$ and $g_0$, the pairing emerges at the boundary between the SYK regime and the regime of free fermions (Fig.~\ref{fig:pd}).
Here, $T_c^{(0)}$ scales with $g^2_0$ and gets suppressed by TRS-breaking disorder, until it vanishes at $ \alpha > \alpha_c$.
We find that the value of $\alpha_c$ is the largest for $N\ll M$ and decreases with increasing $N/M$.
For larger $N/M$ and/or larger $g_0$, the pairing emerges out of the impurity regime.  Here, $T_c^{(0)}$ is independent of $g_0$ for  $\alpha=0$ and decreases with $g_0$ at $\alpha \neq 0$.
For $N=M$, our results agree with Ref.~\onlinecite{HAUCK2020168120}.
Third, we show that the pairing is by itself highly unconventional in that $T^{(0)}_c$ is the largest member of a family of onset temperatures $T^{(n)}_c$ for topologically different pairing states with $n$ zeros in the gap function $\Delta_n (\omega_m)$ along the Matsubara axis.
The condensation energy at $T=0$ then has an infinite set of discrete minima. The deepest one is for the topologically trivial case $n=0$, however the presence of the other solutions enhances fluctuation effects at a finite $T$. We argue that all $T_c^{(n)}$ emerge simultaneously at $\alpha = \alpha_c $.

We also present a detailed comparison between the pairing of dispersion-less fermions in the YSYK model and of dispersion-full fermions at a QCP with an effective, frequency-dependent interaction $V(\Omega) \propto 1/|\Omega|^\gamma$, where the value of the exponent $\gamma$ is specific to the type of a QCP (the $\gamma$ model) \cite{gammamodel1,gammamodel2,gammamodel3,gammamodel4,gammamodel5,PhysRevB.99.144512,PhysRevB.99.180506,PhysRevB.99.014502,PhysRevLett.117.157001,PhysRevB.95.165137,PhysRevB.103.155161,PhysRevB.102.045147,PhysRevB.97.054502,PhysRevB.96.144508}.
 We point out certain similarities between the two models and demonstrate under which circumstances they can be mapped onto each other.
 In particular, we determine the exponent equivalent to $\gamma$, which can be tuned by $N/M$. In both cases there is also an infinite set of topologically distinct pairing states, emerging at its own $T^{(n)}_c$.
Important distinctions come from different effects outside of the low-energy regime, including effects due to the zero-bandwidth in the YSYK model.

The structure of the paper is the following. In the next section, we introduce the model and obtain the equations for fermionic and bosonic self-energies and the pairing vertex. In Sec.~\ref{sec:2}, we present the results  for the normal state assuming a non-superconducting ground state.
We consider $T=0$ in  Sec.~\ref{sec:2_a}  and a finite $T$ in Sec.~\ref{sec:2_b}.
For the latter we analyze separately the SYK regime in Sec.~\ref{sec:2_c} and the impurity regime in Sec.~\ref{sec:2_d}, and obtain the phase diagram for a generic $M/N$ and $T$ in Sec.~\ref{sec:2_e}.  In Sec. \ref{sec:3}, we analyze the pairing instability. We first analyze in Sec. \ref{sec:3_aa} the pairing out of the regime of almost free fermions. Then, in Sec. \ref{sec:3_a}, we obtain the condition for the pairing at $T=0$ and prove that the ground state is a superconductor if TRS-breaking disorder is below a critical value.
In Sec.~\ref{sec:3_b} we argue that there exists an infinite number of solutions for the pairing gap, which differ in how many times the gap function changes sign along the Matsubara axis. In Sec.~\ref{sec:3_c}, we consider finite $T$.
We obtain the critical 
temperatures $T^{(n)}_c$
 and analyze its behavior as function of $M/N$, the coupling $g_0$, and the TRS-breaking disorder $\alpha$.
In Sec.~\ref{sec:4} we present an in-depth comparison between the YSYK model and the $\gamma-$model.  We present our conclusions in Sec.~\ref{sec:5}.  Some technical aspects are discussed in the appendices.

\section{Model}
\label{sec:1}

\subsection{Yukawa-SYK Hamiltonian}
\label{sec:1_a}

We follow earlier works~\cite{PhysRevB.100.115132,HAUCK2020168120, PhysRevLett.124.017002}
and consider the model of $N$ dispersion-less electrons with spin $\s$ and zero chemical potential, randomly coupled to $M$ bosons, which represent, e.g., phonons or a scalar order-parameter field.
The Hamiltonian of the model is
\be
H=\frac{1}{2}\sum_{k=1}^M\left( \pi_k^2+\omega_0^2\phi_k^2 \right) +\frac{1}{\sqrt{MN}}\sum_{i,j=1}^N\sum_{k=1}^M g_{ijk}\phi_k c_{i\s}^\dg c_{j\s}\,,
\label{eq:H}
\ee
where $c_{i\s}^{(\dg)}$ are fermionic operators, and  $\phi_k$ describe bosons with canonical momentum $\pi_k$ and mass $\omega_0$.
Electrons and bosons are both dynamical degrees of freedom with no spatial dependence.  One can view Eq.~\ref{eq:H} as a toy model for strongly interacting fermions with negligible bandwidth. The couplings $g_{ijk} = g_{ijk}'+ig_{ijk}''$ are generally complex and Gaussian-distributed with zero mean value and the second moment given by
\begin{align}
\overline{g_{ijk}'g_{i'j'k'}'}&=\left(1-\frac{\alpha}{2}\right)\overline{g}^2\delta_{k,k'}(\delta_{i,i'}\delta_{j,j'}+\delta_{i,j'}\delta_{j,i'})\nonumber\\
\overline{g_{ijk}''g_{i'j'k'}''}&=\frac{\alpha}{2}\overline{g}^2\delta_{k,k'}(\delta_{i,i'}\delta_{j,j'}-\delta_{i,j'}\delta_{j,i'})\nonumber\\
\overline{g_{ijk}'g_{i'j'k'}''}&=0\,.
\label{eq:H_1}
\end{align}
The dimensionless parameter $\alpha$ determines the strength of TRS-breaking disorder and acts as pair breaking
for superconductivity.

One can also consider a non-random version of Eq. (\ref{eq:H}) with two effective four-fermion interactions, mediated by the propagator of $\phi_k$, with the couplings set by Eq. (\ref{eq:H_1}). In the one-loop approximation, the results for the non-random model are the same as for the model of Eq. (\ref{eq:H}).

\subsection{Eliashberg equations}
\label{sec:1_b}

We use the replica trick to perform the disorder average and derive the set of coupled equations for fermionic self-energy  $\Sigma (\omega_m)$, bosonic polarization $\Pi(\omega_m)$, and the pairing vertex $\Phi(\omega_m)$, as functions of Matsubara frequency  (see, e.g., the Appendix in  Ref.~\onlinecite{PhysRevB.100.115132} for a detailed derivation). 
In the limit $N,M \to \infty$, we obtain
\begin{align}
\Sigma(\omega_m)&=i \bar g^2 T \sum_{\omega'_m}G(\omega'_m)D(\omega_m-\omega'_m) \nonumber \\
\Pi(\omega_m)&=2\bar g^2\frac{N}{M}
 T \sum_{\omega'_m}  \left[G(\omega'_m)G(\omega_m+\omega'_m)\right.\nonumber\\
&\hspace{2.5cm}\left.-(1-\alpha)F^*(\omega'_m)F(\omega_m+\omega'_m) \right] \nonumber \\
\Phi(\omega_m)&=- (1-\alpha)\bar g^2 T \sum_{\omega'_m} F(\omega_m) D(\omega_m-\omega'_m)\, .
\label{eq:eqs}
\end{align}
 The functions $G(\omega_m)$ and  $F (\omega_m)$ on the right hand side of these equations are
  the dressed normal and anomalous electron propagators $G(\omega_m)=-i [\omega_m +\Sigma(\omega_m)]/[(\omega_m +\Sigma(\omega_m))^2+\abs{\Phi(\omega_m)}^2]$ and $F(\omega_m)=-\Phi(\omega_m)/[(\omega_m +\Sigma(\omega_m))^2+\abs{\Phi(\omega)_m}^2]$,  and  the function  $D(\omega_m)$ is the dressed boson propagator $D(\omega_m)=1/[\omega^2_m+\omega_0^2+\Pi(\omega_m)]$.
The equations for $\Sigma (\omega_m)$ and $\Phi(\omega_m)$ have the same functional form as the Eliashberg equations for an electron-phonon interaction, and to shorten notations we will be calling  Eqs.~(\ref{eq:eqs}) Eliashberg equations.

In the following we will be interested in the limit $\Phi (\omega_m) \to 0$, valid when the system is at the verge of developing a pairing instability.
 To leading order in $\Phi (\omega_m)$, we obtain
\begin{align}
\Sigma(\omega_n)&=\bar g^2 T\sum_{\omega_m}\frac{D(\omega_n-\omega_m)}{\omega_m+\Sigma(\omega_m)}\label{eq:EliashbergSigma}\\
\Pi(\omega_n)&=-2\bar g^2\frac{N}{M}T\sum_{\omega_m} \frac{1}{\omega_m+\Sigma(\omega_m)}\frac{1}{\omega_n+\omega_m+\Sigma(\omega_n+\omega_m)}\label{eq:EliashbergPi}\\
\Phi(\omega_n)&=(1-\alpha)\bar g^2 T\sum_{\omega_m} \frac{\Phi(\omega_m)}{(\omega_m+\Sigma(\omega_m))^2}D(\omega_n-\omega_m)\,.
\label{eq:linPhiT}
\end{align}
At $T=0$, the frequency sum is replaced by the integral $T \sum_{\omega_m} =\int d\omega/(2\pi)$, and the equations become
\begin{align}
\Sigma(\omega)&=\bar g^2 \int\! \frac{d\omega'}{2\pi} \frac{D(\omega-\omega')}{\omega'+\Sigma(\omega')} \label{eq:EliashbergSigma_0}
\\
\Pi(\omega)&=-2\bar g^2\frac{N}{M}\int\! \frac{d\omega'}{2\pi} \frac{1}{\omega'+\Sigma(\omega')}\frac{1}{\omega+\omega'+\Sigma(\omega+\omega')} \label{eq:EliashbergPi_0}\\
\Phi(\omega)&=(1-\alpha)\bar g^2 \int\! \frac{d\omega'}{2\pi} \frac{\Phi(\omega')}{(\omega'+\Sigma(\omega'))^2}D(\omega-\omega')~\label{eq:lingap}
\end{align}

The equation for the pairing function $\Phi (\omega_m)$ can be re-expressed as an equation for the gap function
$\Delta(\omega_n)=\Phi(\omega_n)/(1+\Sigma(\omega_n)/\omega_n)$
\be
\Delta(\omega_n)=\bar g^2 T\sum_{m}\frac{D(\omega_n-\omega_m)}{\omega_m+\Sigma(\omega_m)}\left[ (1-\alpha)\frac{\Delta_m}{\omega_m}-\frac{\Delta_n}{\omega_n} \right].
\label{eq:gap}
\ee
 At  $T=0$,
\begin{align}
\Delta(\omega)&=\bar g^2 \int\! \frac{d\omega'}{2\pi}  \frac{D(\omega-\omega')}{\omega'+\Sigma(\omega')}\left[ (1-\alpha)\frac{\Delta (\omega')}{\omega'}-\frac{\Delta}{\omega} \right]
\label{eq:gap_1}
\end{align}
In the rest of the paper, we analyze the solutions of the Eliashberg equations at $T=0$ and finite $T$ for various ratios of fermion and boson numbers $N/M$ and different ${\bar g}$.
 For the latter, it is convenient to introduce a dimensionless coupling constant
  \be
 g_0 = \frac{\bar g}{\omega^{3/2}_0}
 \ee

\section{Normal state analysis}
\label{sec:2}
\subsection{$T=0$: SYK quantum criticality}
\label{sec:2_a}
\begin{figure}[t]
\begin{center}
\includegraphics[width=.8\columnwidth]{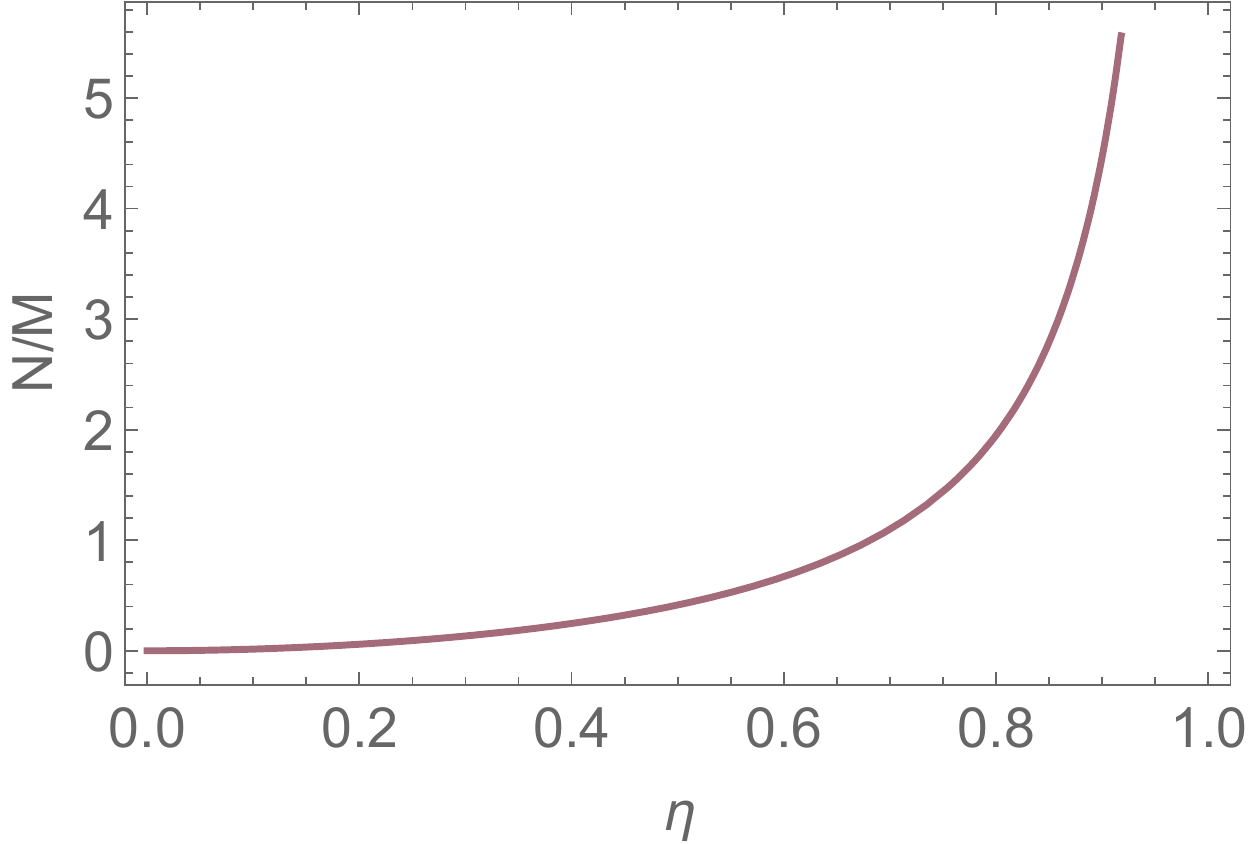}
\end{center}
\caption{Relation between the ratio of electron and boson numbers $N/M$ and the exponent $\eta$ that characterizes the power-law behavior in the SYK regime.}
\label{fig:ratio}
\end{figure}

We first analyze the solutions of the Eliashberg equations for $\Sigma (\omega)$ and $\Phi (\omega)$ at $T=0$. Earlier works~\cite{PhysRevB.100.115132,PhysRevLett.124.017002} demonstrated that boson and fermion fluctuations
balance each other, driving the system towards a critical state with vanishing boson mass for \emph{any} value of the bare mass $\omega_0$. In the critical state, fermion and boson propagators get strongly renormalized so that at low energies $\Sigma(\omega)\gg \omega$ and $\Pi(\omega)\gg\omega^2$.
We
search for self-consistent solutions for $\Sigma(\omega)$ and $\Pi (\omega)$, subject to $\Pi (0) = - \omega^2_0$.
We assume and then verify that the self-energy has a power-law form
 \begin{align}
 \Sigma (\omega) =\sgn{\omega}\bar\omega_\eta^{(1+\eta)/2}\abs{\omega}^{(1-\eta)/2}
 \label{eq:Sigmaqu}
 \end{align}
 as long as $ \Sigma (\omega) > \omega$.
 Substituting this form into Eq.~(\ref{eq:EliashbergPi_0}) and evaluating the frequency dependent part of $\Pi (\omega)$,  we obtain $\Pi (\omega) = - \omega^2_0 + \delta \Pi (\omega)$, where
\begin{align}
\delta \Pi (\omega) = b_\eta g_0^2\frac{\omega_0^3}{\bar\omega_\eta^{1+\eta}}\abs{\omega}^\eta
\label{eq:deltaPiqu}
 \end{align}
 with
 \be
b_\eta=-\frac{1}{\pi^2} \frac{\eta}{1-\eta} \sin (\pi\eta) \Gamma\left(\frac{1+\eta}{2}\right)\Gamma(-\eta)\,.
\label{b_eta}
\ee
Substituting $\Sigma (\omega)$ and $\Pi (\omega)$ into Eq.~(\ref{eq:EliashbergSigma_0}), we find that the
the ansatz for the self-energy
 is reproduced
if the exponent $\eta$ is related to $N/M$ via
\be
\frac{N}{M}=\frac{\eta}{1-\eta}\frac{\tan\left(\pi\frac{\eta}{2}\right)}{\tan\left(\pi\frac{1+\eta}{4}\right)}\,.
\label{eq:eta}
\ee
This equation determines $\eta$ as a universal function of $N/M$.  We plot $\eta$ vs $N/M$ in Fig.~\ref{fig:ratio}.
The same relation was found for a different type of Yukawa coupling in Ref.~\onlinecite{PhysRevLett.124.017002}.
For $N=M$, $\eta=0.6815$, consistent with Ref.~\onlinecite{PhysRevB.100.115132}. For $M\gg N$, $\eta\rightarrow0$, in the opposite limit $M\ll N$,  $\eta\rightarrow1$.

 The prefactor $\bar\omega_\eta$ in Eqs.~(\ref{eq:Sigmaqu}) and (\ref{eq:deltaPiqu}) is then determined from the condition
$ \Pi (0) =-\omega^2_0$.
 Substituting the self-energy again into Eq.~(\ref{eq:EliashbergPi_0}) and evaluating $\Pi (0)$,
 we obtain
  \be
\bar \omega_\eta=a_\eta^{2/(1+\eta)}  g_0^2 \omega_0\,
\label{eq:A}
\ee
 where
 \be
 a_\eta=\left[\frac{2}{\pi} \frac{\eta}{1+\eta} \frac{\tan{\frac{\pi \eta}{2}}}{\tan{\frac{\pi(1+\eta)}{4}}} \Gamma\left(\frac{2\eta}{1+\eta}\right)\Gamma\left(\frac{3+\eta}{1+\eta}\right)\right]^{(1+\eta)/2}\,.
\label{eq:ag}
\ee
In fact, this last expression is only approximately correct because the corresponding integral in Eq.~(\ref{eq:EliashbergPi_0}) is determined by $\omega'$ for which $\Sigma (\omega') \sim \omega'$, while we assumed $\Sigma (\omega') \gg \omega'$ in the calculation
 for $\Sigma$ and $\delta\Pi$.
 We will neglect this subtlety and use Eq.~(\ref{eq:ag}) below.
We present the details of the derivation of Eqs.~(\ref{eq:eta})-(\ref{eq:ag}) in App.~\ref{app:normalSYK}.
For later reference, in the limits $\eta \to 0$ (small $N/M$)  and $\eta \to 1$ (large $N/M$)
\begin{equation}
\begin{aligned}
a_\eta&\xrightarrow{\eta\rightarrow0}\sqrt{\eta}  \\
b_\eta&\xrightarrow{\eta\rightarrow0}\eta
\end{aligned}
\qquad
\begin{aligned}
a_\eta&\xrightarrow{\eta\rightarrow1}\frac{1+\eta}{2\pi(1-\eta)} \\
b_\eta&\xrightarrow{\eta\rightarrow1}\frac{1}{\pi(1-\eta)}
\end{aligned}
\end{equation}
For intermediate values of $\eta$, $a_\eta$ and $b_\eta$ are numbers of order one and ${\bar \omega}_\eta \sim \omega_0 g^2_0$.

\subsection{Finite temperature}
\label{sec:2_b}

At finite temperature, additional thermal effects appear. It was shown in Ref.~\onlinecite{PhysRevB.100.115132} that for $N=M$, the SYK solution, appropriately extended to include a temperature-dependent mass in the boson propagator, holds up to a certain temperature, determined by the interaction.
At larger $T$, the system crosses over to the regime of almost free fermions at $g_0 < 1$ and to another universal regime at $g_0 > 1$. In this last regime, dubbed impurity-like NFL, $\Sigma (\omega)$ is still larger than $\omega$, however, the self-energy is frequency-independent.
Here, we generalize the analysis of both SYK and impurity regimes to arbitrary $N/M$. We show that the boson mass and thermal self-energy depend on $N/M$ and thus on $\eta$. As a result, characteristic scales also vary with $N/M$, and the phase diagram depends not only on the coupling, but also on $N/M$. In particular, we demonstrate that for small $N/M$, there is no direct cross-over from the SYK to the impurity regime. Instead, intermediate regimes appear. In the opposite limit of large $N/M$, the self-energies in the SYK and in the impurity regimes have almost identical forms, and the system's behavior in the two regimes becomes indistinguishable.

\subsubsection{SYK regime}
\label{sec:2_c}
 We define the SYK regime at a finite $T$ as the one where the self-energy still has a power-law form, Eq.~\eqref{eq:Sigmaqu}, with continuous frequency $\omega$ replaced by discrete fermionic Matsubara frequencies $\omega_n = \pi T(2n+1)$, i.e.,
  \begin{align}
 \Sigma (\omega_n) = \omega_n a_\eta \left(\frac{g^2_0 \omega_0}{\abs{\omega_n}}\right)^{(1+\eta)/2}
 \label{eq:Sigmaqu_1}
 \end{align}
 Using this self-energy, we obtain the polarization operator in the form:
 $\Pi(\omega_n,T)= \Pi (0, T) +\delta\Pi(\omega_n, T=0)+
 \hat{\delta \Pi} (\omega_n,T)$, where  $\delta\Pi(\omega_n, T=0)$ is given by Eq.~\eqref{eq:deltaPiqu} with $\omega$ replaced by bosonic $\omega_n = 2\pi T n$
 \begin{align}
\delta \Pi (\omega_n) = \frac{ b_\eta}{a_\eta^2 }\omega_0^{2}\left(\frac{\abs{\omega_n}}{g_0^2\omega_0}\right)^\eta\,,
\label{eq:deltaPi_1}
 \end{align}
 and $\hat{\delta \Pi} (\omega_n,T)=\omega_0^2(T/g_0^2\omega_0)^\eta p_\eta(\omega_n /T)$ (see App.~\ref{app:mT} for details and the expression for $p_\eta(x)$).
 Finally, $\Pi (0,T) = - \omega^2_0 + m^2 (T)$ with the temperature-dependent mass
\be
m^2(T)= \omega^2_0 \frac{c_\eta}{a_\eta^2}  \left(\frac{T}{g_0^{2}\omega_0} \right)^\eta\,,
\label{eq:mass}
\ee
 where
 \be
 c_\eta=-\frac{4}{\pi} \frac{N}{M} \cos\Big(\frac{\pi\eta}{2}\Big)\Gamma(\eta)(2^{1-\eta}-1)\zeta(\eta)
 \ee
 and $\zeta(z)$ denotes the zeta function. The limits are
 $c_\eta\rightarrow \eta$ for $\eta\rightarrow 0$ and $c_\eta\rightarrow \ln 2$ for $\eta\rightarrow 1$.

 To analyze how thermal effects influence the SYK solution for various $\eta$,
 we rewrite $\Sigma (\omega_n)$ by pulling out the thermal contribution (the one with $m=n$ in Eq.~(\ref{eq:EliashbergSigma}))
\begin{align}
\Sigma(\omega_n)&=\bar g^2 T\sum_{m}\frac{D(\omega_n-\omega_m)}{\omega_m+\Sigma(\omega_m)}
 \nonumber \\
&= \frac{ g_0^2 \omega_0^3 T}{m^2(T)} \frac{1}{\omega_n+\Sigma(\omega_n)} \nonumber \\
&+ \frac{ g_0^2 \omega_0^3 T}{m^2(T)}
\sum_{m\neq n} \frac{1}{\omega_m+\Sigma(\omega_m)} \frac{1}{1+{\bar p} \left(|\omega_m-\omega_n|/T\right)}
\label{eq:smallceta}
\end{align}
where
$ {\bar p} (x) = \left(b_\eta x^\eta + a^2_\eta p(x)\right)/c_\eta$.
For
small or intermediate $\eta$, the mass term remains finite and $b_\eta/c_\eta \sim a^2_\eta/ c_\eta \sim 1$, so that for relevant $\omega/T= O(1)$ the thermal piece is of the same order as the terms in the sum over $m \neq n$.
 The thermal piece then can be neglected in an order-of-magnitude calculation. The rest gives the SYK form of the self-energy, Eq.~\eqref{eq:Sigmaqu}.
 However, when $\eta$ tends to one ($N \gg M$),
  the mass terms vanishes in comparison to $\delta\Pi+\delta\hat\Pi$, i.e.
  $ {\bar p} (x \sim 1)$ becomes large because $a^2_\eta/c_\eta \sim 1/(1-\eta)^2$, and the thermal piece in Eq.~(\ref{eq:smallceta}) becomes parametrically larger than the sum over $m \neq n$.   This implies that the system crosses over to a different regime, in which thermal effects dominate. Following Ref.~\onlinecite{PhysRevB.100.115132} we call this an "impurity" regime because the thermal contribution to the self-energy mimics the one for impurity scattering.

For a generic $\eta$, the system also crosses-over to the impurity regime, but this happens at large enough $g_0$.
The argument is that in Eq.~(\ref{eq:smallceta}) we neglected the bare $\omega^2$ in the bosonic propagator. Keeping this term, we get an extra $(\omega_n -\omega_m)^2/m^2$ in the bosonic propagator under the $\sum_{n\neq m}$.
For relevant $|\omega_n - \omega_m| = O(T)$, this last term is of order $g_0^{2\eta} (T/{ \omega_0})^{2-\eta}$, and for large enough $g_0$ it necessary becomes larger than other terms in $\Pi (\omega_m, T)$. Once this happens, the thermal piece in the self-energy again becomes the dominant one.

The route to the impurity regime by increasing the coupling has been discussed in Ref.~\onlinecite{PhysRevB.100.115132} for $M=N$. We emphasize that the impurity regime can be reached already at weak coupling by varying the ratio of electron and boson flavors towards $M \ll N$.

\subsubsection{Impurity regime}
\label{sec:2_d}
We now search for the self-consistent solution of the Eliashberg equations at a finite $T$, assuming that the thermal piece dominates. We label the corresponding variables with subindex $i$.
We express all quantities in terms of $M/N$ instead of $\eta$, because the exponent $\eta$ is a characteristic of the power-law behavior of the self-energy, which, as we will see, no longer holds in the impurity regime.

 We still assume that  $\Sigma (\omega_m) \gg \omega_m$. Keeping only the thermal contribution in the Eliashberg equation for the self-energy (the $m=n$ term in the sum in Eq.~(\ref{eq:EliashbergSigma})), we obtain
\be
\Sigma_i(\omega_n)\approx \frac{g^2_0 \omega^3_0 T}{m^2_i (T)} \frac{1}{\Sigma_i(\omega_n)}\,,
\label{eq:Sigmatherm}
\ee
The boson mass $m_i^2= D^{-1}(0)$ is determined self-consistently from $\Pi (0,T) = -\omega^2_0 + m^2_i$. Evaluating $\Pi (0,T)$ with the self-energy from (\ref{eq:Sigmatherm}), we obtain
\be
m^2_i =\left(\frac{\pi}{2}\frac{M}{N}\right)^2\frac{\omega_0 T}{g_0^2}\,.
\label{eq:mi}
\ee
Substituting Eq.~\eqref{eq:mi} into Eq.~\eqref{eq:Sigmatherm}, we find that the self-energy is independent of temperature
\be
\Sigma_i (\omega_n)=\text{sgn}(\omega_n)\frac{2g_0^2}{\pi}\frac{N}{M} \omega_0\,.
\label{eq:Sigmaimp}
\ee
Using this $\Sigma_i$, we find that the full boson polarization is given by $\Pi_i (\omega_m, T) = -\omega^2_0 + m^2_i + \delta \Pi_i (\omega_m, T)$, where
\be
\delta\Pi_i(\omega_m, T)=\frac{\pi}{2g_0^2}\frac{M}{N} \omega_0\abs{\omega_m}
\label{eq:Piimp}
\ee
Self-consistency in the impurity regime requires that $m^2 (T)$ is smaller than $\delta\Pi_i(\omega_m) + \omega^2_m$ for relevant $\omega_m \sim T$. For $N \gg M$ this holds for any $g_0$.
 In the opposite limit $N \ll M$, this holds for large enough
$g_0^2 > (M/N)^2 (\omega_0/T)$.    Because the limits $N \gg M$ and $N \ll M$ correspond to $\eta \approx 1$ and $\eta \ll 1$, these results are consistent with the ones in Sec. \ref{sec:2_c}. We address the crossover between the SYK and the impurity regimes in Sec.~\ref{sec:2_e} below, when we discuss the phase diagram.

\begin{figure*}[t]
\begin{center}
\includegraphics[trim = 160mm 80mm 180mm 60mm,clip,scale=0.17]{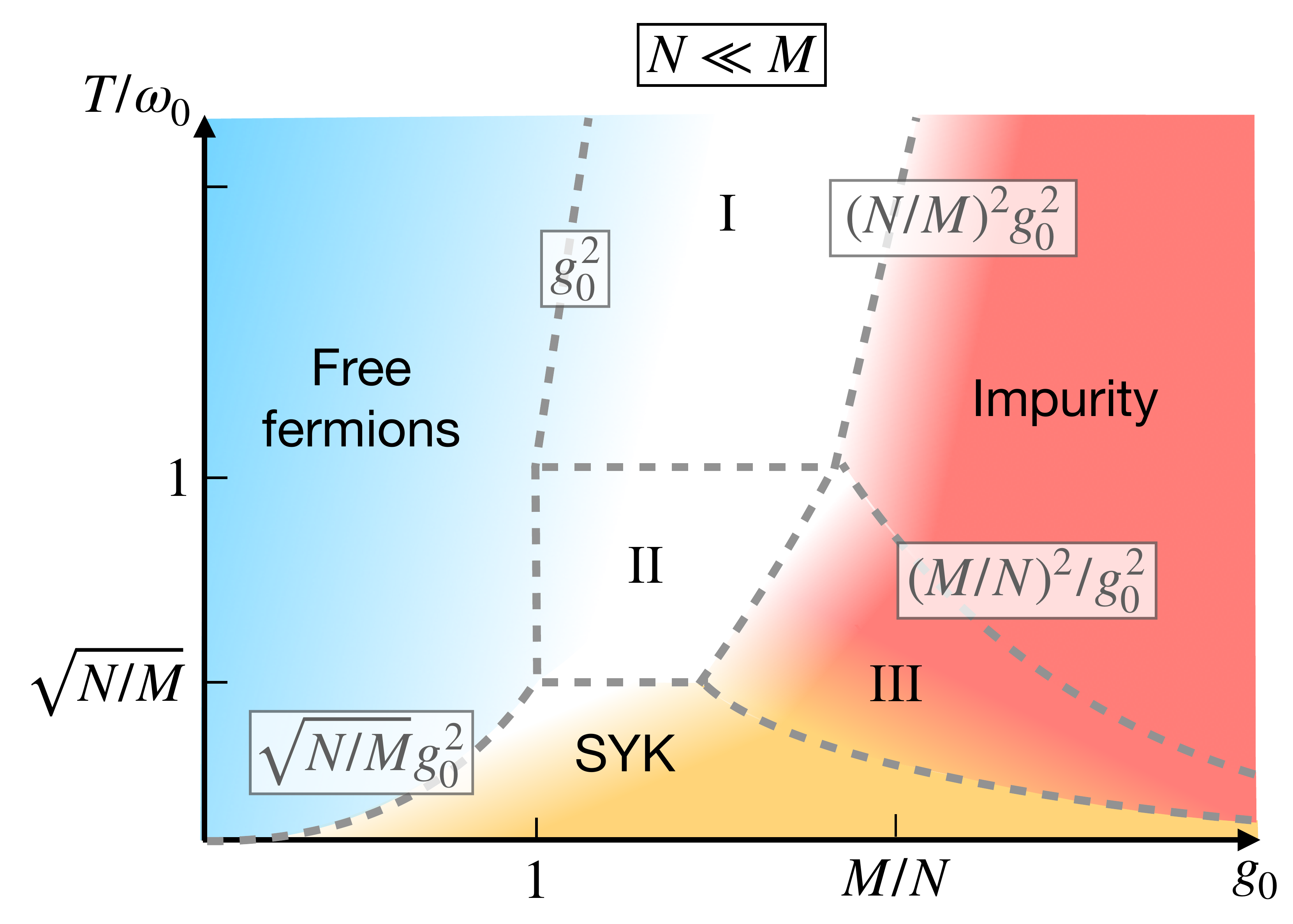}
\includegraphics[trim = 160mm 80mm 180mm 60mm,clip,scale=0.17]{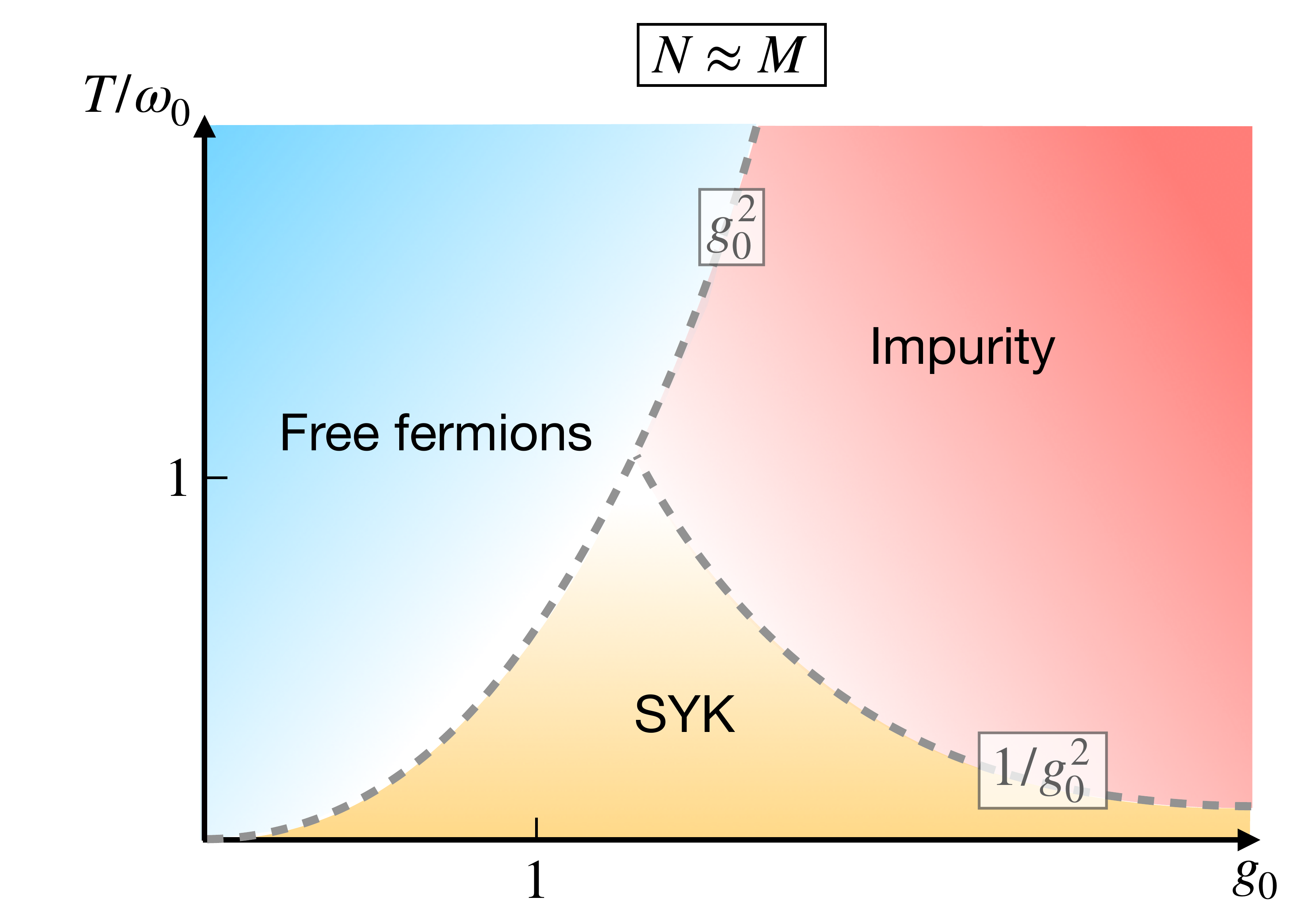}
\includegraphics[trim = 160mm 80mm 180mm 60mm,clip,scale=0.17]{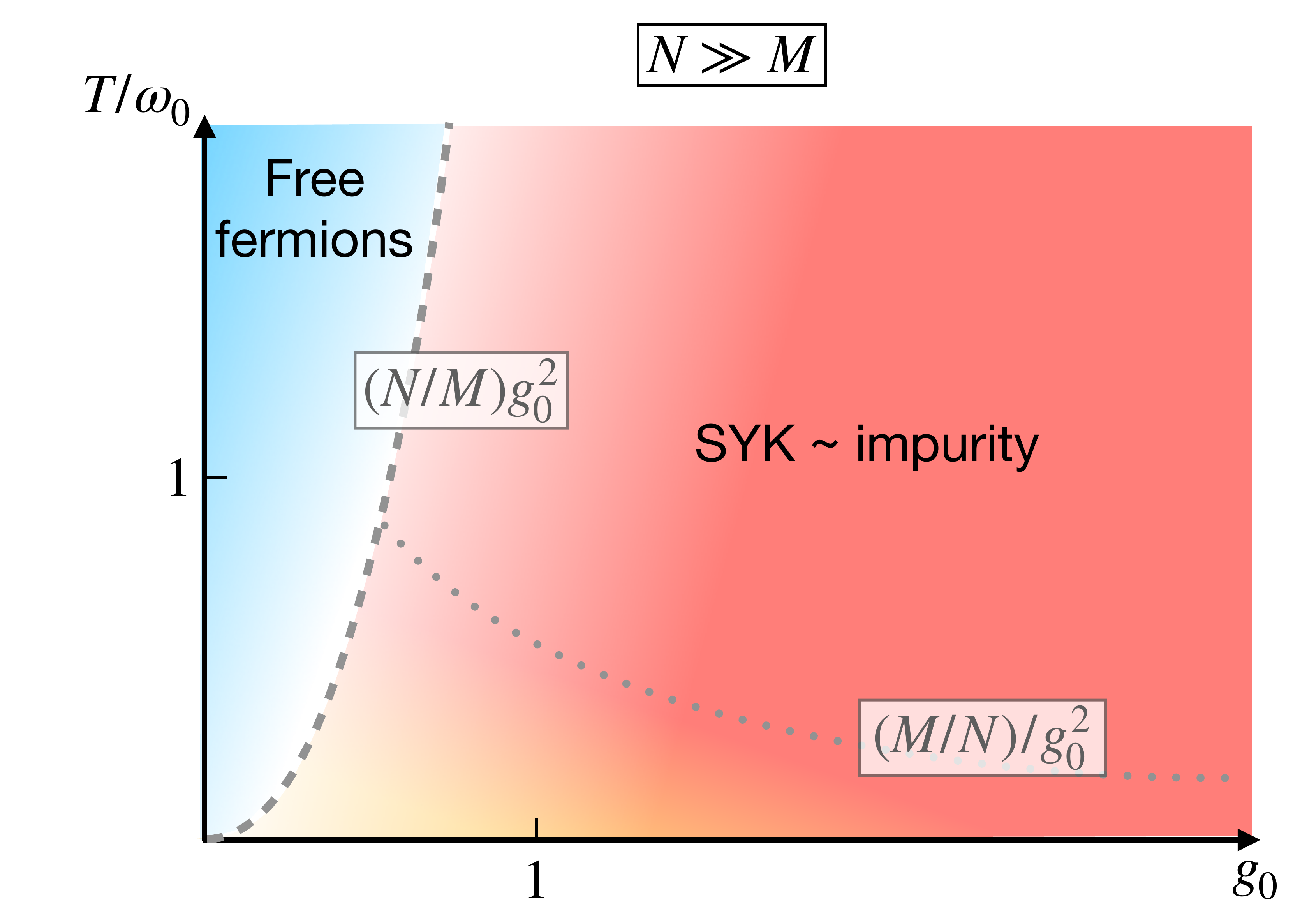}
\end{center}
\caption{Schematic phase diagrams if $T\sim \omega$ and the dimensionless coupling $g_0$ are varied for different ratios of fermion and boson flavors $N\ll M$ (left), $N=M$ (middle) and $N\gg M$ (right) assuming a non-superconducting ground state. Three universal phases are present in all three cases: a regime of free fermions (blue), a NFL regime with impurity-like self-energy (red) and a NFL SYK-like regime with power-law self-energy. Dashed lines represent crossovers between the regimes, white boxes specify their functional form. For $N\ll M$, intermediate phases I-III appear (see text).
 For $N\gg M$, the spectral properties of SYK and impurity regime are indistinguishable. Below the dotted line, the induced frequency dependence of the polarization operator becomes larger than the bare frequency dependence $\delta\Pi>\omega^2$.}
\label{fig:pd}
\end{figure*}

\subsubsection{Regime of free fermions}
\label{sec:2_f}

A finite $T$ also opens up a window in the parameter range, where both fermionic and bosonic self-energies $\Sigma$ and $\Pi$ are small compared to the bare $\omega_n$ and the bare $\omega^2_0 + \omega^2_m$ in the inverse fermion and boson propagator, respectively. Thus, fermions remain weakly interacting quasiparticles, and bosons retain an excitation gap  close to the bare $\omega_0$. The appearance of a regime of free fermions can be understood by comparing the SYK expressions for the self-energy, Eq.~\eqref{eq:Sigmaqu_1}, and boson mass, Eq.~\eqref{eq:mass}, with the bare $\omega_n$ and the bare $\omega^2_0$. For $\omega_n\sim T$, free-fermion behavior holds as long as $\Sigma (T) \leq T$ and $m^2 (T)  \approx  \omega^2_0$.
We find that these conditions are satisfied for $T \geq a_\eta^{2/(1+\eta)} \omega_0^2 g_0^2$. This becomes $T \geq \sqrt{N/M}\omega_0 g_0^2$ for $\eta\approx 0$, and $T \geq (N/M)\omega_0g_0^2$ for $\eta\approx 1$. For generic $0<\eta<1$, i.e. $M\approx N$, $a_\eta^{2/(1+\eta)}$ is of order one, and the condition simplifies to $T \geq \omega_0 g_0^2$.

\subsubsection{Phase diagram}
\label{sec:2_e}
We now discuss the phase diagram for a generic $M/N$ and $T\sim\omega$. It is instructive to separately consider the cases $N \approx M$, $N\gg M$ and $N\ll M$. We sketch the corresponding phase diagrams in Fig.~\ref{fig:pd}.

The phase diagram for $N\approx M$ has been obtained in Ref.~\onlinecite{PhysRevB.100.115132}. It contains the regime of free fermions up to $T/\omega_0\sim g_0^2$, 
and a cross-over to the SYK (impurity) regime occurs for small (large) temperatures $T<\omega_0$ ($T>\omega_0$).

For  $N\gg M$ ($\eta \approx 1$), we found in Secs. \ref{sec:2_c}, \ref{sec:2_d} that the impurity regime emerges already at
small  $T$ and $g_0$. At the same time, we note that the forms of $\Sigma_i$ and $m_i$ in Eqs.~\eqref{eq:Sigmaimp}, \eqref{eq:mi} differ from the corresponding expressions in the SYK regime, Eqs.~\eqref{eq:Sigmaqu_1}, \eqref{eq:mass}, only by multiplicative factors $\mathcal O(1)$.  This implies that the two regimes are essentially indistinguishable. We use the label ''SYK$\sim$impurity" in the right panel in Fig.~\ref{fig:pd} to reflect this. The crossover line in this panel is between the SYK$\sim$impurity regime and the regime of free fermions. The crossover is at $T/\omega_0  \sim g^2_0 (N/M)$, where $\Sigma \sim T$.  Right on the crossover line, $m^2/\omega^2_0$ is still small by a factor $M/N$.
However, to the left of this line, $m$ rapidly increases in a narrow range of $g_0$ of order
 $\sqrt{M/N}$, and approaches $\omega_0$. Simultaneously, the self-energy drops and becomes smaller than $\omega_n$, i.e., the system almost instantly crosses over to the the free fermion  regime.

In the opposite limit $N\ll M$ ($\eta\ll1$), the forms of the self-energy in the SYK and the impurity regimes differ substantially. We argue that in this case the system cannot cross-over directly from one regime to the other.
This can be seen by comparing the electronic self-energies $ \Sigma_{i}$ and $\Sigma (T)$ at the lower boundary of the impurity regime at $T /\omega_0\sim (M/N)^2/g_0^2$, where $ m_i^2\sim T^2$.  For a direct cross-over, the two self-energies must be of the same order. However, $\Sigma_{i}\sim \omega_0  (N/M)g_0^2$, while $\Sigma(T) \sim \omega_0 (M/N)^{3/4} g^{2\eta}_0$.
 Because $\Sigma_{i}$ and $\Sigma (T)$ are
 so
 different, an intermediate regime develops between the SYK and the impurity regime (region III in Fig.~\ref{fig:pd}), where the mass and self-energy gradually evolve from their values in the impurity regime.
For  $g_0 \gg 1$,
such a regime exists even for $M \sim N$ because at $T \sim \omega_0/g^2_0$,
$\Sigma_i\sim g_0^2 \omega_0$ is larger than $\Sigma (T)\sim g_0^{2\eta}\omega_0$ by $g_0^{2(1-\eta)}$.
Similarly, there is no direct crossover between the impurity regime and free-fermion regime. Indeed, at
the other boundary of the impurity region, at $T/\omega_0\sim (N/M)^2 g_0^2 >1$, where $m_i^2$ becomes of order $\omega_0^2$, $\Sigma_i$ still exceeds the bare $\omega_n \sim T$.
This opens up intermediate region I in Fig.~\ref{fig:pd}. For a given $T/\omega_0 >1$, this region holds between larger $g^2_0 \sim (T/\omega_0) (M/N)^2$ and smaller $g^2_0 \sim  (T/\omega_0)$. Within this region, the bosonic propagator retains its bare value, but the self-energy remains larger than $T$.
It is still given by the thermal contribution in the Eliashberg equation as in the impurity regime but with the temperature-independent, bare boson mass:
  $\Sigma_{\text{I}}\sim \sqrt{ g^2_0 \omega_0 T}$.
 At $T/\omega_0 <1$ and $g_0 >1$, there is another intermediate region II between the intermediate region III and the free-fermion region,
where the thermal contribution to the self-energy becomes comparable to the rest.
 In region II, the bosonic propagator retains its bare value, like in region I, and $\Sigma_{\text{II}}\sim g_0 \omega_n > T$.

\section{Superconductivity}
\label{sec:3}

NFL behavior in the normal state is one effect of the system's self-tuning to criticality.
Another one is the appearance of a singular pairing interaction mediated by a gapless boson.
In our YSYK model, this interaction scales as $(1-\alpha)$ and is attractive for $\alpha <1$, see  Eq. (\ref{eq:linPhiT}). We recall that $\alpha$ is the strength of TRS-breaking disorder.
The singular pairing interaction opens a possibility that the NFL ground state is unstable towards binding of fermions into pairs.
To verify whether this is the case, we study the gap equation and check whether it has a solution below a certain $T_c$.
To what extent pairing of dispersion-less fermions leads to superconductivity is a separate issue that we do not address in this paper. To simplify the presentation, we just call a state below a pairing instability a superconducting state.

\subsection{Superconductivity in the regime of free fermions}
\label{sec:3_aa}

We begin by analyzing superconductivity in the range of parameters where in the normal state the self-energy for fermions and the polarization operator for bosons are small compared with bare inverse fermionic and bosonic propagators, i.e., both fermions and bosons can be approximated as free particles. This regime emerges most naturally at finite temperature in the limit $N \ll M$ ($\eta \ll 1$) and $g_0 < 1$.
We assume and then verify that in this regime a superconducting instability occurs at $T_c \sim {g_0^2 \omega_0}$, and relevant internal $\omega_m$ and external $\omega_n$ in Eq.~(\ref{eq:linPhiT}) for the pairing vertex are of order $T_c$. One can easily verify that in this situation $\omega^2_0\gg |\omega_n-\omega_m|^2$, hence, to first approximation, the bosonic propagator can be viewed as frequency-independent. The pairing vertex $\Phi (\omega_n)$  then also becomes frequency independent.  
Cancelling 
$\Phi (\omega_m) = \Phi$ in  the r.h.s. and l.h.s. of Eq.~(\ref{eq:linPhiT}), we obtain the equation for $T_c$:
\be
T_c = {g_0^2 \omega_0} \frac{1-\alpha}{4\pi^2}
 \sum_m\frac{1}{(m+1/2)^2}=
\frac{1-\alpha}{4} g^2_0 \omega_0
 \label{eq:ap_5}
\ee
Note that the Matsubara sum converges because the pairing kernel contains the square of the fermionic $G(\omega_m)$.
The convergence justifies our assumption that typical $\omega_m \sim T_c$. Then, relevant $\omega_n$ are also of order $T_c$.

We see from Eq.~(\ref{eq:ap_5}) that $T_c$ depends quadratically on the coupling constant and scales as $1-\alpha$, i.e., $T_c$ is non-zero for all $\alpha <1$, where the pairing interaction is attractive.
We will see, however, that the actual situation is more complex, and for any $\eta >0$ there is a threshold for superconductivity at $\alpha_{c} <1$. The reason is that Eq. (\ref{eq:ap_5}) has been derived assuming that for relevant $\omega_n \sim T$, the self-energy is smaller than $T$. This holds for $T_c \sim {g_0^2 \omega_0}$ at small/intermediate $\alpha$, but not for $T_c \to 0$ at $\alpha \to 1$.
We will also see that there exist other solutions for the pairing, for which $\Phi (\omega_n)$  changes sign several times as a function of Matsubara frequency. These solutions appear at temperatures for which $\Sigma (\omega_n \sim T) >T$.
To demonstrate these results, in the next section we first analyze the gap equation at $T=0$ with fully dressed fermion and boson propagators, and then extend the analysis to finite $T$

\subsection{Superconductivity from a NFL --  zero temperature}
\label{sec:3_a}

\begin{figure}[t]
\begin{center}
\includegraphics[width=.8\columnwidth]{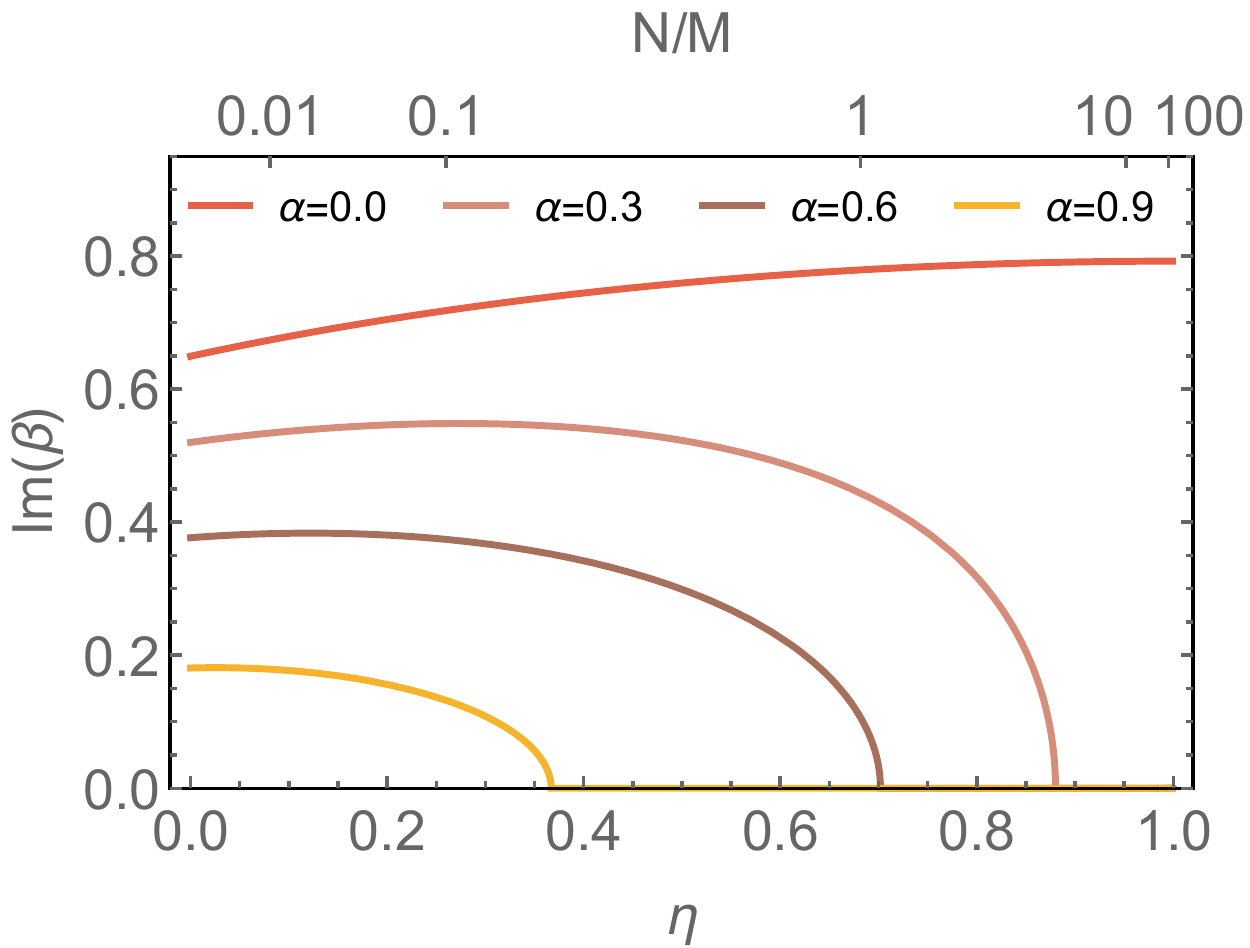}\\
\vspace{.5\baselineskip}
\includegraphics[width=.8\columnwidth]{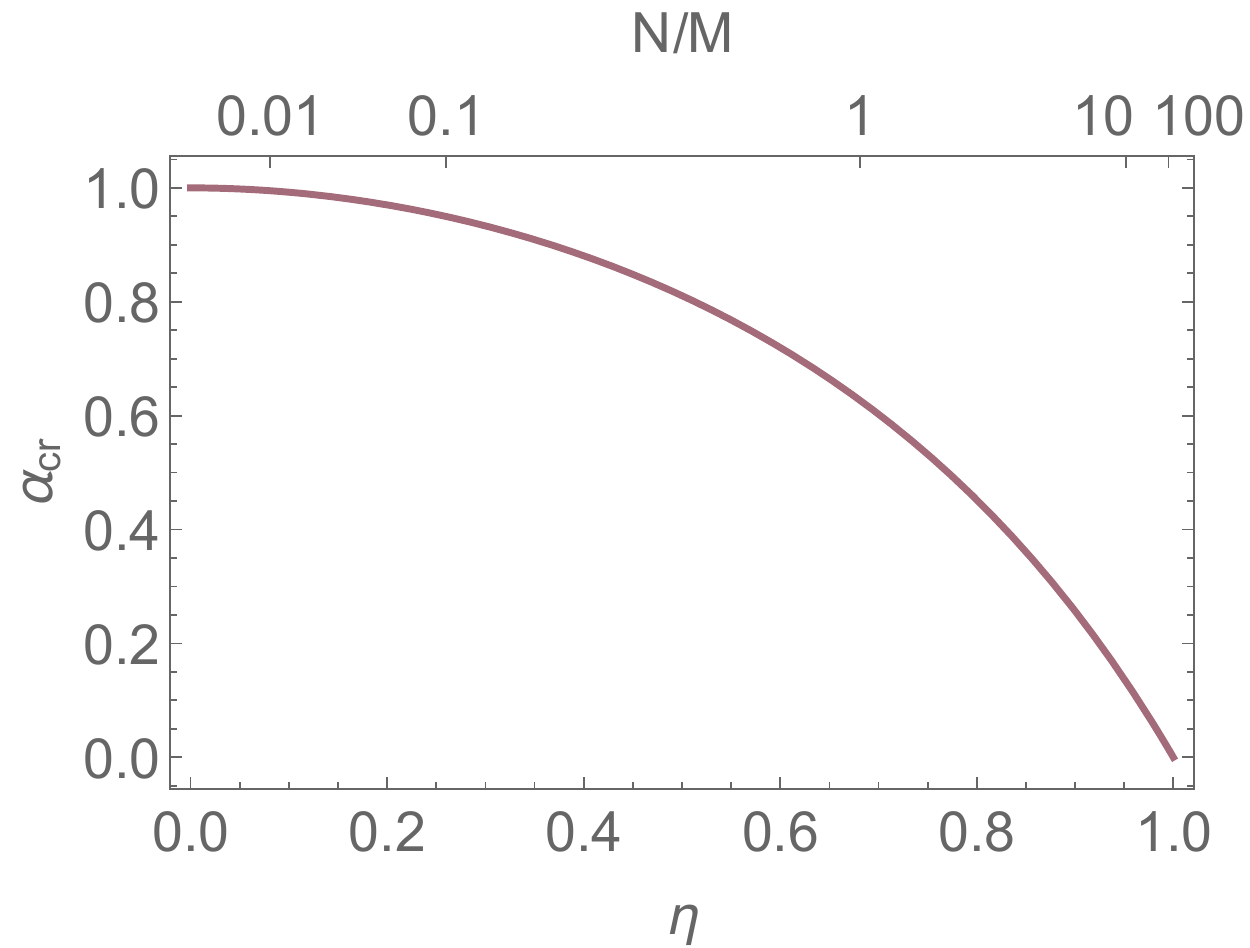}
\end{center}
\caption{Top: 
Imaginary part of the exponent $\beta=\eta/2+i\bar\kappa$ of the pairing solution $\phi(\omega)\propto\omega^\beta$ obtained from Eq.~\eqref{eq:SCcondT0}. It vanishes above a critical $\alpha_c$. For $\alpha\leq\alpha_c$ the solution of the linear gap equation at zero temperature oscillates, which is interpreted as a sign for a superconducting ground state. The parameter $\alpha$ acts as pair breaking and prevents an oscillating solution for $\alpha>\alpha_c$. Bottom: Critical $\alpha_c$ for different $\eta$.}
\label{fig:alphac}
\end{figure}

At $T=0$, the ground state without superconductivity is the SYK-like NFL with scaling forms of fermionic and bosonic propagators. We  recall that $\Sigma (\omega) \sim \omega^{(1-\eta)/2} {\bar \omega}_\eta^{(1+\eta)/2}$.
Consider first the linearized equation for the pairing vertex $\Phi (\omega)$ at $\omega \ll {\bar \omega}_\eta$, where $\Sigma (\omega) \gg \omega$.  We will see below that relevant internal $\omega'$ in Eq.~(\ref{eq:lingap})
 are comparable to $\omega$.  For such $\omega'$ we can neglect the bare $\omega'$ in the fermionic propagator and keep only $\Sigma (\omega')$. We can also neglect $\omega^2$ in the bosonic propagator compared to $\delta \Pi$ as long as $g_0\lesssim 1$, which we assume to hold.
 Substituting the SYK forms of $\Sigma$ and $\delta \Pi$ into Eq.~(\ref{eq:lingap}), we express it as
\begin{align}
\label{eq:lingap1}
\Phi(\omega)=\frac{1-\alpha}{2\pi b_\eta}\int\!d\omega'\frac{\Phi(\omega')}{\abs{\omega'}^{1-\eta}\abs{\omega-\omega'}^\eta}\,.
\end{align}
The kernel of Eq.~\eqref{eq:lingap1} is marginal (its scaling dimension is -1), hence the solution is a power-law function $\Phi(\omega)\propto1/\omega^\beta$. Substituting into Eq.~\eqref{eq:lingap1}, we obtain the condition for the exponent
\begin{align}
1=\frac{1-\alpha}{2\pi b_\eta}\psi_\eta(\beta)
\label{eq:SCcondT0}
\end{align}
with $\psi_\eta(\beta)=B(\beta,\eta-\beta)+B(1-\eta,\beta)+B(1-\eta,\eta-\beta)$ and the beta function $B(x,y)=\Gamma(x)\Gamma(y)/\Gamma(x+y)$.
A solution of Eq.~(\ref{eq:SCcondT0}) in the form $\beta_{1,2} = \eta/2 \pm \kappa$ exists for any $\eta$ (i.e., any $N/M$),
but it changes  qualitatively with  $\alpha$:
the exponent $\kappa$ is real if $\alpha$ is larger than a certain $\alpha_c <1$ (defined in Eq. (\ref{eq:alphac}) below) and imaginary $\kappa=i\bar \kappa$ if $\alpha \leq \alpha_c$.
  We argue below that this change implies that superconductivity develops at $\alpha < \alpha_c$, but not at $\alpha > \alpha_c$.
We show Im$(\beta)={\bar \kappa}$ as a function of $\eta$ for various $\alpha$ in Fig.~\ref{fig:alphac}.
The critical $\alpha_c$, which separates the power-law behavior with real and complex exponent, depends on $\eta$ as
\begin{align}
\alpha_c&=1-\frac{2\pi b_\eta}{\psi_\eta\left(\frac{\eta}{2}\right)},
\label{eq:alphac}
\end{align}
We plot $\alpha_c$ as a function of $\eta$ in Fig.~\ref{fig:alphac}. It is smaller than $1$ for all non-zero $\eta$. For $M=N$, $\eta=0.6815$ and $\alpha_c=0.6265$, consistent with Ref.~\cite{HAUCK2020168120}. For $\eta\rightarrow 0$, we find
\begin{align}
\alpha_c\approx 1-\frac{\pi }{4}\eta^2
\end{align}
In the opposite limit $\eta\rightarrow 1$, the critical $\alpha_c$ tends to zero
\begin{align}
\alpha_c\approx \frac{1}{2}(\pi+\ln16)(1-\eta)\,,
\label{e_1}
\end{align}
We can see this analytically by expanding  Eq.~\eqref{eq:SCcondT0} in $1-\eta$:
\begin{align}
1&=\frac{1-\alpha}{2}(1-\eta)\left[ \frac{2}{1-\eta} + \frac{\pi}{\sin(\pi\beta)} - H_{\beta-1} -H_{-\beta} \right]\nonumber \\
&+ \mathcal{O}\left((1-\eta)^2\right)\,
\label{n_1}
\end{align}
where $H_x$ is the harmonic number.
This reproduces Eq.~\eqref{e_1} for $\beta = \eta/2 \approx 1/2$.
We note in this regard that while $\alpha_c \to 0$ at $\eta \to 1$, the value of ${\bar \kappa}$ remains finite at $\alpha =0$. Indeed, setting $\alpha=0$ in (\ref{n_1}) and solving $\frac{\pi}{\sin(\pi\beta)} - H_{\beta-1} -H_{-\beta}=0$, we obtain $\beta =1/2 \pm 0.792i $ in good agreement with the numerical solution in Fig.~\ref{fig:alphac}.

We now associate $\alpha_c $ with the critical point of superconductivity.
We argue, following Ref. \cite{gammamodel1}, that a solution of the non-linear gap equation can be found for $0<\alpha<\alpha_c$, but not for $\alpha_c<\alpha<1$.

For $\alpha > \alpha_c$, the normalized power-law solution of the linearized gap equation corresponds to $\beta =\eta/2 - |\kappa|$, and this solution can be obtained perturbatively, starting from an infinitesimally small frequency-independent $\Phi_0$ (see Ref. \cite{gammamodel1}).
 However, the perturbative solution does not yield a divergent pairing susceptibility \footnote{The pairing susceptibility $\propto \Phi_0 (\omega_0/\omega)^{
\eta/2-|\kappa|} $ for two electrons with (running) frequencies $\omega_1=-\omega_2=\omega$ is finite for all $\omega\neq 0$ even though the total frequency $\omega_1+\omega_2=0$. This is in contrast to, e.g., the BCS susceptibility, which diverges logarithmically with the total frequency $\ln(\omega_1+\omega_2)$.}
  This indicates that the non-superconducting, NFL ground state is stable against pairing.
  The result is expected as TRS pairing breaking disorder acts against superconductivity even when the pairing interaction is attractive.

 At  $\alpha = \alpha_c$, the two exponents $\beta_{1,2} = \eta/2$ coincide.  A more careful analysis shows that at this $\alpha$,
   \be
   \Phi(\omega) \propto  \frac{1 - c \log|\omega|}{|\omega|^{\eta/2}}
   \label{n_2}
   \ee
   where $c$  is arbitrary.
 For $\alpha < \alpha_c$, $\beta_{1,2} =\eta/2\pm i\bar \kappa$ are complex, and $\Phi(\omega)$ oscillates as a function of frequency
\be
\Phi(\omega) \propto \frac{1}{|\omega|^{\eta/2}} \cos{ \left({\bar \kappa} \log{\frac{\omega_0}{|\omega|}} + \phi\right)}
\label{n_4}
\ee
These oscillations cannot be obtained perturbatively starting from $\Phi_0$ because the kernel in \eqref{eq:lingap1} is entirely positive so that an initially positive function cannot change sign in perturbation theory.
 The emergence of an oscillatory solution enables the construction of a solution of the non-linear gap equation using the same reasoning as in BCS/Eliashberg theory that one can approximate
the solution of the non-linear gap equation
$\Phi_{nl}(\omega)$ by some constant $\Phi_{nl}(0)$ up to $\omega \sim \Phi_{nl}(0)$, and by the solution of the linear gap equation at higher frequencies.
This introduces a boundary condition at $\omega \sim \Phi_{nl}(0)$, which in our case can be satisfied by choosing $\Phi_{nl}(0)$ to match the extrema of $\Phi (\omega)$ (see next Section). This suggests that the NFL ground state may now be unstable against pairing.

 To show that this is indeed the case, i.e., that $\alpha_c$ is a critical strength of TRS-breaking disorder which separates superconducting and NFL ground states, we need to show that at  $\alpha = \alpha_c$, the linear equation for $\Phi (\omega)$ has a non-trivial solution for
\emph{all} $\omega$, not only for $\omega < {\bar \omega}_\eta$, which we considered so far.  Such a solution, if it exists,  should match Eq.~(\ref{n_2}) at $\omega \ll {\bar \omega}_\eta$.
 Keeping $\omega'$ along with $\Sigma (\omega')$, we express Eq.
~\eqref{eq:lingap} as
\begin{align}
\Phi(\omega)=(1-\alpha_c)\frac{{\bar \omega}_\eta^{1+\eta}}{2\pi b_\eta}\int_0^\infty\! &d\omega'  \frac{\Phi(\omega')}{\left(\omega'+{\bar \omega}_\eta^{(1+\eta)/2}
{\omega'}^{(1-\eta)/2}\right)^2}\nonumber\\*
&\times\left[\frac{1}{\abs{\omega-\omega'}^\eta}+\frac{1}{\abs{\omega+\omega'}^\eta}\right]\,.
\label{eq:lingap2}
\end{align}
 For convenience, we restricted the integral to positive frequencies.
 At large $\omega \gg {\bar \omega}_\eta$, $\Phi (\omega)$ must decay as $1/|\omega|^\eta$,  as one can verify by direct substitution of this form into Eq.~(\ref{eq:lingap2}).

\begin{figure}[t]
\begin{center}
\includegraphics[width=.8\columnwidth]{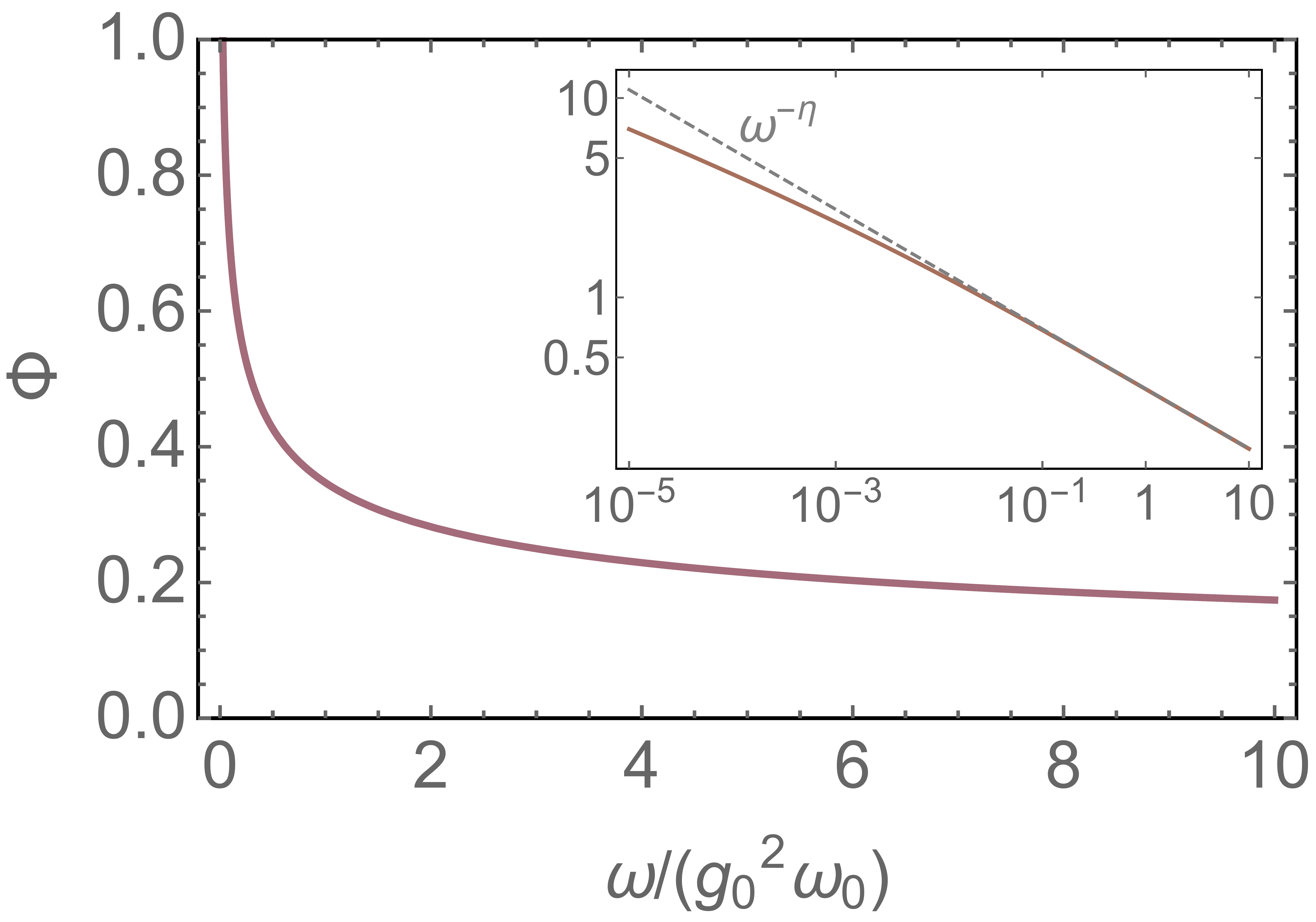}
\end{center}
\caption{The pairing vertex $\Phi(\omega)$ obtained from the differential equation \eqref{eq:dgl} at $\alpha=\alpha_c$ for $\eta=0.3$. Its low-frequency behavior follows Eq.~\eqref{n_2}. At larger frequencies, it decays as $\omega^{-\eta}$ as shown in the double logarithmic plot in the inset.}
\label{fig:Phic}
\end{figure}

To proceed further, we convert Eq. (\ref{eq:lingap2}) into an approximate differential equation.
 For this, we assume that $\eta$ is small, in which case the largest contributions to the r.h.s. come from either $\omega' \ll \omega$ or from $\omega' \gg \omega$.  Then
 $\int_0^\infty d \omega' f(\omega')(1/|\omega - \omega'|^\eta + 1/|\omega + \omega'|^\eta)$
 can be approximated
 by
 $2/|\omega|^\eta \int_0^{|\omega|} d \omega' f (\omega') + 2 \int_{|\omega|}^\infty  d \omega' f(\omega')/{\omega'}^\eta$.
 Using this in
 Eq.~(\ref{eq:lingap2}), we convert the integral equation for $\Phi (\omega)$ into
\begin{align}
&\frac{d}{d\omega}\left[\left(\frac{\omega}{\bar\omega_\eta}\right)^{1+\eta}\frac{d}{d\omega}\Phi(\omega)\right]\nonumber\\
&=-(1-\alpha_c) \frac{\eta}{\pi b_\eta}\frac{\Phi(\omega)}{\left(\omega+{\bar \omega_\eta}^{(1+\eta)/2}\omega^{(1-\eta)/2}\right)^2}\,.
\label{eq:dgl}
\end{align}
Eq.~(\ref{eq:dgl}) has two independent solutions
$\Phi_{1,2}(\omega)$, i.e.,
 $\Phi(\omega)=c_1\Phi_1(\omega)+c_2\Phi_2(\omega)$.
 The overall factor is irrelevant for the linearized gap equation, so there is in fact one parameter $c_1/c_2$.
 We have found the solutions
  $\Phi_{1,2}$
  analytically in terms of hypergeometric functions. The expressions are rather complex and we present them in App.~\ref{app:DGL}.
At $\omega \gg {\bar \omega}_\eta$, we use the expansion of a hypergeometric function at large argument and obtain
\be
\Phi(\omega)\rightarrow \left(\frac{{\bar \omega}_\eta}{\omega}\right)^\eta + \text{const.}
\ee
The constant is cancelled out by choosing a particular ratio of $c_2/c_1$. With this choice, we recover the required high-frequency behavior  $\Phi (\omega) \propto 1/|\omega|^\eta$. In the low-frequency limit, we recover Eq. (\ref{n_2}) with
$c$ related to $c_1/c_2$.
The crossover between Eq.~(\ref{n_2}) and the $1/|\omega|^\eta$  decay is at $\omega \sim {\bar \omega}_\eta /b_\eta \sim (a^{2/(1+\eta)}_\eta/b_\eta) g_0^2\omega_0$. For small $\eta$ this scale is $g_0^2\omega_0$.
We emphasize that the crossover scale remains finite at $\eta \to 0$  despite that the pairing interaction becomes almost a constant in this limit, and the theory has no external cutoff. The reason is the same as we named in the last section:
we do not get an unconstrained BCS equation at $\eta \to 0$  because the  kernel of the  gap equation contains the square of the electron propagator instead of the first power.
We plot $\Phi (\omega)$  at $\alpha = \alpha_c$ in Fig.~\ref{fig:Phic}.
The $1/\omega^\eta$ behavior indeed holds down to $\omega \sim \omega_0 g_0^2$.

 We see therefore that the  linearized  gap equation does have a non-trivial solution at $\alpha = \alpha_c$.
This is consistent with the idea that $\alpha_c$ is the critical value of TRS-breaking disorder, separating a NFL ground state at $\alpha_c<\alpha<1$ and a superconducting ground state at $\alpha < \alpha_c$.  Ref.~\onlinecite{HAUCK2020168120} reached a similar conclusion by requiring the low-frequency solution of the gap equation, Eq.~(\ref{n_2}), to vanish at a particular $\omega \sim g_0^2\omega_0$ and choosing $c$ to impose this condition.
 We also note that at $\omega \sim {g_0^2\omega_0}$, $\delta \Pi \sim \omega^2_0$ is larger than $(g_0^2 \omega_0)^2$, as long as $g_0\ll 1$. This justifies our assumption that the bare $\omega^2$ term in the bosonic propagator can be neglected. At  $g_0 = O(1)$ this assumption is still valid for qualitative reasoning, but $\omega^2$  should be kept in $D(\omega)$ in numerical calculations.

\subsection{An infinite number of solutions for the pairing} 
\label{sec:3_b}

 We now consider the case $\alpha \leq \alpha_c$ when we expect the ground state to be a superconductor in more detail.
A naive expectation would be that for these $\alpha$, the solution of the linearized gap equation disappears, and a finite  $\Phi_{nl}(\omega)$ emerges as the solution of the non-linear gap equation.
In our case, this is partly true. Namely, a finite $\Phi_{nl} (\omega)$ does appear, however the
solution of the linearized gap equation
does not disappear. To see this, we keep $\eta \ll 1$ and  analyze the linearized gap equation (\ref{eq:dgl}) for $\alpha < \alpha_c$. Applying the same reasoning as before, we again find the normalized $\Phi (\omega)$ as the sum of the two hypergeometric functions, $\Phi (\omega) = c_1 \Phi_1 (\omega) + c_2 \Phi_2 (\omega)$.  The boundary condition at high frequencies, $\Phi (\omega) \propto 1/|\omega|^\eta$, is satisfied if we choose a particular $c_1/c_2$ (which depends on $\alpha$) to cancel out a parasitic constant. At small $\omega$, $\Phi (\omega)$ oscillates in the same way as we found in the previous section
\be
\Phi (\omega)\rightarrow \frac{1}{|\omega|^{\eta/2}} \cos\left({ \bar \kappa} \log\frac{\omega_0}{|\omega|} + \phi\right)\,,
\label{n_8}
\ee
where $\phi$ is expressed via $c_2/c_1$
\footnote{Strictly speaking,  $\bar \kappa_{DE}$, obtained by solving the differential equation, differs by a numerical factor from $\bar \kappa_{IE}$, extracted from the integral equation. This difference does not play any role in our analysis and we neglect it in our analytic reasoning. For small $\eta$ and $\alpha\lesssim \alpha_c$, $\bar \kappa_{DE}$ and $\bar \kappa_{IE}$ coincide:
 $\bar \kappa_{DE}=\bar \kappa_{IE}=\sqrt{(\alpha_c-\alpha)/\pi}$.}.

 We now argue that the solution of the linearized gap equation is the end point of an infinite set of solutions of the non-linear gap equation. For this, we recall the reasoning in the previous section that one can construct an approximate solution of the non-linear gap equation by setting $\Phi_{nl} (\omega)$ to be some constant $\Phi_{nl} (0)$ up to a boundary frequency
$\omega \sim \Phi_{nl}(0)$, taking $\Phi_{nl} (\omega)$ to be proportional to $\Phi (\omega)$ at higher frequencies.
The constant $\Phi_{nl} (0)$ has to be chosen such that $\Phi_{nl} (\omega)$ is smooth at the boundary.
In our case, a natural candidate for the boundary frequency is the position of an extrema in $\Phi (\omega)$.
However, because of the oscillatory behavior of $\Phi (\omega)$, there is a set of extrema at $\omega_n^* \propto \exp(-\pi n/{\bar \kappa})$. 
Following this logic, we obtain not one but an infinite set of candidate solutions of the non-linear gap equation with gap amplitudes $\Phi^{(n)}_{nl} (0) \sim \omega_n^*$.
The function $\Phi_{nl}^{(0)} (\omega)$ is sign-preserving, $\Phi_{nl}^{(1)} (\omega)$ changes sign one time, and so on. In this nomenclature, the solution of the linearized equation
 $\Phi (\omega) = \Phi_{nl}^{(\infty)} (\omega)$.
 Each zero of $\Phi_n (\omega)$ on the Matsubara axis is a center of a vortex in the complex frequency plane \cite{gammamodel4}. Thus, the solutions with different $n$ have different number of vortices in the upper half-plane of frequency and are therefore topologically distinct.

\begin{figure}[t]
\begin{center}
\includegraphics[width=.8\columnwidth]{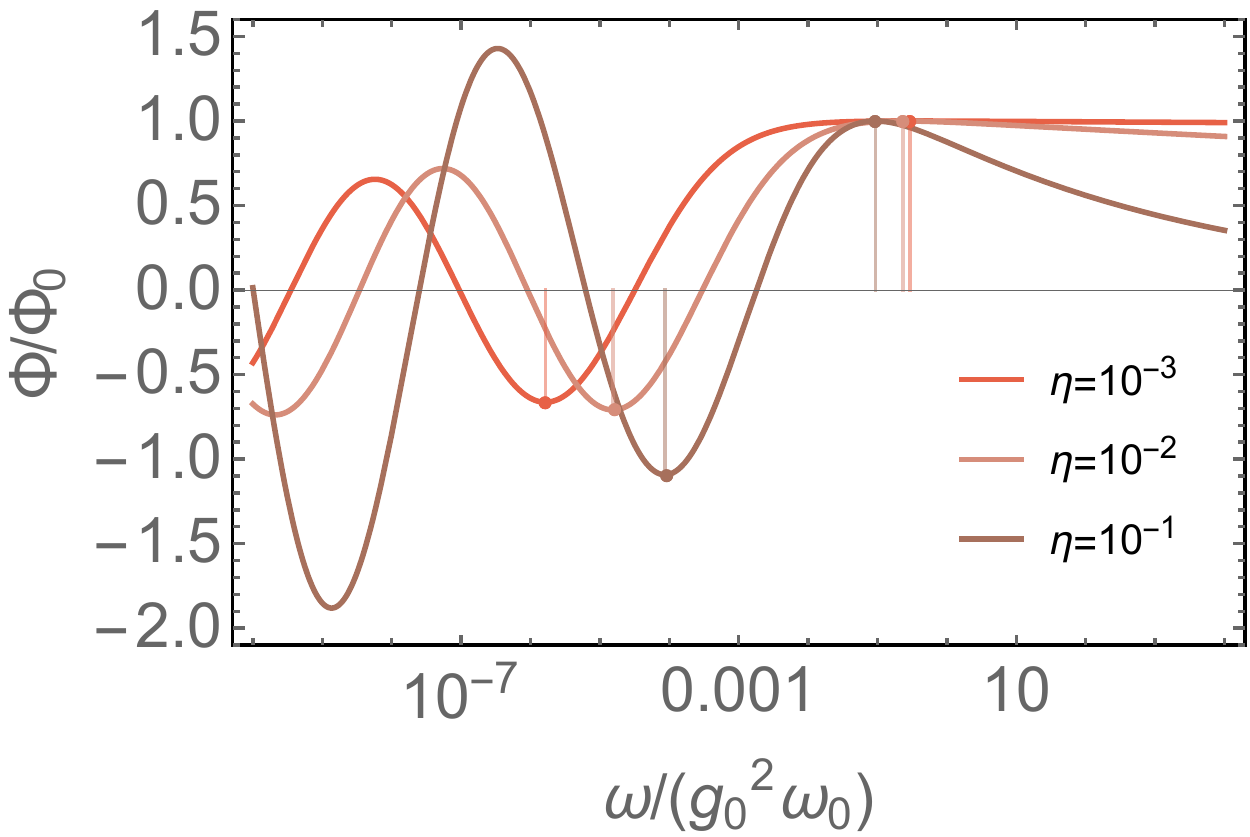}
\end{center}
\caption{The pairing vertex $\Phi(\omega)$ obtained from the differential equation Eq.~\eqref{eq:dgl} for $\alpha<\alpha_c$. $\Phi$ oscillates at low frequencies on an exponential scale. One can approximately identify the extrema $\omega^*_n$ with the set of onset temperatures for pairing $T_c^{(n)}$ (see discussion below Eq.~\eqref{n_8} and Sec.~\ref{sec:3_c}). For $\eta\to 0$, the largest $\omega^*_0$ behaves qualitatively different from the rest: $\omega_0^*$ tends to a constant $\sim\omega_og_0^2$, while all other $\omega_n^*$ with $n\geq1$ vanish with $\eta$. This can be clearly seen in the figure, where $w_0^*$ and $\omega_1^*$ are marked by vertical lines.
For a better comparison, $\Phi(\omega)$ is rescaled by $\Phi_0=\Phi(\omega_0^*)$.}
\label{fig:phidgl}
\end{figure}

For $\alpha$ slightly below $\alpha_c$ and small $\eta$ we find  ${\bar \kappa}=\sqrt{(\alpha_c-\alpha)/\pi}$, hence
 $\Phi_{nl}^{(n)}(0) \propto \exp[-(n\pi-\phi)\sqrt{\pi/(\alpha_c-\alpha)}]$ are exponentially small.  At smaller $\alpha$, ${\bar \kappa} = O(1)$, and the magnitudes of the candidate solutions $\Phi_{nl}^{(n)} (0)$ are set by characteristic scales in the problem. Analyzing the analytical solution of the linearized gap equation at small $\eta$ and a generic $\alpha < \alpha_c$,  we  find two such scales.  One is $\omega \sim g_0^2\omega_0$, where $\Phi (\omega)$ reaches a first maximum upon decreasing $\omega$. The second is $\omega \sim {\bar \omega}_\eta \approx \eta g_0^2\omega_0$, below which $\Phi (\omega)$ displays log-oscillations. We plot $\Phi (\omega)$ in Fig.~\ref{fig:phidgl} for $\alpha =0$ and several small $\eta$.
 We mark the positions of the largest maximum and the onset of log-oscillations (the largest minimum). We clearly see that
 their difference increases as $\eta$. This means that the magnitude of the $n=0$ candidate solution of the non-linear equation is $\Phi_{nl}^{(0)} (0) \sim g_0^2\omega_0$, while the magnitudes of the other solutions with $n \geq 1$ are $\Phi_{nl}^{(n)} (0) \sim \eta {g_0^2 \omega_0} \exp[{-\pi n/{ \bar \kappa}}]$.
Thus, in the limit $\eta \to 0$, $\Phi_NL^{(0)}$ remains finite, consistent with Eq. (\ref{eq:ap_5}),  but all other $\Phi_n$ vanish.

\subsection{
 Superconductivity from a NFL - the onset temperatures}
\label{sec:3_c}

\begin{figure}[t!]
\begin{center}
\includegraphics[width=.8\columnwidth]{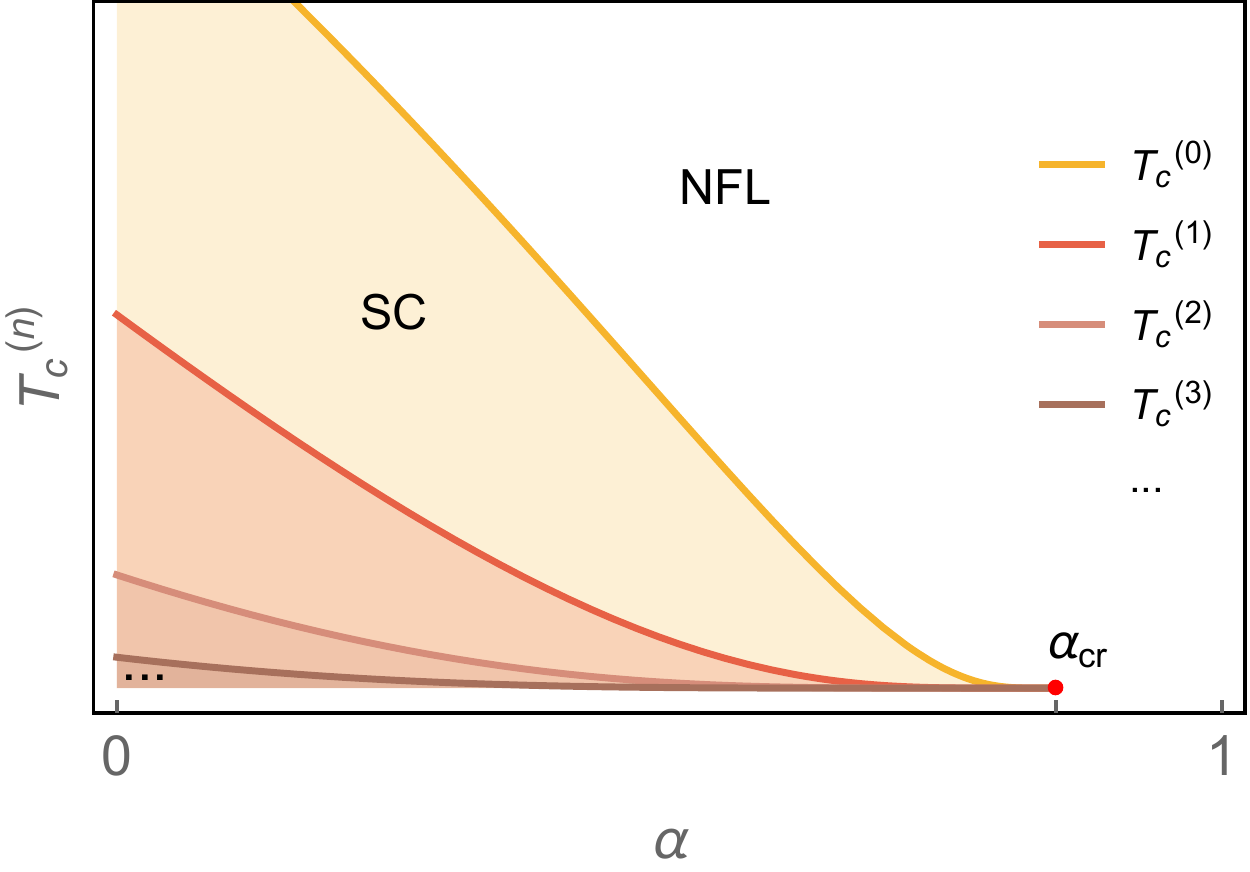}
\end{center}
\caption{Sketch of the infinite set of critical temperatures $T_c{(1)}>T_c{(2)}>T_c{(3)}\ldots$ as a function of the pair-breaking parameter $\alpha$ that all vanish in the same critical $\alpha_c$.}
\label{fig:Tcschematic}
\end{figure}

\begin{figure}[t!]
\begin{center}
\includegraphics[width=.8\columnwidth]{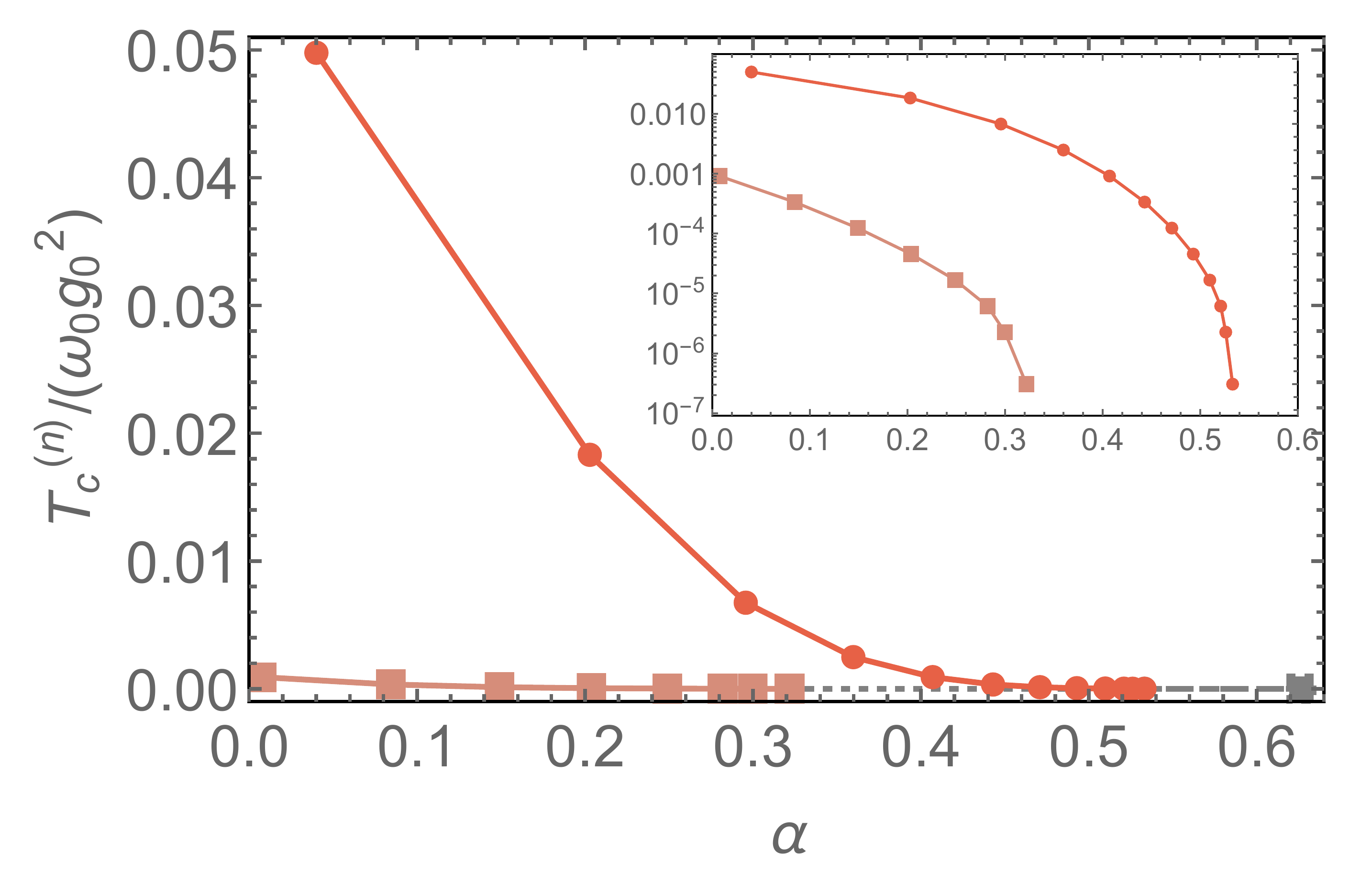}
\end{center}
\caption{The two largest critical temperatures $T_c^{(0)}$ (red) and $T_c^{(1)}$ (rose) for $\eta=0.68$ ($M\approx N$), see App.~\ref{app:SCT} for other values of $\eta$. The gray symbol marks the critical $\alpha_c\approx0.63$, where we expect $T_c^{(n)}$ to vanish. It is not reached completely due to numerical restrictions. Inset shows the same data on a logarithmic scale. We show the corresponding gap functions for $\alpha=0$ in Fig.~\ref{fig:phi}.}
\label{fig:Tcalpha}
\end{figure}

\begin{figure}[t]
\begin{center}
\includegraphics[width=.8\columnwidth]{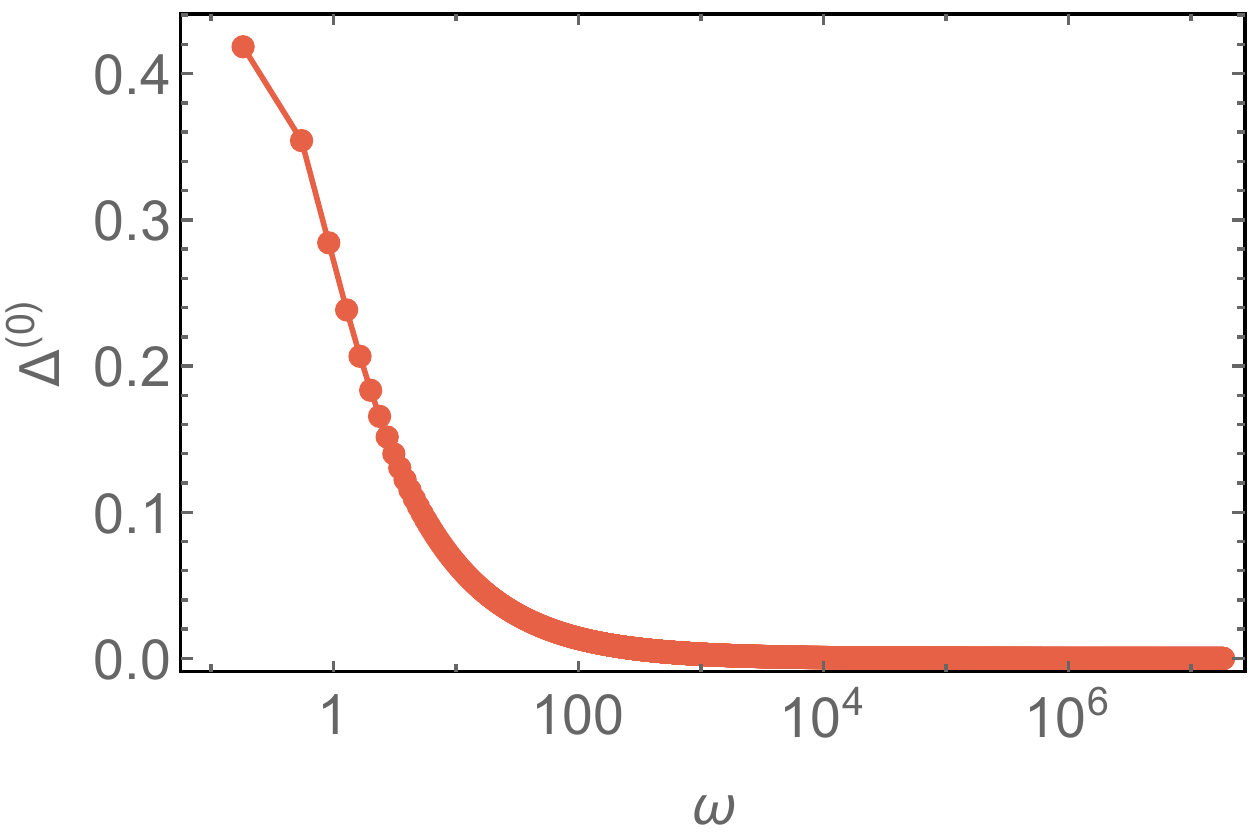}
\includegraphics[width=.8\columnwidth]{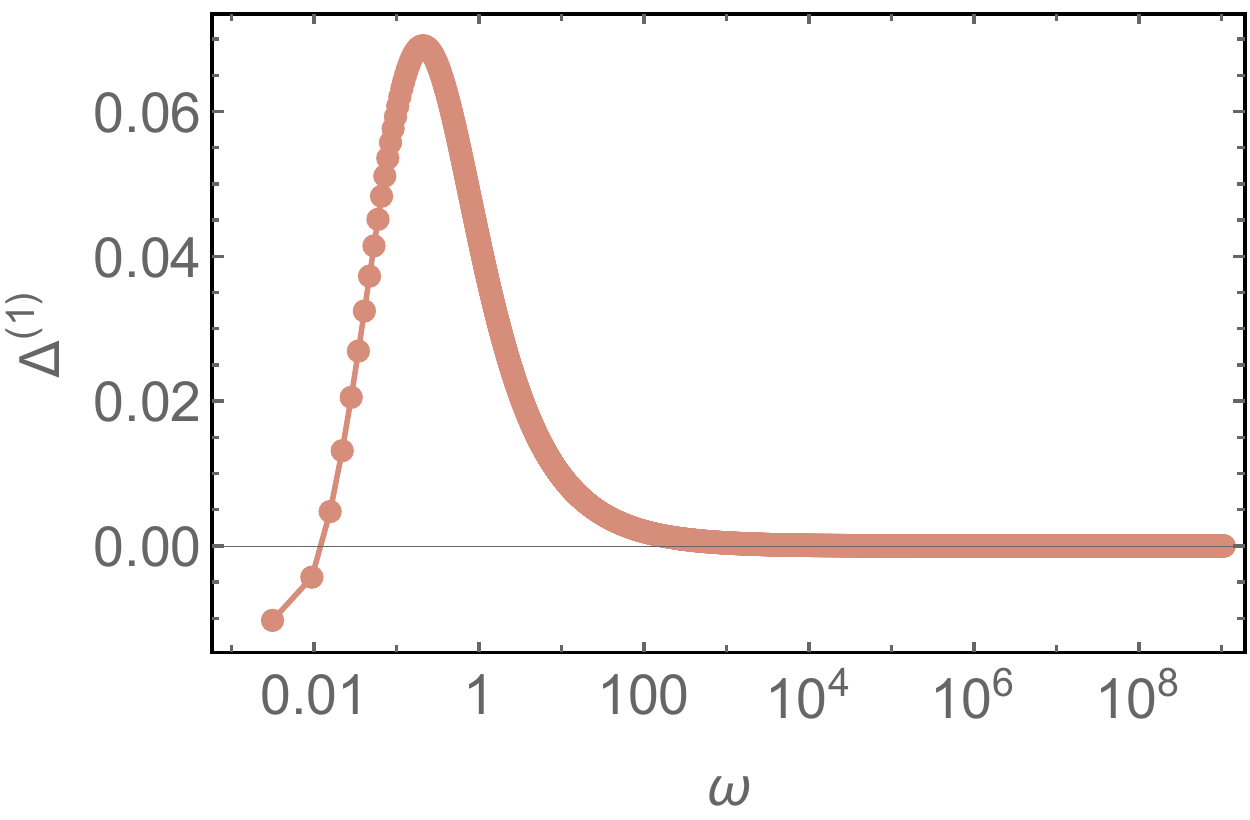}
\includegraphics[width=.8\columnwidth]{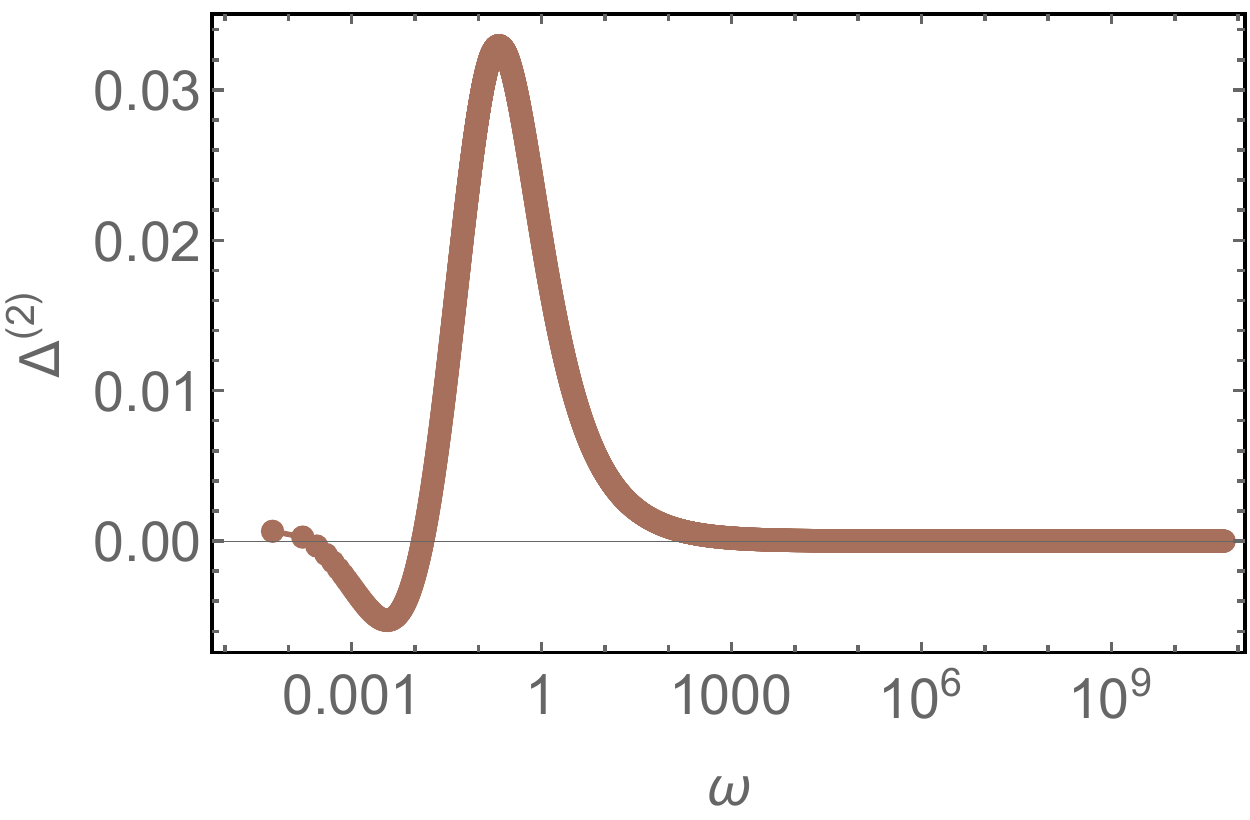}
\end{center}
\caption{The eigenfunction $\Delta(\omega_n)$ corresponding to the three largest $T_c^{(n)}$ for $\eta=0.68$ ($M\approx N$) and $\alpha=0$ ($T_c^{(0)}\approx5.835\times 10^{-2}g_0^2\omega_0$, $T_c^{(1)}\approx0.994\times 10^{-3}g_0^2\omega_0$ and $T_c^{(2)}\approx1.79\times 10^{-5}g_0^2\omega_0$). It is obtained at discrete Matsubara frequencies, lines are guides to the eye. The number of zeros increases for the different gap functions from zero to three.
}
\label{fig:phi}
\end{figure}

As a next step, we verify our reasoning for multiple solutions by extending the analysis of the gap equation to finite $T$.
If the candidate $\Phi_{nl}^{(n)} (\omega)$ exist, they must give rise to a set of critical temperatures $T^{(n)}_{c}\propto \Phi_{nl}^{(n)} (0)$, which all must emerge simultaneously once $\alpha$ becomes smaller than $\alpha_c$.
We illustrate this in Fig.~\ref{fig:Tcschematic}. Below we present strong numerical evidence that for any finite $\eta$ there indeed exists a set of  $T^{(n)}_c$.
We obtain these temperatures by solving the gap equation Eq.~\eqref{eq:gap_1} as an eigenvalue problem utilizing the hybrid-frequency technique introduced in Ref.~\onlinecite{PhysRevB.102.024525} to perform the Matsubara sum down to very low T.
As an input for these calculations, we use the expressions for the electron self-energy and boson polarization that we have determined in Sec.~\ref{sec:2}. 
As we have noted previously, the explicit appearance of the self-energy is special to the YSYK model and rooted in its zero-dimensionality.  We consider separately the pairing out of the SYK regime and out of the impurity regime, and obtain the first few 
 $T^{(n)}_c$ as a function of $M/N$ and the coupling $g_0$.

\subsubsection{Pairing in the SYK regime}
\label{sec:3_d}
We start with the analysis of the gap equation \eqref{eq:gap} in the SYK regime for various values of $\eta$ ($M/N$). We use  Eq.~\eqref{eq:Sigmaqu_1} for the electron self-energy and Eqs.~\eqref{eq:deltaPi_1} and \eqref{eq:mass} for the boson propagator. In explicit form the equation for $\Delta (\omega_n) = \Delta_n$ is
  \begin{align}
\Delta_n&\approx\frac{1}{\pi t^{\frac{1+\eta}{2}}}\sum_m \frac{\sgn{\omega_m}}{|2m+1|\pi t^{\frac{1+\eta}{2}}+a_\eta(|2m+1|\pi)^{\frac{1-\eta}{2}}} \nonumber \\
&\times\frac{a_\eta^2}{c_\eta+b_\eta(2\pi)^\eta\abs{n-m}^\eta}\left[(1-\alpha)\frac{\Delta_m}{2m+1} - \frac{\Delta_n}{2n+1} \right]\,.
\label{eq:DeltaSYK}
\end{align}
 where $t = T/(\omega_0 g^2_0)$.

We solve this equation numerically and
indeed 
find a hierarchy of critical onset temperatures $T_c^{(0)}>T_c^{(1)}>T_c^{(1)}>\ldots$ for given $\alpha$ and $\eta$. We show $T_c^{(0)}$ and $T_c^{(1)}$ as function of $\alpha$ for an exemplary value of $\eta$ in Fig.~\ref{fig:Tcalpha} (see App.~\ref{app:SCT} for other $\eta$).
The critical temperatures are suppressed when $\alpha$ increases and within our numerical accuracy all vanish at $\alpha_c$, in agreement with our analytical analysis at $T=0$. The computational requirements to reach $T_c^{(n)}\propto \exp(-\sqrt{\pi} n/\sqrt{\alpha_c-\alpha})$ near $\alpha_c$ drastically increase as an exponentially large number of Matsubara points $m_{max}\gtrsim \omega_0/T_c^{(n)}$ needs to be kept \cite{PhysRevB.102.024525}.
 Although the existence of $\alpha_c$ is a zero-temperature result, the very suppression of $T_c^{(n)}$ with $\alpha$
 can be
 traced to the term with $\omega_n=\omega_m$ in Eq. (\ref{eq:DeltaSYK}). This term cancels out in the r.h.s. of Eq.~(\ref{eq:gap}) at $\alpha =0$, but does not cancel at $\alpha\neq 0$. Because $\alpha$ measures the strength of the disorder that breaks time-reversal symmetry, the suppression of $T_c^{(n)}$ with $\alpha$ in the YSYK model at $\alpha \neq 0$ can be considered a NFL counterpart of the suppression of $T_c$ by pair-breaking magnetic impurities in Abrikosov-Gor'kov theory of pairing in a dirty Fermi liquid~\cite{ag}

In Fig.~\ref{fig:phi}, we show the gap functions $\Delta_n$ with $n=0,1,2$, corresponding to the three largest onset temperatures.
We see that $\Delta_n$ changes sign $n$ times as a function of Matsubara frequency, again in agreement with our zero-temperature analysis.

\begin{figure}[t]
\begin{center}
\includegraphics[width=.8\columnwidth]{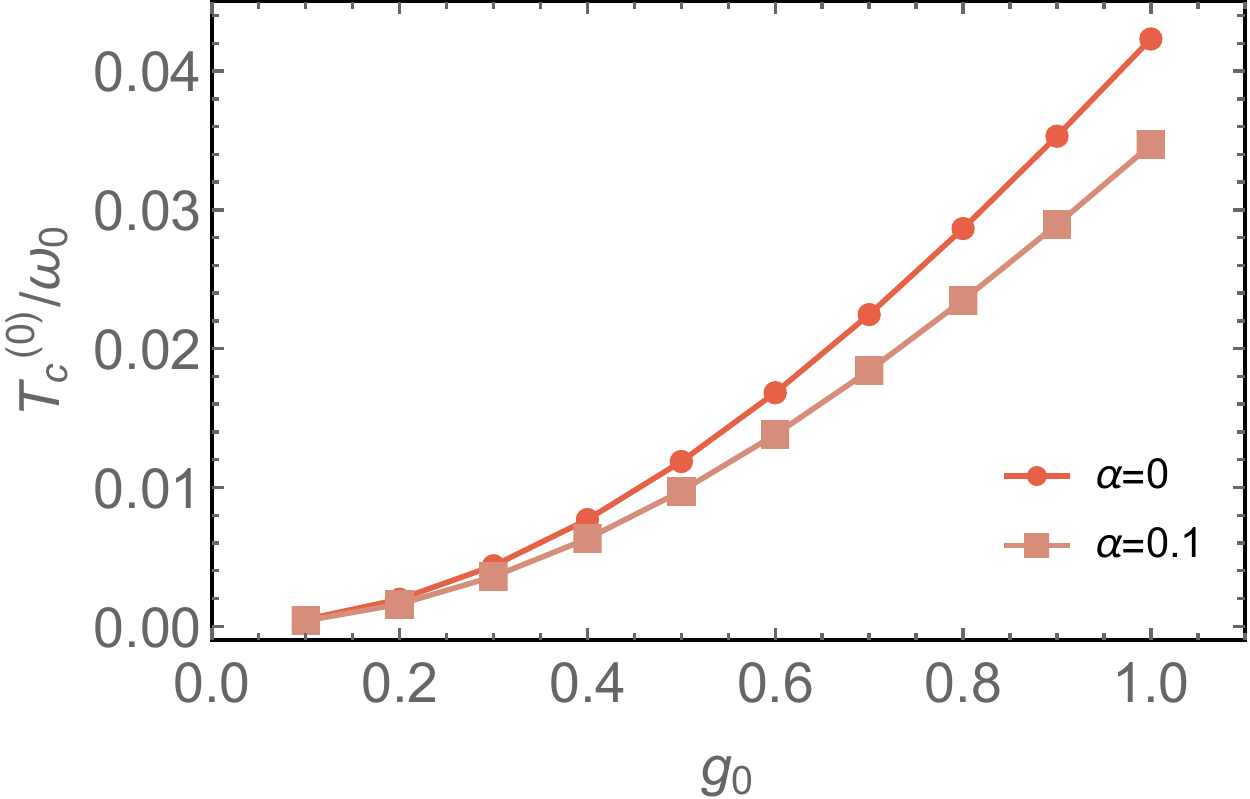}\\
\end{center}
\caption{The largest onset temperature $T_c^{(0)}$ for pairing out of the SYK regime for $\eta=0.3$ and small $\alpha$. It follows a quadratic dependence on the coupling $g_0$.
}
\label{fig:Tcg0SYK}
\end{figure}

\begin{figure}[t]
\begin{center}
\includegraphics[width=.8\columnwidth]{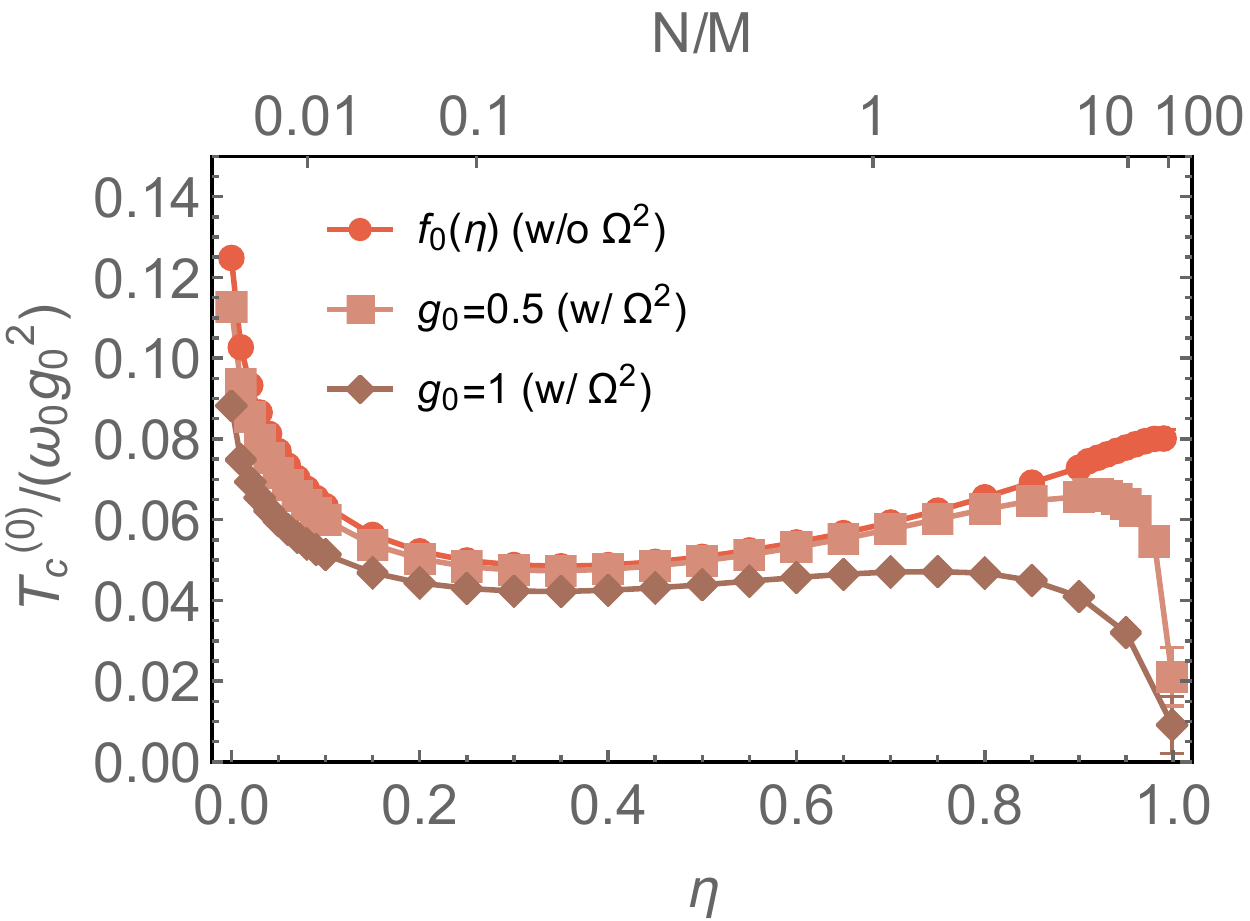}\\
\vspace{.5\baselineskip}
\includegraphics[width=.8\columnwidth]{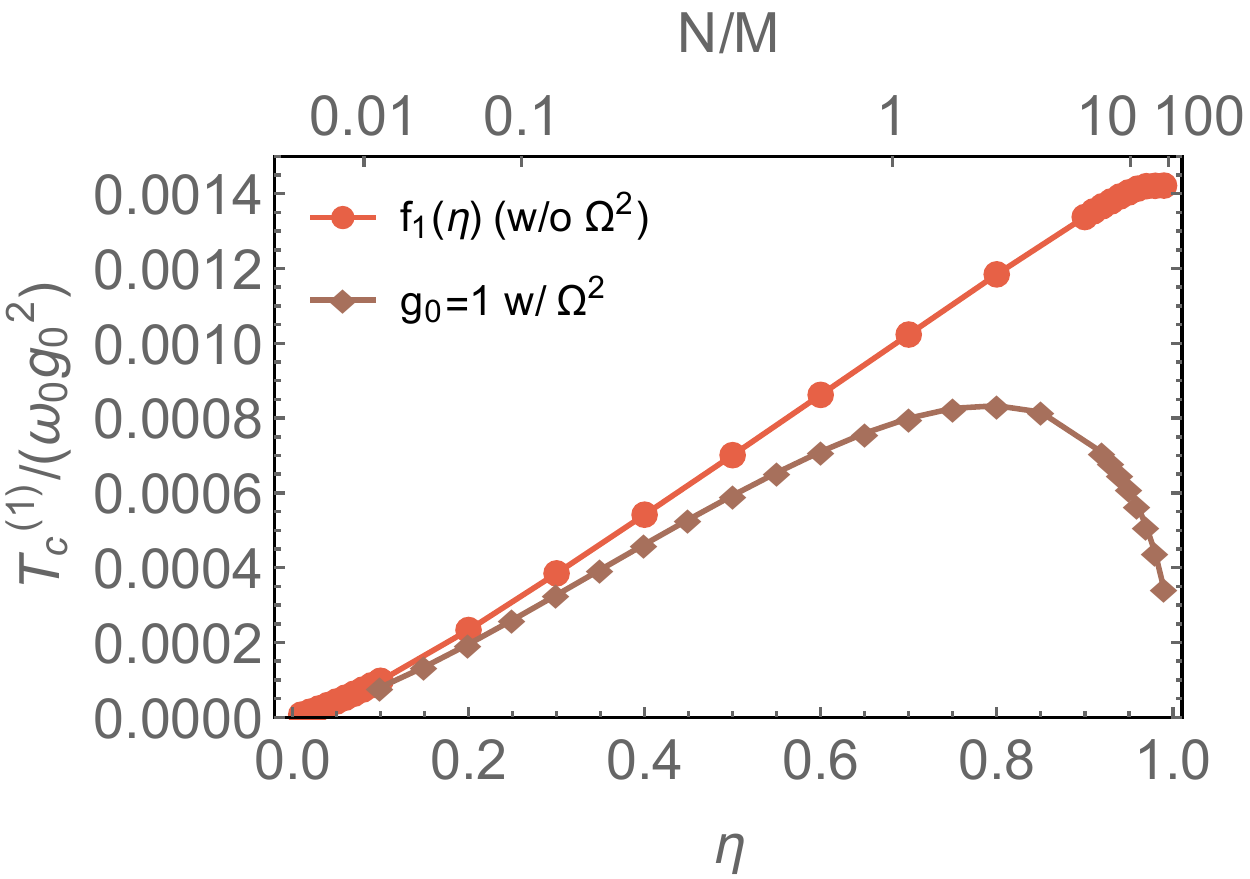}
\end{center}
\caption{The largest $T_c^{(0)}$ (top) and second largest $T_c^{(1)}$ (bottom) critical temperature for $\alpha=0$ and varying $\eta$ in the SYK regime, where $T_c^{(n)}=f_n(\eta)\omega_0g_0^2$ (see Eq.~\eqref{eq:feta}) as long as the bare $\Omega^2$-term in the boson propagator is negligible. For larger $g_0$ or $\eta\to1$, $T_c^{(n)}$ starts to deviate.
}
\label{fig:Tceta}
\end{figure}

Further, it follows from Eq.~(\ref{eq:DeltaSYK}) that the onset temperatures in the SYK regime scale with $\omega_0 g_0^2$,
 as only the combination $t=T/(\omega_0g_0^2)$ appears,
 i.e.,
\be
T_c^{(n)}=f_n(\eta) \omega_0 g_0^2
\label{eq:feta}
\ee
where $f_n(\eta)$ does not depend on the coupling.
The corrections to Eq.~(\ref{eq:feta}) come from the bare $\omega^2_m$-term in the boson propagator, which we neglected in Eq.~(\ref{eq:DeltaSYK}). For small $\eta$, this approximation is justified (see  Sec.~\ref{sec:2_c}), but for larger $\eta$ (or large couplings), the bare $\omega^2_m$-term cannot be neglected. In Fig.~\ref{fig:Tcg0SYK}, we show our numerical solution of the linearized gap equation with the full bosonic propagator for $\eta =0.3$. We clearly see
  the quadratic dependence on the coupling as in Eq.~(\ref{eq:feta}). In Fig.~\ref{fig:Tceta}, we show $T_c^{(n)}/(\omega_0 g_0^2)$ along with $f_n(\eta)$ for $n=0,1$. We see that for small $\eta$, $T_c^{(n)}$ follows Eq.~(\ref{eq:feta}), but for larger $\eta$ the critical temperature deviates. 

We also see from Fig. ~\ref{fig:Tceta} that for small
and
 intermediate $\eta$, $T_c^{(0)}$ and $T_c^{(1)}$ behave differently as functions of $\eta$: $T_c^{(0)} \sim \omega_0 g^2_0$ is almost independent on $\eta$ and remains finite at $\eta =0$, while $T_c^{(1)}$ decreases approximately linearly with $\eta$. This agrees with our zero-temperature analysis, where we argued that the characteristic scale for $\Delta_{n=0}$ is independent on $\eta$ while for $\Delta_{n=1}$ the corresponding scale is proportional to $\eta$.
Note that the prefactor in $T^{(0)}_c \sim \omega_0 g^2_0$ at small $\eta$, obtained using Eq. (\ref{eq:DeltaSYK}),
is correct only by order of magnitude as at such temperature the system is in the free-fermion regime, where $T^{(0)}_c$ is given by Eq. (\ref{eq:ap_5}).  The temperature $T^{(1)}_c$ and all other $T^{(n)}_c$ with $n\geq 2$ remain inside the SYK regime. We emphasize that the vanishing of all $T_c^{(n)}$ with $n\geq 1$ at $\eta=0$ implies the necessity of a dynamical pairing interaction in order to get an infinite set of pairing solutions.

\subsubsection{Pairing in the impurity regime}
\label{sec:3_e}

\begin{figure}[t]
\begin{center}
\includegraphics[width=.8\columnwidth]{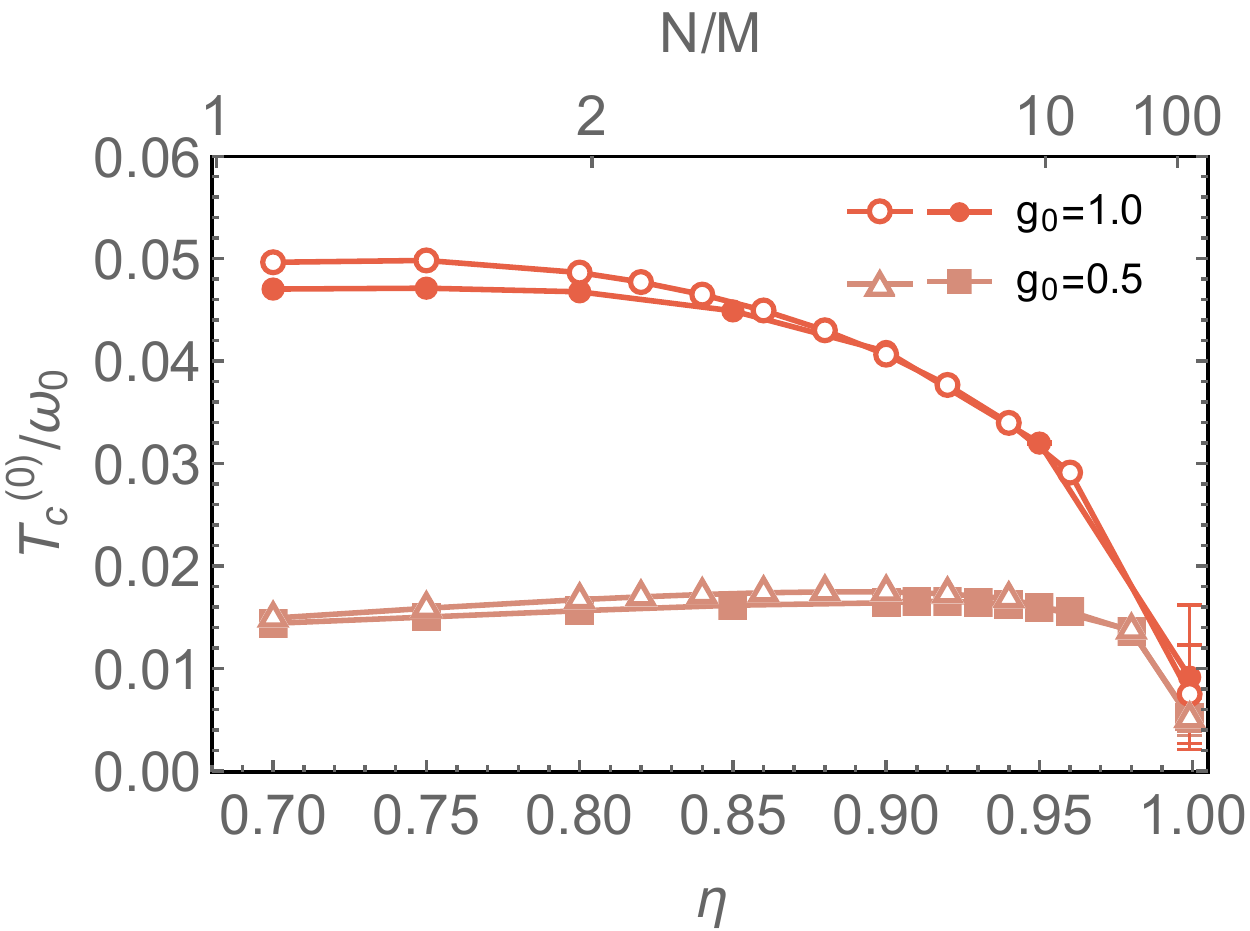}\\
\vspace{.5\baselineskip}
\includegraphics[width=.8\columnwidth]{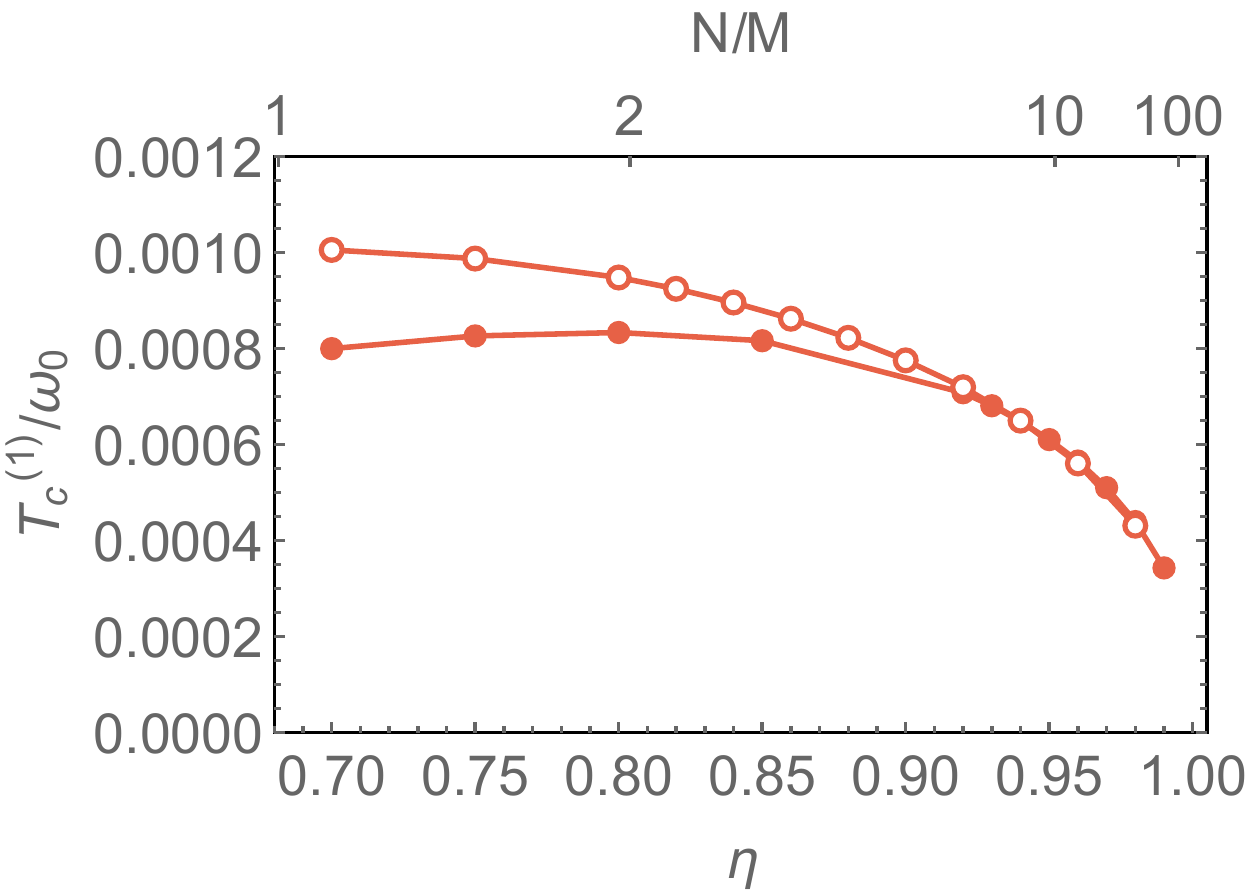}
\end{center}
\caption{The largest (top) and second largest (bottom) critical temperature for $\alpha=0$ and varying $\eta$ obtained with the SYK (filled symbols) and impurity (open symbols) self-energy for different $g_0$ (bare $\Omega^2$-term included).
}
\label{fig:Tcquimp}
\end{figure}

Next, we analyze the gap equation using as input the expressions for the self-energy and the bosonic polarization in the impurity regime, Eqs.~\eqref{eq:mi}-\eqref{eq:Piimp}. As we demonstrated in Sec.~\ref{sec:2_d}, the impurity regime appears at large coupling $g_0$ and/or at large $N/M$.
The linearized gap equation in the impurity regime has the form
  \begin{align}
&\Delta(\omega_n)=g_0^2\omega_0^3T\sum_{m}\left[ (1-\alpha)\frac{\Delta_m}{\omega_m}-\frac{\Delta_n}{\omega_n} \right] \frac{\text{sgn}(\omega_m)}{\abs{\omega_m}+\frac{2}{\pi}\frac{N}{M}g_0^2\omega_0} \nonumber\\
&\hspace{1.1cm}\times\frac{1}{(\omega_n-\omega_m)^2+\frac{\pi}{2}\frac{M}{N}\frac{\omega_0}{g_0^2}
\abs{\omega_n-\omega_m}+\left(\frac{\pi}{2}\frac{M}{N}\right)^2\frac{\omega_0}{g_0^2}T}
\label{eq:gapimp}
\end{align}
We first 
recall
that we found in Sec.~\ref{sec:2_d} that at large $N/M$, the critical behavior in the impurity regime is almost indistinguishable from that in the SYK regime. Hence, $T_c$, obtained using Eqs. (\ref{eq:DeltaSYK}) and (\ref{eq:gapimp}) must coincide. In Fig.~\ref{fig:Tcquimp}, we compare $T^{(0)}_c$ and $T^{(1)}_c$ obtained from
 Eqs.~
 (\ref{eq:DeltaSYK}) and (\ref{eq:gapimp}). We see that for both $n=0$ and $n=1$, the critical temperatures obtained from the two gap equations approach each other for $\eta\rightarrow 1$, which, we remind, corresponds to large $N/M \approx 1/(1-\eta)$. We emphasize that a set of $T^{(n)}_c$ emerges in both SYK and impurity regimes.

\begin{figure}[t]
\begin{center}
\includegraphics[width=.8\columnwidth]{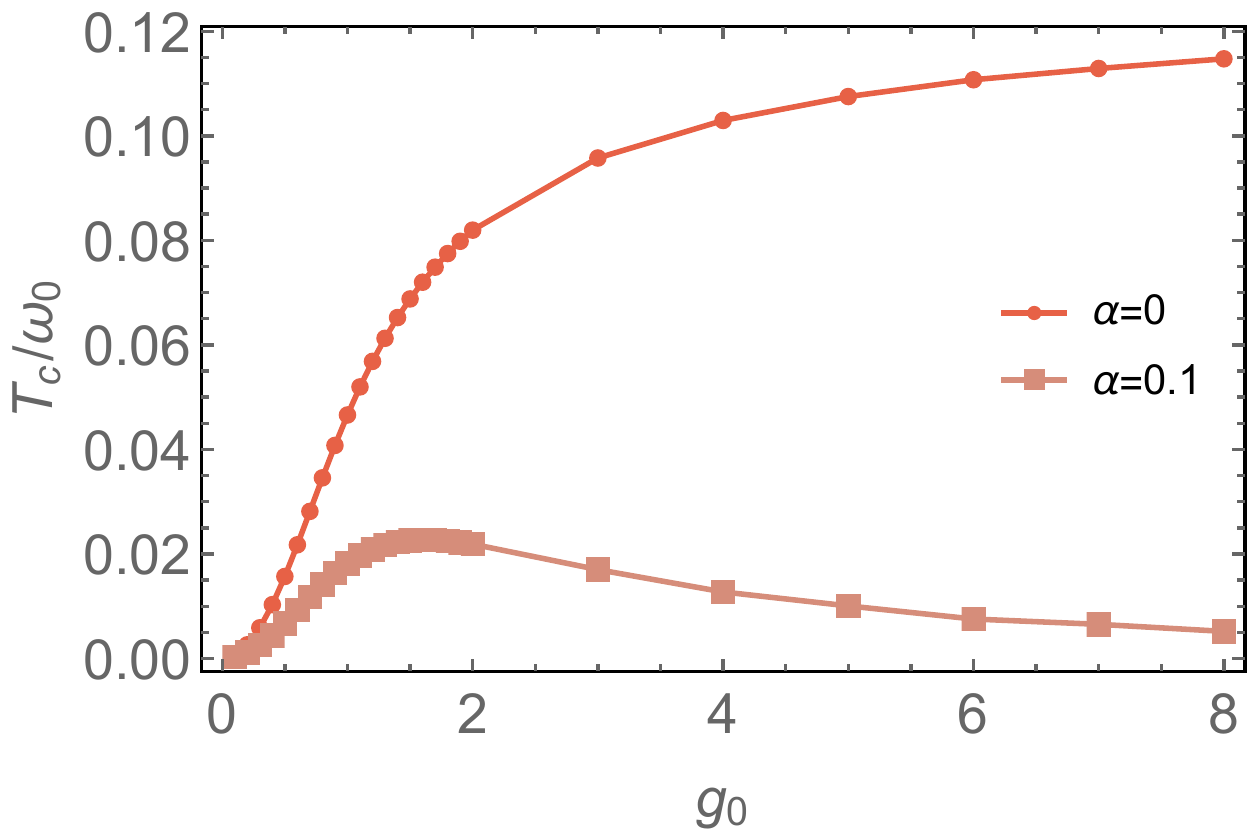}
\end{center}
\caption{The largest critical temperature $T_c^{(0)}$ as function of the coupling strength $g_0$ for $\eta=0.8$ ($N/M\approx 2$). The data is obtained using the SYK expressions as input in the gap equation. However, the expressions of the impurity regime are comparable for this $\eta$ (see Fig.~\ref{fig:Tcquimp} and Sec.~\ref{sec:2_d}). For small $g_0$, $T_c^{(0)}$ increases quadratically. For large $g_0$, the behavior depends on the presence of pair-breaking disorder. If $\alpha=0$, $T_c^{(0)}$  keeps increasing and then saturates, if $\alpha\neq0$ it rapidly drops down. }
\label{fig:Tcg0}
\end{figure}

Next, we analyze the form of $T^{(0)}_c$ in the impurity regime in more detail.  For simplicity, we call this temperature simply 
$T_c$. Deep in the impurity regime, at $Ng_0^2/M\gg1$,  we assume and then verify that for relevant $\omega_n, \omega_m \sim T_c$, the fermionic self-energy is much larger than the bare $\omega_m$, and the $(\omega_n-\omega_m)^2$ term is the largest in the bosonic propagator. Keeping only these terms, we reduce the gap equation (\ref{eq:gapimp}) to
 \begin{align}
&\Delta(\omega_n) \nonumber \\
&\approx \frac{1}{8\pi^2}\frac{M}{N} \frac{\omega_0^2}{T^2_c}\sum_{m}\left[ (1-\alpha)\frac{\Delta_m}{2m+1}-\frac{\Delta_n}{2n+1} \right]\frac{ \text{sgn}(\omega_m)}{(n-m)^2}\,.
\label{eq:eta1}
\end{align}
 For $\alpha =0$, Eq. (\ref{eq:eta1}) yields
\be
T_c \sim \omega_0\sqrt{M/N}
\ee
independent of the coupling constant $g_0$, i.e. $T_c$ is solely determined by the ratio of electron and boson flavors.  In the limit $N/M \gg 1$, this result can be equivalently expressed as $T_c \sim \omega_0 \sqrt{1-\eta}$.
Our numerical calculations using the full gap equation confirm these findings. Fig.~\ref{fig:Tcquimp} shows that $T_c$  drops at $\eta \to 1$, and Fig.~\ref{fig:Tcg0} shows that $T_c$ becomes independent on $g_0$ at large $g_0$ and fixed $N/M$.
 Using the explicit form of $T_c$,  we can also verify that the assumptions that led to Eq.~\eqref{eq:eta1} are valid.
  On a more careful look, we find that the coupling-independent behavior of $T_c$ is rooted in the electronic self-energy growing large when $g_0^2 N/M\gg 1$. We recall that such self-energy is generated by the thermal piece in the self-energy equation.

For $\alpha \neq 0$, $T_c$ extracted from (\ref{eq:eta1}) vanishes as the pair-breaking $n=m$ contribution diverges. Keeping the subleading term in the bosonic propagator in (\ref{eq:gapimp}) to regularize the divergence, we obtain a more accurate gap equation in the form
\begin{align}
&\Delta(\omega_n)\approx\notag \\
& \frac{1}{8\pi^2}\frac{M}{N} \frac{\omega_0^2}{T^2}\sum_{m\neq n}\left[ (1-\alpha)\frac{\Delta_m}{2m+1}-\frac{\Delta_n}{2n+1} \right]\frac{ \text{sgn}(\omega_m)}{(n-m)^2}\nonumber\\
&-\frac{2}{\pi^2}\alpha \frac{N}{M} g_0^2\frac{\omega_0}{T} \frac{\Delta_n}{|2n+1|}
\label{n_11}
\end{align}
Expanding in $\alpha$, we find that $\alpha$ reintroduces the dependence on the coupling and suppresses the critical temperature by
\be
\delta T_c/T_c \propto\alpha g_0^2 (N/M)^{3/2}\,.
\ee
For $g_0 = O(1)$, the corrections due to $\alpha$ are amplified by large $(N/M)^{3/2}$.
We compare $T_c$ for $\alpha=0$ and a representative $\alpha\neq0$ in Fig.~\ref{fig:Tcg0}. We clearly see that for $\alpha =0$, $T_c$ saturates at large $g_0$, while for $\alpha \neq 0$, $T_c$  decreases with increasing $g_0$, in line with our analytical reasoning.

We note that for $\alpha =0$, the gap equation (\ref{n_11}), or, equivalently (\ref{eq:eta1}), is the same as for the well-studied case of electron-phonon pairing in the limit of vanishing Debye frequency. We discuss this case further in Sec. \ref{sec:4} below in light of the comparison with the $\gamma$-model.
 In this case, the set of solutions $\Delta_n$ becomes continuous and strong fluctuations are expected.

\subsubsection{Phase diagram including superconductivity}

\begin{figure*}[t]
\begin{center}
\includegraphics[trim = 160mm 80mm 180mm 60mm,clip,scale=0.17]{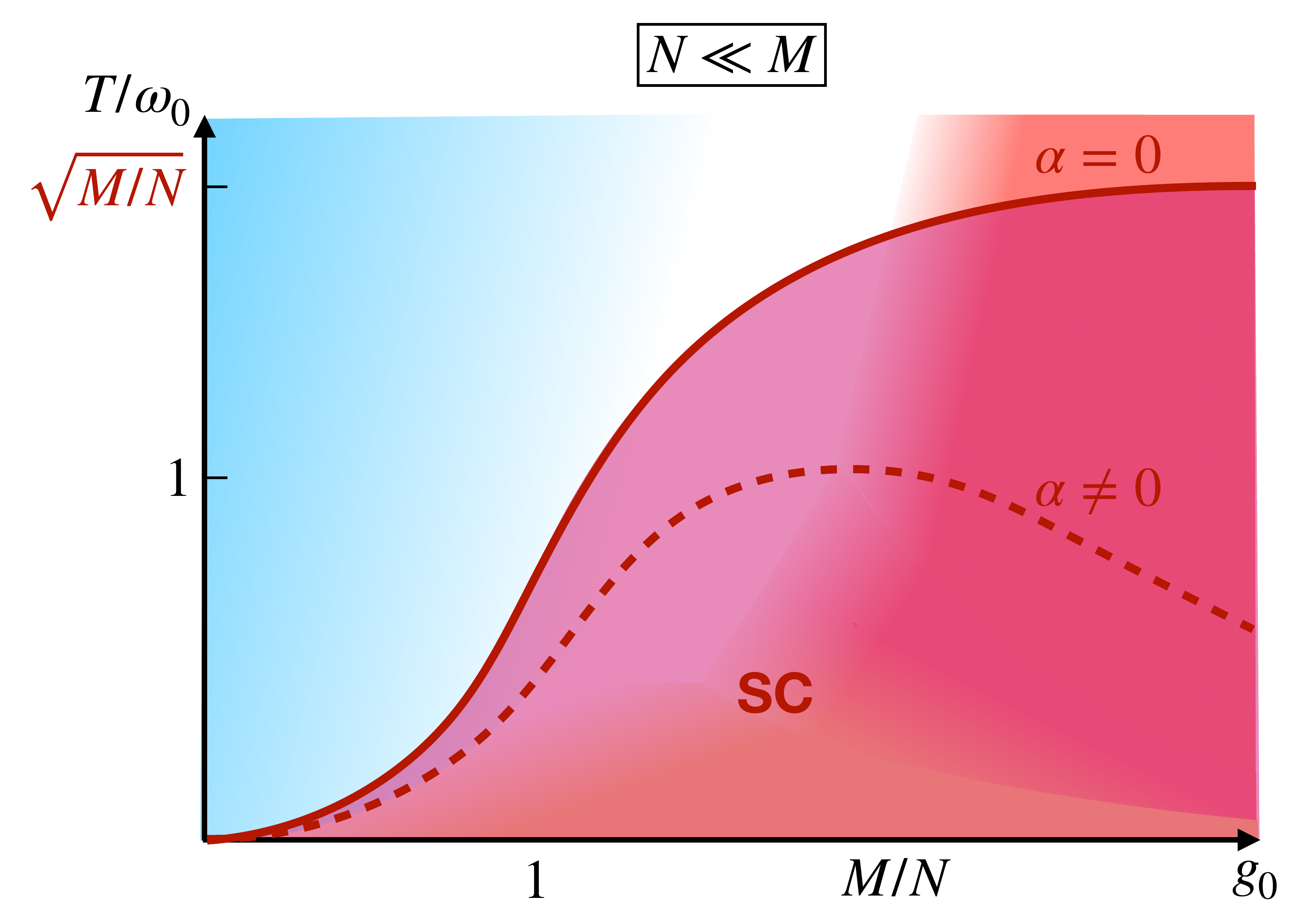}
\includegraphics[trim = 160mm 80mm 180mm 60mm,clip,scale=0.17]{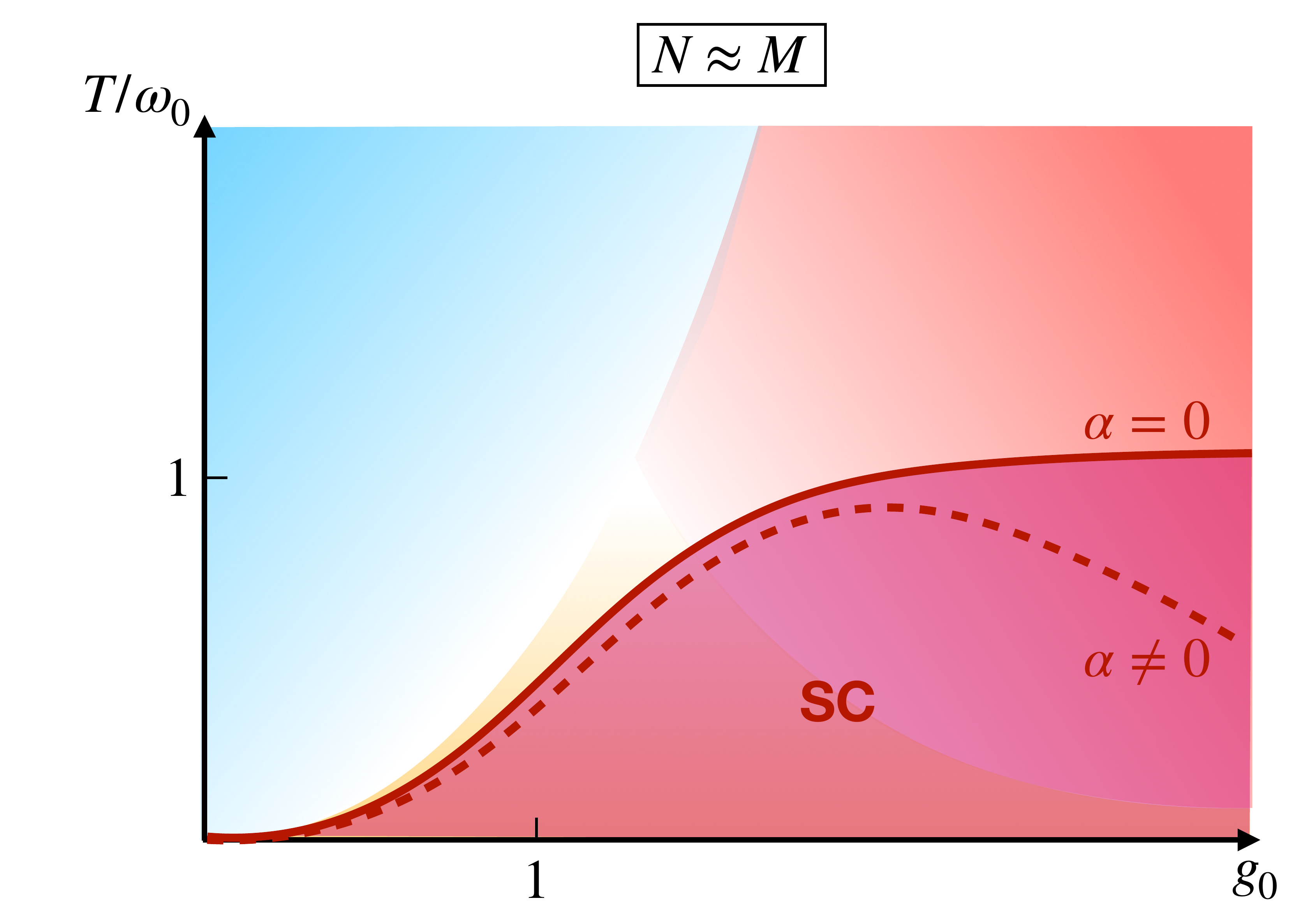}
\includegraphics[trim = 160mm 80mm 180mm 60mm,clip,scale=0.17]{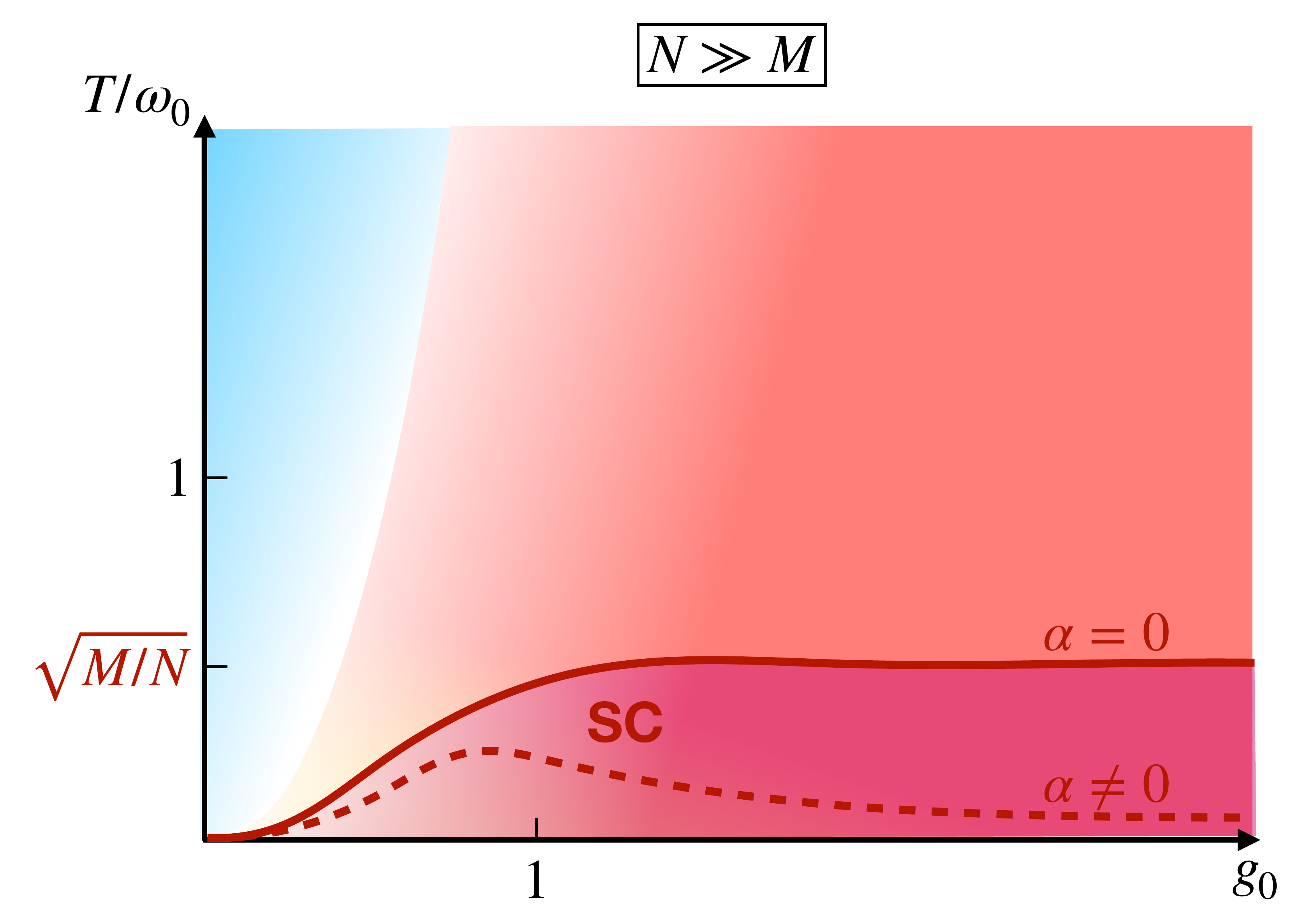}
\end{center}
\caption{
Schematic phase diagrams for $N\ll M$ (left), $N=M$ (middle) and $N\gg M$ (right) including superconductivity. The background coloring corresponds to the normal-state phase diagrams in Fig.~\ref{fig:pd}. The dark red line marks the largest onset temperature for pairing without (solid) and with (dashed) TRS-breaking disorder. Below this line, a pairing state replaces the different NFL phases of the normal-state phase diagram. We do not address here the if the pairing of dispersion-less fermions leads to a superconducting condensate.
}
\label{fig:pdSC}
\end{figure*}

Combining the results of the previous sections, we obtain how the onset temperature of pairing evolves as function of temperature and coupling for different $N/M$. We focus on the largest $T_c^{(0)}=T_c$ and consider the evolution of $T_c$ in the three exemplary cases $N\ll M$, $N\approx M$ and $N\gg M$. We show the results in Fig.~\ref{fig:pdSC}. The behavior for $N\approx M$ (the middle panel in Fig.~\ref{fig:pdSC}) is consistent with Ref.~\onlinecite{HAUCK2020168120}.

Let us first discuss the common features for very small or very large $g_0$.
Generally, for small $g_0$ and $\alpha=0$, we have found that $T_c$ increases as $\omega_0 g_0^2$. At $\alpha \neq 0$, the initial slope is reduced by $1-\alpha$. This is true when superconductivity develops out of the free-fermion or out of the SYK regime.  The combination of our analytical and numerical results shows that for $N\ll M$,  $T_c$ is inside the free-fermion regime, for  $N\approx M$, $T_c$ is parameter-wise at the boundary between the free-fermion and the SYK regimes, but numerically well inside the SYK regime, and for $N\gg M$, $T_c$ is deep inside the SYK$\sim$impurity regime.

For very large $g_0$, pairing always develops out of the impurity regime. In this case, the onset temperature $T_c \sim \omega_0 \sqrt{M/N}$ saturates at a coupling-independent value if $\alpha=0$. The ratio $T_c/\omega_0$ is parametrically large for $N\ll M$ and small for $N\gg M$. If $\alpha\neq0$, $T_c$ is reduced by a correction that scales as $\alpha g^2_0$.

The behavior at intermediate couplings changes qualitatively depending on $N/M$. For $N\approx M$, the change from small- to large- coupling behavior occurs around $g_0\approx 1$, when the SYK regime crosses over to the impurity regime. For $N\gg M$, where SYK and impurity regimes are indistinguishable, it occurs  around smaller $g_0\approx (M/N)^{1/4}$ when $\delta\Pi(\Omega)\sim\Omega^2$. For $N\ll M$, the evolution of $T_c$  depends on $\alpha$. For $\alpha =0$, $T_c$ passes through the intermediate range I, where it increases from $T_c \sim \omega_0$
  at $g_0 = O(1)$  to $T_c \sim\sqrt{M/N}\omega_0 \gg \omega_0$ at the boundary of the impurity regime at
  $g_0\sim (M/N)^{3/4}$. The form of $T_c$ in regime I roughly follows $T_c \sim \omega_0 g^{2/5}_0$.
For $\alpha \neq 0$, pair-breaking effects reduce  $T_c$ for $g_0 >1$ to  $T_c \sim \omega_0/\sqrt{\alpha} f (\alpha g^{4/5}_0)$, where $f(x \ll 1) \sim \sqrt{x}$ and $f(x \gg 1)$ is finite.
  In this case, $T_c$ passes through regime II 
  before entering the impurity regime.

Finally, let us note that, even if the largest onset temperature $T_c = T_c^{(0)}$ is outside of the SYK regime, 
 the $T_c^{(n)}$ of other pairing states are much smaller and
 can remain within the SYK regime. When $\alpha$ approaches $\alpha_c$, all $T^{(n)}_c$ emerge out of the SYK regime.

\section{A comparison with the $\gamma$ model}
\label{sec:4}

It is instructive to compare the results that we obtained for the YSYK model with the ones for a set of models of
\emph{dispersion-full} fermions with a large bandwidth brought to a QCP by some external parameter (see e.g., the list of literature in Ref. \cite{gammamodel1}). Like in YSYK models, it is assumed that the same interaction, mediated by a soft critical boson, gives rise to a NFL and to pairing with spatial symmetry specific to a given  QCP (e.g., a $d-$wave at an antiferromagnetic QCP).

For a subset of such models, it is further assumed that bosons are slow modes compared to fermions, for one reason or the other.  In this situation, one can neglect vertex corrections and obtain the set of closed equations for the fermionic self-energy, the pairing vertex, and the bosonic polarization operator, similar to Eqs.~(\ref{eq:eqs}), but with additional integrals over momentum:
\begin{align}
\Sigma(\omega_m,k)&=i  T {\bar g}^2 \sum_{\omega'_m} \int \!\!\! \frac{d^2 k'}{(2\pi)^2}
G(\omega'_m,k')D(\omega_m-\omega'_m, k-k') \nonumber \\
\Pi(\Omega_m,q)&=2 {\bar g}^2
 T \sum_{\omega'_m} \int \!\!\!\frac{d^2 k'}{(2\pi)^2}
  \left[G(\omega'_m, k')G(\omega_m+\omega'_m, k+k')\right.\nonumber\\
&\left. -F^*(\omega'_m,k')F(\omega_m+\omega'_m,k+k') \right] \nonumber \\
\Phi(\omega_m,k)&= \nonumber \\
&- {\bar g}^2
T \sum_{\omega'_m} \int \!\!\! \frac{d^2 k'}{(2\pi)^2} F(\omega'_m, k')  D(\omega_m-\omega'_m, k-k')
\label{eq:eqs_1}
\end{align}
where
$G(\omega_m, k)=- (i(\omega_m +\Sigma(\omega_m) + \epsilon_k)/[(\omega_m +\Sigma(\omega_m))^2+\abs{\Phi(\omega_m)}^2 + \epsilon^2_k]$,
   $F(\omega_m, k)=-\Phi(\omega_m)/[(\omega_m +\Sigma(\omega_m))^2+\abs{\Phi(\omega)_m}^2 + \epsilon^2_k]$, and
   $D(\omega_m, k) =1/(\omega^2_m +  \omega^2_0 + (k-k_0)^2 +\Pi(\omega_m, k))$. In these expressions
   $\epsilon_k$ is a fermionic dispersion and $k_0$ is the momentum with which density or spin order develops.
    For a metallic QCP, low-energy physics comes from fermions near the Fermi surface, for which
    $\epsilon_k = v_F (k-k_F)$, and the dominant contribution to $\Pi(\omega_m, k)$ comes from Landau damping of a boson into a particle-hole pair.
There is no self-consistent tuning to a QCP,  i.e., the dressed mass vanishes only for a certain bare $\omega_0$.

The condition that bosons are slow compared to fermions further allows one to compute $\Pi (\omega_m, k)$, substitute it into the two other equations, select the pairing channel, and integrate over momentum components by factorizing this integration into integration over the component transverse to the Fermi surface (i.e., over $\epsilon_k$) in the fermionic propagator and over the component along the Fermi surface in the bosonic propagator, whose momentum is confined to be between points on the Fermi surface. This gives rise to a set of two coupled equations for the "local" fermionic self-energy $\Sigma (\omega)$ and the "local" pairing vertex $\Phi (\omega)$. The resulting local theory is not exact as in several cases momentum dependence can be integrated out only if one neglects logarithmic corrections~\cite{acs,PhysRevB.82.075128,sslee_2018,Wang2013}, but it captures 
  power-law frequency scaling in the normal state and the interplay between NFL and superconductivity.

The equations for $\Sigma (\omega)$ and $\Phi (\omega)$ have the same form for different metallic models and differ only in the form of an effective dynamical interaction between fermions $V(\Omega_n)$.  At a QCP, $V(\Omega_n)$ is singular at $T=0$ and scales as $1/|\Omega_n|^\gamma$, where the value of $\gamma$  is model-specific, e,g., $\gamma =1/2$ for an antiferromagnetic QCP, $\gamma =2/3$ at a QCP towards Ising-nematic order, and so on (see Ref. \cite{gammamodel1}
for the list of $\gamma$ for different models). This set of models has been nicknamed "the $\gamma$-model". We use this abbreviation and label the corresponding self-energy and pairing vertex as $\Sigma_\gamma$ and $\Phi_\gamma$. Note that there is no $N/M$ ratio in (\ref{eq:eqs_1}), hence no  analog of Eq. (\ref{eq:eta}). Instead, different values of the exponent $\gamma$ come from different underlying microscopic models.

 In the normal state, the equation for the self-energy in the $\gamma-$model takes the form
\begin{align}
 \Sigma_{\gamma}(\omega_n)&=g^\gamma \pi T \sum_{m}\frac{\text{sgn} (\omega_m)}{\left(|\omega_n-\omega_m|^2 + m^2_\gamma(T)\right)^{\gamma/2}}
\label{n_6}
\end{align}
where $m_\gamma (0) =0$ and $g$ is the effective fermion-boson coupling ($g^\gamma \propto {\bar g}^2$).
It is similar to the corresponding Eq.~(\ref{eq:EliashbergSigma}) in the YSYK model, modulo a different definition of $m_\gamma$, but without the electron propagator $G(\omega')=[\omega'_m + \Sigma (\omega')]^{-1}$ in the r.h.s. of (\ref{n_6}).
The difference can be traced back to additional integration over spatial momenta in the $\gamma$-model.
Despite the difference, the self-energy $\Sigma_\gamma$ at $T=0$ has a power-law frequency dependence,
 $\Sigma_{\gamma}(\omega) \propto \text{sgn} (\omega) |\omega|^{1-\gamma}$, like in YSYK models.
Furthermore, at a finite $T$ and larger coupling $g$, thermal fluctuations become dominant and the system also crosses over to the impurity regime.
The thermal self-energy in this impurity regime becomes similar to the one in the YSYK model in a parameter range where one has to abandon the factorization of momentum integration for the thermal piece and integrate over both components of momentum ${\bf k}$ in the bosonic propagator.

Our goal in this section is to compare the onset of pairing in the two models. For this purpose, it is instructive to compare the YSYK model with the extended $\gamma-$model, in which the interaction in the particle-particle channel has an extra factor $1/N$ compared to that in the particle-hole channel.
 \footnote{This can be rigorously justified by extending the original $\gamma-$model to a matrix SU$(N)$ model.
  In Refs.~\onlinecite{gammamodel1,gammamodel2,gammamodel3,gammamodel4,gammamodel5,PhysRevB.99.144512,PhysRevB.99.180506,PhysRevB.99.014502,PhysRevLett.117.157001,PhysRevB.95.165137,PhysRevB.103.155161,PhysRevB.102.045147,PhysRevB.97.054502,PhysRevB.96.144508} $N$ has been used as a phenomenological continuous parameter, which can have any value.}
 The linearized gap equation in the extended $\gamma$-model takes the form
  \begin{align}
 &\Phi_{\gamma}(\omega_n)= \nonumber \\
  &\frac{g^\gamma}{N}  \pi T \sum_{m} \frac{\Phi_{\gamma} (\omega_m)}{\abs{\omega_m + \Sigma_\gamma (\omega_m)}} \frac{1}{\left[(\omega_n-\omega_m)^2 + m^2_\gamma (T)\right]^{\gamma/2}}
 \label{n_9}
 \end{align}
 or, equivalently,
 \begin{align}
 \Delta_{\gamma}(\omega_n)&=  \frac{g^\gamma}{N}  \pi T \sum_{m} \left[\frac{\Delta_{\gamma} (\omega_m)}{\omega_m} - N
 \frac{\Delta_{\gamma} (\omega_n)}{\omega_n}\right]\nonumber \\
   &\times\frac{\text{sgn} (\omega_m)}{\left[(\omega_n-\omega_m)^2 + m^2_\gamma (T)\right]^{\gamma/2}}\,.
 \label{n_10}
 \end{align}

We first discuss the $T=0$ case, where the equation on $\Phi_\gamma (\omega)$ becomes
\begin{align}
\Phi_{\gamma}(\omega)&=\frac{1-\gamma}{2N}\int\!d\omega'\frac{\Phi_{\gamma}(\omega')}{\abs{\omega'}^{1-\gamma}
\abs{\omega-\omega'}^\gamma} \frac{1}{1 + (|\omega|/\omega_{0\gamma})^\gamma}\,,
\label{n_99}
\end{align}
 where $\omega_{0\gamma} = g/(1-\gamma)^{1/\gamma}$. At small frequencies $\omega \ll \omega_{0\gamma}$, this equation has the same form as Eq.~\eqref{eq:lingap1}. In the limit $\eta\rightarrow 1$ when $1/(\pi b_\eta)\rightarrow (1-\eta)$, we can even exactly identify $\eta$ with $\gamma$ and $(1-\alpha)$ with $1/N$.
The general solution of Eq.~\eqref{n_99} at small frequencies is a combination of two power-laws with complex exponents $\Phi _\gamma(\omega) \propto (1/|\omega|^{\gamma/2}) \cos{({\bar \kappa}_\gamma |\omega|/\omega_{0\gamma} + \phi_\gamma)}$, similar to Eq. (\ref{n_8}) of the YSYK model for $\alpha < \alpha_c$.  The parameter ${\bar \kappa}_\gamma$ is real for
$\gamma >0$ (up to $\gamma \approx 2.8$ for $N=1$). Like for the YSYK model, the appearance of an oscillating solution implies that the ground state is a superconductor. The  effect of $N \neq 1$ mimics the one from a finite $\alpha$ in the YSYK model. Namely,  the  oscillating solution for $\Phi_\gamma (\omega)$, which we associate with superconductivity, holds only for $N < N_{cr}$, while  for larger $N$, the system remains in a NFL state.  In this respect, $N_{cr}$ plays the same role as $\alpha_c$.

Furthermore, at small $\gamma$, one can reduce the integral equation (\ref{n_99}) to a differential equation and solve it for all $\omega$
\cite{gammamodel1},
like we did in Sec. \ref{sec:3_a}. This equation has a normalized solution, which behaves properly at large $\omega$.  Like for the YSYK model, this implies that there exists an infinite set of topologically different pairing states, specified by the number of vortices on the Matsubara axis. For the $\gamma$-model, this reasoning is corroborated by the finding of the exact solution of the original, integral linearized gap equation
\cite{gammamodel1}
 and by the detection of an infinite set of solutions of a non-linear differential equation
 \cite{PhysRevB.99.144512}.

 At finite $T$, the linearized gap equation for the $\gamma$-model has a solution at a discrete set of $T^{(n)}_c$, very similar to what we found for the YSYK model\cite{gammamodel2}. A distinction between the two models arises at $\gamma \to 0$, where the $\gamma$-model reduces to the unconstrained BCS model and $T^{(0)}_c$ tends to infinity, while in the YSYK model, $T^{(0)}_c$ remains finite at $\eta \to 0$ due to the additional $1/\omega'$
from the fermion propagator in the r.h.s. of the gap equation (\ref{eq:gap}), which ensures the convergence of the frequency sum.

The relation between $N$ and $\alpha$ is a bit more tricky\ at $T\neq 0$ than at $T=0$ and depends on how the extension to $N \neq 1$ is made. On a surface, $1/N$ still plays the same role as $1-\alpha$, i.e., the thermal $m=n$ contribution to the gap equation  (\ref{n_10}) in the $\gamma$-model vanishes at $N=1$ due to vanishing of the numerator, and does not vanish at $N \neq 1$.
Beneath the surface, there is a difference: $\alpha$  in the YSYK model is a physically-relevant parameter, which measures the magnitude of TRS-breaking disorder, while in the $\gamma-$model, the extension to $N \neq 1$ is a mathematical trick, and there are two points of view on how to treat the extension. The authors of Refs.~\onlinecite{PhysRevB.103.155161,PhysRevB.102.045147,PhysRevB.97.054502,PhysRevB.96.144508} treated $N$ as potentially physically-relevant parameter and used the extended $\gamma$-model with $N \neq 1$ as the point of departure for a finite-$T$ analysis. In this case, the onset temperature for the pairing  includes the contribution from thermal fluctuations, and it is essential to keep the mass term $m_\gamma (T)$  in the bosonic propagator.  Then $1/N$ and $1-\alpha$ play the same role, and the YSYK and the extended $\gamma$ model are
very similar
also at finite $T$. The authors of Refs.~\onlinecite{gammamodel1,gammamodel2,gammamodel3,gammamodel4,gammamodel5,
PhysRevB.99.144512,PhysRevB.99.180506,PhysRevB.99.014502,PhysRevLett.117.157001} argued that
the known physical QCP models with different $\gamma$ correspond to $N=1$, even when the pairing is in a non-s-wave channel~\cite{msv,acn},
and one should extend to $N \neq 1$ in a way that does not introduce new physics, not present in the $N=1$ model. These authors then departed from the original model with $N=1$, re-expressed the gap equation by eliminating the $m=n$ term in the frequency sum in Eq.~(\ref{n_10}), and only then extended the model to $N \neq 1$. In this case, there is no thermal contribution to $T_c$, and one can neglect the mass term in the bosonic propagator. For this last extension of the $\gamma-$model, a finite-$T$ analysis differs from the one in the YSYK model in that the largest $T^{(0)}_c$ remains finite for any $N$, and in particular, does not disappear at $N_{cr}$ with the smaller $T^{(n)}_c$.
The reason is that in the $\gamma-$model, $\Sigma_\gamma (\omega_n)$ vanishes at Matsubara frequencies $\omega_m  = \pm \pi T$. This numerically enhances $T^{(0)}_c$, but not the other $T^{(n)}_c$.
In the YSYK model, there is no such effect as $\Sigma (\pm \pi T)$ remains finite due to presence of the fermionic propagator in the r.h.s. of the equation for $\Sigma (\omega_n)$. This difference persists even between the models with $N=1$ and $\alpha =0$ as it is a consequence of different NFL behavior in the normal state (where $\alpha$ and $N$ do not appear in the Eliashberg equations).

Finally, we note that for $\alpha =0$,  the gap equation in the YSYK model in the impurity regime Eq.~(\ref{eq:eta1}) is equivalent to the gap equation in the $\gamma$-model for $\gamma =2$ and $N=1$.  For this model, the highest onset temperature of pairing $T^{(0)}_c$ is finite, yet at $T=0$  the set of gap functions $\Delta_n (\omega)$ becomes continuous rather than discrete \cite{gammamodel5}. The authors of Ref.~\onlinecite{gammamodel5} argued that
 in this case fluctuations between different solutions destroy a superconducting order down to $T=0$, and at $0< T < T^{(0)}_c$ the systems displays pseudogap behavior associated with preformed pairs.

\section{Conclusions}
\label{sec:5}

 In this communication we have addressed the interplay between fermionic incoherence and pairing in the YSYK model consisting of $N$ fermions in a quantum dot, randomly coupled to $M$ bosons via a disorder-induced complex interaction. Such a system can be viewed as a toy model to study quantum-critical phenomena in fermionic systems with a bandwidth that is smaller than the strength of the interaction.
In the YSYK model, the same coupling is responsible for NFL behavior and an attractive pairing interaction, similar to what happens in a conventional metal with finite bandwidth brought to a QCP. In contrast to conventional scenarios, no fine-tuning is required to induce quantum-critical behavior. Instead, the YSYK model can reach a quantum-critical ground state and develop superconductivity for any values of bare parameters.
We have shown that the interplay between NFL behavior and superconductivity sensitively depends on the ratio
of fermion and boson flavors $N/M$, the (dimensionless) fermion-boson coupling $g_0$ and the strength of TRS-breaking disorder $\alpha$.

The starting point of our analysis is a set of coupled equations for the electron and boson self-energy, and the pairing vertex
(Eliashberg equations).
 These equations are obtained through disorder averaging, assuming replica-diagonal solutions, and become exact in the limit $N,M\rightarrow \infty$.
 We found and analyzed the solutions of these equations for arbitrary $N/M$, extending earlier results for $N=M$, and uncovered a rich normal-state phase diagram with several strongly-correlated regimes, some of which are not present at $N=M$.
At small temperatures and couplings, the self-consistent solution displays characteristic power-law frequency dependence in fermion and boson self-energies, $|\Sigma|=a_\eta g_0^{1+\eta}\omega_0^{(1+\eta)/2}|\omega|^{(1-\eta)/2}$ and $\delta\Pi=(b_\eta/a_\eta^2)g_0^{-2\eta}\omega_0^{2-\eta}|\omega|^\eta$, similar to the behavior in the original SYK model with random four-fermion interaction. The exponent $\eta$ is uniquely determined by the ratio $N/M$ with $\eta\to0$ for $N/M\to0$ and $\eta\to1$ for $N/M\to\infty$. At larger temperatures and/or couplings, thermal contributions induce another self-consistent, impurity-like solution with frequency-independent fermionic self-energy $|\Sigma_i|\sim \omega_0g_0^2 N/M$, which in turn gives rise to a term linear in frequency in the bosonic propagator $\delta\Pi_i\sim M/(g_0^2N)\omega_0|\Omega|$.
 The SYK and the impurity regimes have been detected in the earlier study of the case $N=M$ in Ref.~\onlinecite{PhysRevB.100.115132}.
We found new physics for $N \neq M$, which becomes most revealing when either $N \gg M$ or $M \gg N$.  For $N \gg M$, we found that SYK and impurity phases become almost identical (i.e., the fermionic self-energy has almost identical form in the two regimes). For $M \gg N$, we found additional intermediate regimes between the SYK, the impurity, and the free-fermion
 regime (in which fermionic and bosonic propagators are close to their bare values).

As a next step, we obtained the onset temperature for superconductivity, $T_c$, that develops out of the SYK or the impurity regime and investigated the impact of TRS-breaking disorder on $T_c$. At smaller values of the coupling, superconductivity sets in at the boundary between the SYK and the free-fermion regime, while for larger $g_0$, it develops out of the impurity regime.
In the first case, the onset temperature increases quadratically as function of $g_0$, and the prefactor depends on $\eta$ 
(and thus $N/M$) and on the strength of TRS-breaking disorder $\alpha$. The $\eta$-dependence is such that onset temperature remains finite at $\eta \to 0$ ($N\ll M$) and vanishes at $\eta\to 1$ ($N\gg M$).
TRS-breaking disorder suppresses the onset temperature, and it vanishes at a critical $\alpha_c$. We have shown that $\alpha_c$ monotonically decreases with increasing $\eta$ (increasing $N/M$). This means that the superconducting state becomes fragile for large $N/M$. When superconductivity develops out of the impurity regime, $T_c$ is independent of the coupling for TRS-preserving disorder ($\alpha =0$) and scales with $\sqrt{M/N}$. A finite $\alpha$ again suppresses $T_c$ and reintroduces the dependence on $g_0$ so that $T_c$ decreases for larger couplings.
 Generally, the superconducting state seems to be more robust (larger $T_c$ and larger $\alpha_c$) when it develops out of the SYK rather than the impurity regime.
 
We also found that the superconducting state is highly unconventional in that there is an infinite, discrete set of topologically different gap functions with $n$ sign changes on the Matsubara axis. They give rise to a set of onset temperatures $T_c^{(n)}$, which all emerge at the same critical $\alpha_c$ in an infinite-order BKT-transition. The different gap functions exist as long as the pairing interaction is dynamical.

Finally, we presented a detailed comparison of the YSYK model and the $\gamma$-model of dynamically interacting dispersion-full fermions near a conventional QCP. We focused on the pairing behavior and argued that it is very similar in the two models. In particular, the infinite set of topologically different pairing solutions arises in both models. Subtle distinctions appear outside of the low-energy regime due to a different role of bandwidth and of pair breaking. Overall, this comparison reveals that the deep relation between the YSYK model and quantum critical behavior of interacting fermionic systems extends beyond the NFL behavior in the normal state and involves superconducting properties.

\subsection*{Acknowledgments}
We thank Artem Abanov, Avraham Klein, Joerg Schmalian, Gonzalo Torroba, Yuxuan Wang, Yiming Wu, and Shang-Shun Zhang  for valuable discussions.
LC was supported by the U.S. Department of Energy (DOE), Office of Basic Energy Sciences, under Contract No. DE-SC0012704. AVC was  supported by the Office of Basic Energy Sciences, U.S. Department of Energy, under award DE-SC0014402.


\begin{appendix}
\bw
\section{T=0 analysis}
\subsection{Self-consistent solution in the normal state}
\label{app:normalSYK}
We start with the ansatz $\Sigma=A\sgn{\omega}\abs{\omega}^{(1-\eta)/2}$, which we use to calculate $\delta\Pi(\omega)=\Pi(\omega)-\Pi(0)$
\begin{align}
\delta\Pi(\omega)&=-2\bar g^2\frac{N}{M}\int\! \frac{d\omega'}{2\pi} \frac{1}{\Sigma(\omega')}\left[\frac{1}{\Sigma(\omega+\omega')}-\frac{1}{\Sigma(\omega')}\right]\nonumber\\
&=\frac{\bar g^2}{A^2}b_\eta\abs{\omega}^\eta
\label{eq:appdeltaPi}
\end{align}
with
\be
b_\eta=-\frac{4}{\pi^2}\frac{N}{M}\cos^2\left(\pi\frac{1-\eta}{4}\right)\cos\left(\pi\frac{\eta}{2}\right)
\Gamma\left(\frac{1+\eta}{2}\right)\Gamma(-\eta)\,.
\ee
Using this in the equation for $\Sigma$ yields
\begin{align}
\Sigma(\omega)&=\bar g^2\int\! \frac{d\omega'}{2\pi} \frac{1}{\Sigma(\omega')}\frac{1}{\delta\Pi(\omega-\omega')}\nonumber \\
&=\frac{\eta}{1-\eta}\frac{M}{N}\frac{\tan\left(\pi\frac{\eta}{2}\right)}{\tan\left(\pi\frac{1+\eta}{4}\right)} A\sgn{\omega}\abs{\omega}^{(1-\eta)/2}\,,
\end{align}
where we have also assumed that the phonon mass is renormalized to zero by $\Pi(0)$. Under this assumption, which we confirm below, we indeed find a self-consistent solution if
\be
1=\frac{\eta}{1-\eta}\frac{M}{N}\frac{\tan\left(\pi\frac{\eta}{2}\right)}{\tan\left(\pi\frac{1+\eta}{4}\right)}\,.
\ee
Next, we have to ensure that $\Pi(0)$ cancels $\omega_0^2$ in the phonon propagator. We find
\begin{align}
\Pi(0)&=-2\bar g^2\frac{N}{M}\int\! \frac{d\omega'}{2\pi} \frac{1}{(\omega'+\Sigma(\omega'))^2}\nonumber\\
&=-\frac{2}{\pi}\bar g^2\frac{N}{M}\Gamma\left(\frac{2\eta}{1+\eta}\right)\Gamma\left(\frac{3+\eta}{1+\eta}\right)A^{-2/(1+\eta)}\,,
\end{align}
and from the condition $\omega_0^2+\Pi(0)=0$, we obtain the coefficient $A$
\be
A=a_\eta\left(\frac{\bar g^2}{\omega_0^2}\right)^{(1+
\eta)/2}
\ee
with
\be
a_\eta=\left[\frac{2}{\pi}\frac{N}{M}\Gamma\left(\frac{2\eta}{1+\eta}\right)
\Gamma\left(\frac{3+\eta}{1+\eta}\right)\right]^{(1+\eta)/2}\,.
\label{eq:appag}
\ee
To calculate $\Pi(0)$, we have to keep the bare $\omega$ in the electron propagator to avoid a UV divergence.
In summary, Eqs.~\eqref{eq:appdeltaPi}-\eqref{eq:appag} yield a self-consistent solution of the non-superconducting Eliashberg equations.

\subsection{Gap equation as differential equation}
\label{app:DGL}
The gap equation reads
\begin{align}
\Phi(\omega)=(1-\alpha)\frac{a_\eta^2}{2\pi b_\eta}\bar\omega^{1+\eta}\int_0^\infty\!&d\omega' \frac{\Phi(\omega')}{\left(\omega'+a_\eta\bar\omega{\omega'}^{(1-\eta)/2}\right)^2}
\left[\frac{1}{\abs{\omega-\omega'}^\eta}+\frac{1}{\abs{\omega+\omega'}^\eta}\right]
\end{align}
where we restricted the integral to positive frequencies and defined $\bar\omega=\bar g^2/\omega_0^2$.
We rewrite it approximately as a differential equation using the limits $\omega\gg\omega'$ and $\omega\ll\omega'$. The result is valid for $\eta\ll1$ when the contribution $\omega\sim\omega'$ can be neglected.
We write approximately
\begin{align}
\Phi(\omega)\approx(1-\alpha)f_\eta\bar\omega^{1+\eta}\left[ \int_0^\omega\!d\omega'\frac{\Phi(\omega')}{\left(\omega'+a_\eta\bar\omega^{(1+\eta)/2}\omega'^{(1-\eta)/2}\right)^2}\frac{2}{\omega^\eta} + \int_\omega^\infty\!d\omega'\frac{\Phi(\omega')}{\left(\omega'+a_\eta\bar\omega^{(1+\eta)/2}\omega'^{(1-\eta)/2}\right)^2}\frac{2}{\omega'^\eta} \right]
\end{align}
with $f_\eta=a_\eta^2/(2\pi b_\eta)$. We then obtain for the derivative
\begin{align}
\frac{d}{d\omega}\left[\left(\frac{\omega}{\bar\omega}\right)^{1+\eta}\frac{d}{d\omega}\Phi(\omega)\right]=-2(1-\alpha)\eta f_\eta\frac{\Phi(\omega)}{\left(\omega+a_\eta\bar\omega^{(1+\eta)/2}\omega^{(1-\eta)/2}\right)^2}\,.
\end{align}
We rewrite the differential equation in terms of $z=\omega^{(1+\eta)/2}$ (analogous for $\bar z$)
\be
(z+a_\eta\bar z)^2\left[ (\eta+1)z^2\Phi''(z)+(3\eta+1)z\Phi'(z) \right]+\frac{4\eta a_\eta^2}{(\eta+1)\pi b_\eta}(1-\alpha)\bar z^2\Phi(z)=0\,.
\ee
The solution of this equation is of the form
\begin{align}
\Phi(z)&=C_1 (z+a_\eta\bar z)^{\frac{1+\eta-b}{2(1+\eta)}}z^{-\frac{\eta+a}{1+\eta}}{}_2F_1\left(\frac{1-\eta-2a-b}{1(1+\eta)},\frac{1+3\eta-2a-b}{1(1+\eta)},1-\frac{2a}{1+\eta},-\frac{z}{a_\eta\bar z}\right)\nonumber\\
&+C_2 (z+a_\eta\bar z)^{\frac{1+\eta+b}{2(1+\eta)}}z^{-\frac{\eta-a}{1+\eta}}{}_2F_1\left(\frac{1-\eta+2a+b}{1(1+\eta)},\frac{1+3\eta+2a+b}{1(1+\eta)},1+\frac{2a}{1+\eta},-\frac{z}{a_\eta\bar z}\right)
\label{eq:phidgl}
\end{align}
with the hypergeometric function ${}_2F_1$ and we defined $a=\sqrt{\eta(\eta-4(1-\alpha)/(\pi b_\eta))}$ and $b=\sqrt{(1+\eta)^2-16\eta(1-\alpha)/(\pi b_\eta)}$. Note that $b\rightarrow-b$ in the expression $\Phi(z)$ is also a solution (but not linearly independent). If $z\ll a_\eta\bar z$ or $\omega\ll a_\eta^{2/1(+\eta)}\bar\omega$, we recover the power law that we found in our analysis of the gap equation for small frequencies in the the main text
\be
\Phi(\omega)\rightarrow C_1 a_\eta^{\frac{1+\eta-b}{2(1+\eta)}}\bar\omega^{\frac{1+\eta-b}{4}}\omega^{-\frac{\eta+a}{2}}+C_2 a_\eta^{\frac{1+\eta+b}{2(1+\eta)}}\bar\omega^{\frac{1+\eta+b}{4}}\omega^{-\frac{\eta-a}{2}}\,.
\ee
For large frequencies, we see that the gap equations yields $\Phi(\omega)\propto\omega^{-\eta}$.
The leading term of the solution obtained from the differential equation for $z\gg a_\eta\bar z$ is a constant and we request it to vanish, which leads to a condition for $C_1,C_2$
\be
C_1(a_\eta\bar z)^{\frac{1-\eta-2a-b}{2(1+\eta)}}\frac{\Gamma\left(\frac{2\eta}{1+\eta}\right)
\Gamma\left(1-\frac{2a}{1+\eta}\right)}{\Gamma\left(\frac{1+3\eta-2a+b}{2(1+\eta)}\right)
\Gamma\left(\frac{1+3\eta-2a-b}{2(1+\eta)}\right)} + C_2(a_\eta\bar z)^{\frac{1-\eta+2a+b}{2(1+\eta)}}\frac{\Gamma\left(\frac{2\eta}{1+\eta}\right)
\Gamma\left(1+\frac{2a}{1+\eta}\right)}{\Gamma\left(\frac{1+3\eta+2a+b}{2(1+\eta)}\right)
\Gamma\left(\frac{1+3\eta+2a-b}{2(1+\eta)}\right)}=0\,.
\ee
The next-to-leading term is then proportional to $\omega^{-\eta}$
\be
\Phi(z)\rightarrow \left[C_1(a_\eta\bar z)^{\frac{1-\eta-2a-b}{2(1+\eta)}}\frac{\Gamma\left(-\frac{2\eta}{1+\eta}\right)
\Gamma\left(1-\frac{2a}{1+\eta}\right)}{\Gamma\left(\frac{1-\eta-2a+b}{2(1+\eta)}\right)
\Gamma\left(\frac{1-\eta-2a-b}{2(1+\eta)}\right)} + C_2(a_\eta\bar z)^{\frac{1-\eta+2a+b}{2(1+\eta)}}\frac{\Gamma\left(-\frac{2\eta}{1+\eta}\right)
\Gamma\left(1+\frac{2a}{1+\eta}\right)}{\Gamma\left(\frac{1-\eta+2a+b}{2(1+\eta)}\right)
\Gamma\left(\frac{1-\eta+2a-b}{2(1+\eta)}\right)}\right] \left(\frac{a_\eta\bar \omega}{\omega}\right)^\eta
\ee
in agreement with our expectation based on the integral gap equation.

\section{Finite temperature}

\subsection{Temperature-dependent mass}
\label{app:mT}

The thermal contribution does not lead to a divergence, because the phonon propagator develops a temperature-dependent mass as shown in Ref.~\onlinecite{PhysRevB.100.115132}.
The temperature- and frequency-dependent polarization operator in the SYK regime is given by
\begin{align}
\Pi(\omega_n,T)&=-2\bar g^2 \frac{N}{M}\sum_m \frac{1}{\Sigma(\omega_m)\Sigma(\omega_m+\omega_n)}\notag \\
&=-2\bar g^2 \frac{N}{M}\sum_k e^{i\pi k}\int \frac{d\omega'}{2\pi} \frac{e^{-ik\omega'/T}}{\Sigma(\omega')\Sigma(\omega'+\omega_n)}\notag \\
&=\Pi(\omega_n,T=0)-2\bar g^2\frac{N}{M}\sum_{k\neq0} e^{i\pi k}\int \frac{d\omega'}{2\pi} \frac{e^{-ik\omega'/T}}{\Sigma(\omega')\Sigma(\omega'+\omega_n)}\notag \\
&=\delta\Pi(\omega_n,T=0)+\Pi(0,0)+\Pi_T(\omega_n,T)\,,
\end{align}
where we have made use of the Poisson summation formula. With its help we can separate the contribution to the polarization operator that we found at zero temperature. We can simplify the thermal contribution further
\begin{align}
\Pi_T(\omega_n,T)&=-2\bar g^2\frac{N}{M}\sum_{k\neq0} e^{i\pi k}\int \frac{d\omega'}{2\pi} \frac{e^{-ik\omega'/T}}{\Sigma(\omega')\Sigma(\omega'+\omega_n)}\notag \\
&=-2\bar g^2\frac{N}{M}\sum_{k\neq0} e^{i\pi k}\int d\tau\frac{e^{-i\omega_n\tau}}{\Sigma(\tau+k/T)\Sigma(-\tau)}\notag \\
&=2\bar g^2\frac{N}{M}\sum_{k\neq0} e^{i\pi k}\frac{\Gamma^2\big(\frac{1+\eta}{2}\big)\cos^2\big(\pi\frac{1-\eta}{4}\big)}{\pi^2 A^2}\int d\tau e^{-i\omega_n\tau}\frac{\text{sgn}(\tau+k/T)\text{sgn}(-\tau)}{|\tau+k/T|^{\frac{1+\eta}{2}}|\tau|^{\frac{1+\eta}{2}}}\notag \\
&=-4\bar g^2\frac{N}{M}\frac{\Gamma^2\big(\frac{1+\eta}{2}\big)\cos^2\big(\pi\frac{1-\eta}{4}\big)}{\pi^2 A^2} \sum_{k=1}^\infty \int_0^\infty\! d\tau\frac{(-1)^k\cos(\omega_n\tau)}{\tau^{\frac{1+\eta}{2}}} \left[\frac{1}{(\tau+\frac{k}{T})^{\frac{1+\eta}{2}}} +\frac{\text{sgn}(\tau-\frac{k}{T})}{(\tau-\frac{k}{T})^{\frac{1+\eta}{2}}} \right]\notag \\
&=-4 \omega_0^2\frac{N}{M}\frac{\Gamma^2\big(\frac{1+\eta}{2}\big)\cos^2\big(\pi\frac{1-\eta}{4}\big)}{\pi^2 a_\eta^2g_0^{2\eta}\omega_0^{\eta}} \notag \\
&\times \sum_{k=1}^\infty (-1)^k\left(\frac{T}{k}\right)^\eta \left[\int_0^\infty\! d\tau\frac{\cos\big( k \tau\frac{\omega_n}{T}\big)}{\tau^{\frac{1+\eta}{2}}(\tau+1)^{\frac{1+\eta}{2}}} - \int_0^1\! d\tau\frac{\cos\big( k \tau\frac{\omega_n}{T}\big)}{\tau^{\frac{1+\eta}{2}}(1-\tau)^{\frac{1+\eta}{2}}} + \int_1^\infty\! d\tau\frac{\cos\big( k \tau\frac{\omega_n}{T}\big)}{\tau^{\frac{1+\eta}{2}}(\tau-1)^{\frac{1+\eta}{2}}} \right]
\label{eq:appPiT1}
\end{align}
where we have first used Fourier transformation, then rewritten the sum and integral over positive $k$ and $\tau$, and finally rescaled $\tau\rightarrow k\tau/T$.
If the sum and integral commute, we can perform the summation and obtain
\begin{align}
&\Pi_T(\omega_n,T)=-4 \omega_0^2\frac{N}{M}\frac{\Gamma^2\big(\frac{1+\eta}{2}\big)\cos^2\big(\pi\frac{1-\eta}{4}\big)}{\pi^2 a_\eta^2g_0^{2\eta}\omega_0^{\eta}}  \frac{T^\eta}{2} \notag \\
&\times\left[\int_0^\infty\! d\tau\frac{\text{Li}_\eta(-e^{-i\frac{\omega}{T}\tau})+\text{Li}_\eta(-e^{-i\frac{\omega}{T}\tau})}{\tau^{\frac{1+\eta}{2}}(\tau+1)^{\frac{1+\eta}{2}}} - \int_0^1\! d\tau\frac{\text{Li}_\eta(-e^{-i\frac{\omega}{T}\tau})+\text{Li}_\eta(-e^{-i\frac{\omega}{T}\tau})}{\tau^{\frac{1+\eta}{2}}(1-\tau)^{\frac{1+\eta}{2}}} + \int_1^\infty\! d\tau\frac{\text{Li}_\eta(-e^{-i\frac{\omega}{T}\tau})+\text{Li}_\eta(-e^{-i\frac{\omega}{T}\tau})}{\tau^{\frac{1+\eta}{2}}(\tau-1)^{\frac{1+\eta}{2}}} \right]
\label{eq:appPiT}
\end{align}
with the polylogarithm Li$_\eta(z)$, which takes special forms for $\eta=0$ or $\eta=1$, Li$_0(z)=z/(1-z)$ and Li$_1(z)=-\ln(1-z)$.

In Eqs.~\eqref{eq:appPiT1} and \eqref{eq:appPiT}, it is obvious that $\Pi_T(\omega_n, T)$ is a function of $\omega_n/T$ multiplied by $\omega_0^2(T/g_0^2\omega_0)^\eta$, as we stated in the main text. We also separate the frequency-independent part $\Pi_T(\omega_n, T)=\Pi_T(0,T)+\delta\hat\Pi(\omega_n,T)$, so that we can write
\be
\delta\hat\Pi(\omega_n,T)=\Pi_T(\omega_n, T)-\Pi_T(0,T)=\omega_0^2\left(\frac{T}{g_0^2\omega_0}\right)^\eta p_\eta\left(\frac{\omega_n}{T}\right)
\ee
This defines the function $p_\eta$. In summary, we can write the polarization operator as $\Pi(\omega_n,T)=\Pi(0,T)+\delta\Pi(\omega_n,0)+\delta\hat\Pi(\omega_n,T)$ with $\Pi(0,T)=\Pi(0,0)+\Pi_T(0,T)$, which we used in the main text.
The frequency-independent part adds a temperature-dependent mass to the propagator $\Pi(0,T)=-\omega_0^2=m^2(T)$, which we can calculate analytically
\begin{align}
m^2(T):=\Pi_T(0,T)=\frac{c_\eta}{a_\eta^2g_0^{2\eta}}\omega_0^2\left(\frac{T}{\omega_0}\right)^\eta
\end{align}
with $c_\eta=-4\cos(\pi\eta/2)\Gamma(\eta)(2^{1-\eta}-1)\zeta(\eta)N/(\pi M)$ and the zeta function $\zeta$.

$\delta\hat\Pi$ describes the thermal corrections added to the frequency-dependent zero-temperature expressions and the temperature-dependent mass. We can get an estimate for the temperature dependence by expanding $\delta\hat\Pi$ for $\omega_n\ll T$.
The leading order in the expansion is given by
\begin{align}
\delta\hat\Pi(\omega_n,T)=   d_\eta \left(\frac{T}{\omega_0}\right)^\eta \frac{\omega_n^2}{T^2} \omega_0^2 +\mathcal O\left(\left(\frac{\omega_n}{T}\right)^3\right)
\end{align}
with
\be
d_\eta=-2 \frac{N}{M}\frac{\Gamma^2\big(\frac{1+\eta}{2}\big)\cos^2\big(\pi\frac{1-\eta}{4}\big)}{\pi^2 a_\eta^2g_0^{2\eta}} (1-2^{3-\eta})\zeta(\eta-2)\left[ B(\frac{5-\eta}{2},-2+\eta)- B(\frac{5-\eta}{2},\frac{1-\eta}{2}) +B(\frac{1-\eta}{2},-2+\eta) \right]
\ee
For $\eta\rightarrow0$ and finite $T$, the corrections vanish $d_\eta=\mathcal O(\eta)$. With the help of Eq.~\eqref{eq:appPiT}, we can even show that $\Pi_T(\omega_n,T)=m^2(T)$ exactly for $\eta= 0$ without expanding in $\omega_n/T$. However, for $\eta\rightarrow 1$, $m^2\rightarrow 0$ and corrections are of order one.
In summary, for small to moderate $\eta$, we can approximate the boson propagator in the SYK regime including leading thermal corrections as
\be
D^{-1}(\omega_n,T)=\omega_n^2+\delta\Pi(\omega_n)+m^2(T)\,.
\ee
This expression is exact for $\eta\rightarrow0$. Let us also note that it was shown in Ref.~\onlinecite{PhysRevB.100.115132} that the $T=0$ SYK-like expressions plus the thermal mass $m^2(T)$ fit well to the numerical solution at finite, small temperatures for $M=N$ ($\eta\approx0.68$). In addition, the Monte-Carlo computations in Ref.~\cite{2020arXiv200106586P} demonstrated critical SYK-like scaling for finite temperatures and finite $N,M$ for a similar YSYK model.

\subsection{Superconductivity}
\label{app:SCT}

\begin{figure}[t!]
\begin{center}
\includegraphics[width=.3\columnwidth]{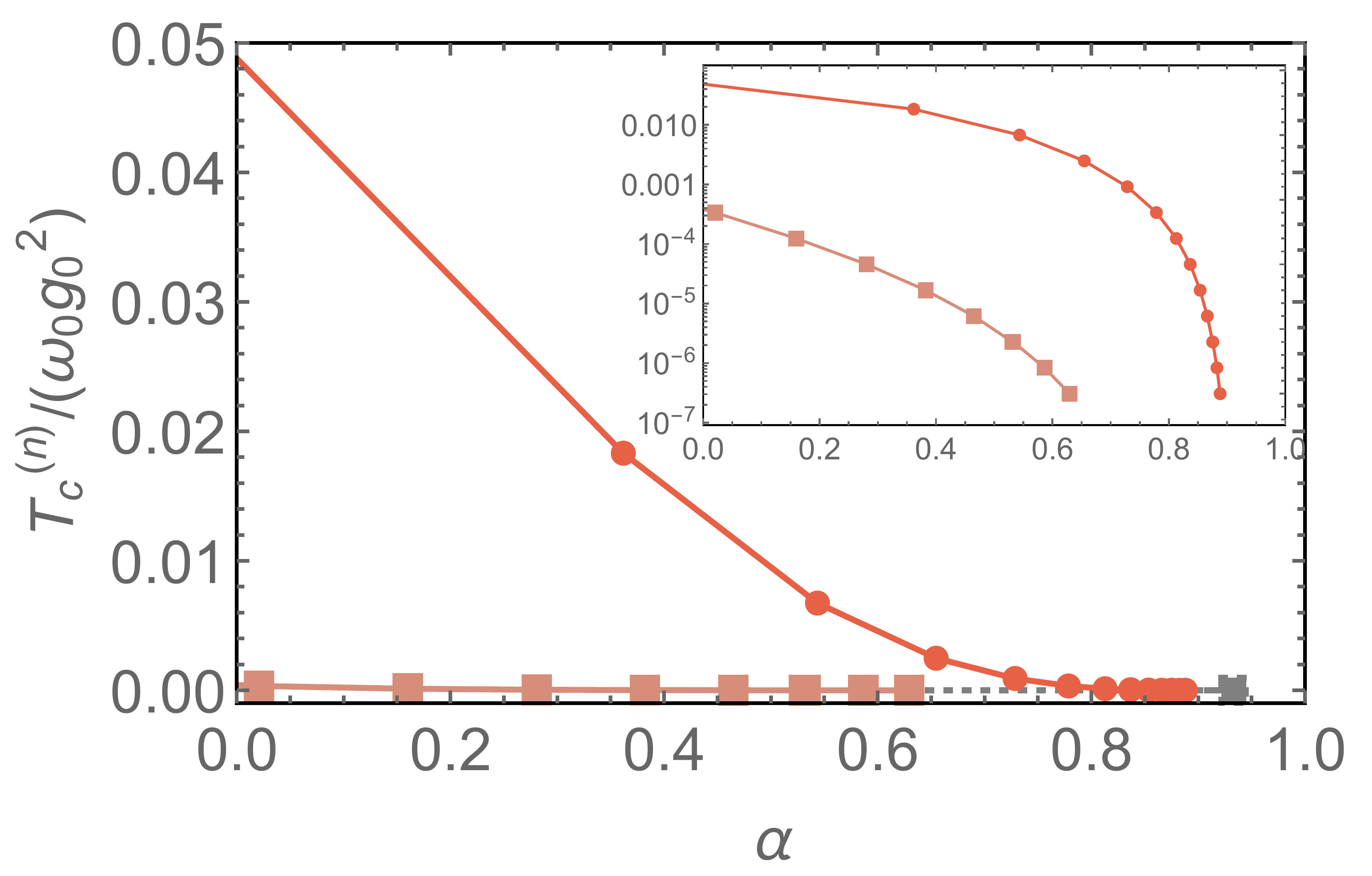}\qquad
\includegraphics[width=.28\columnwidth]{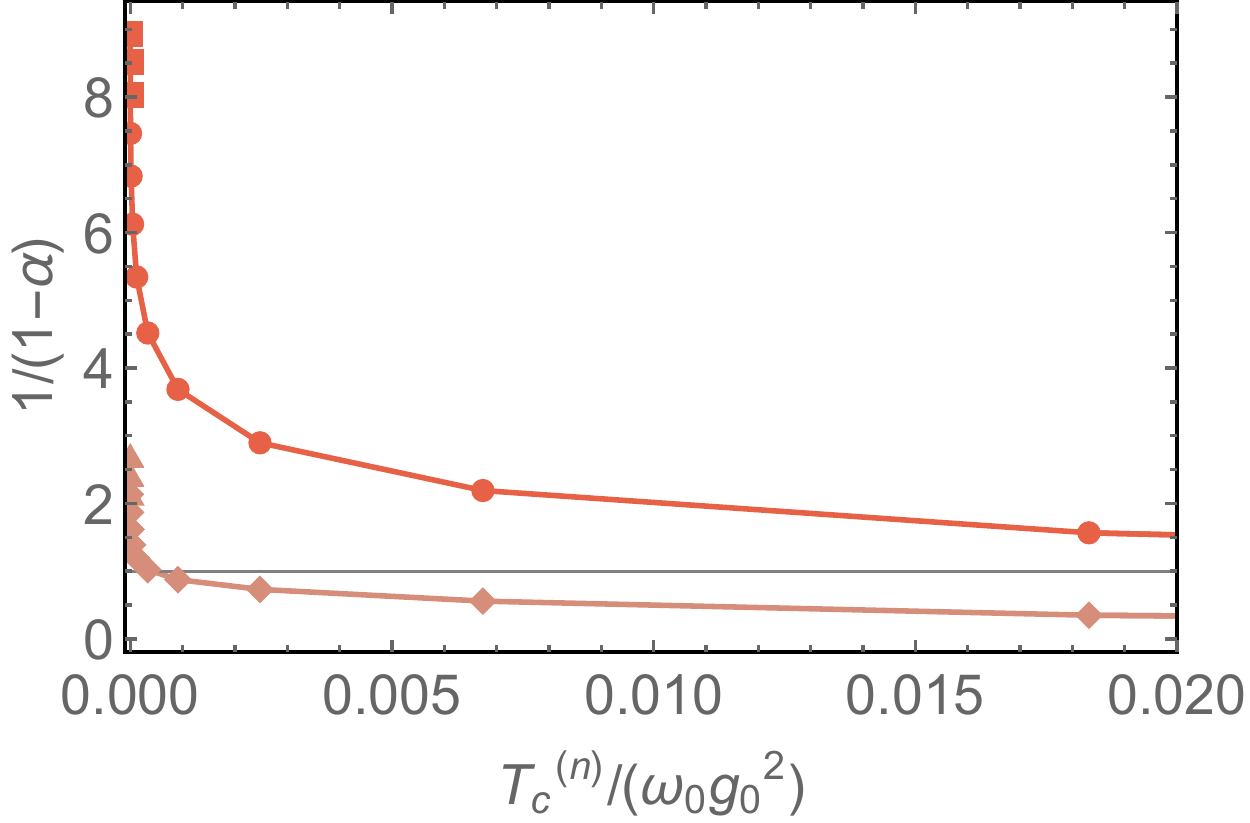}\\
\includegraphics[width=.3\columnwidth]{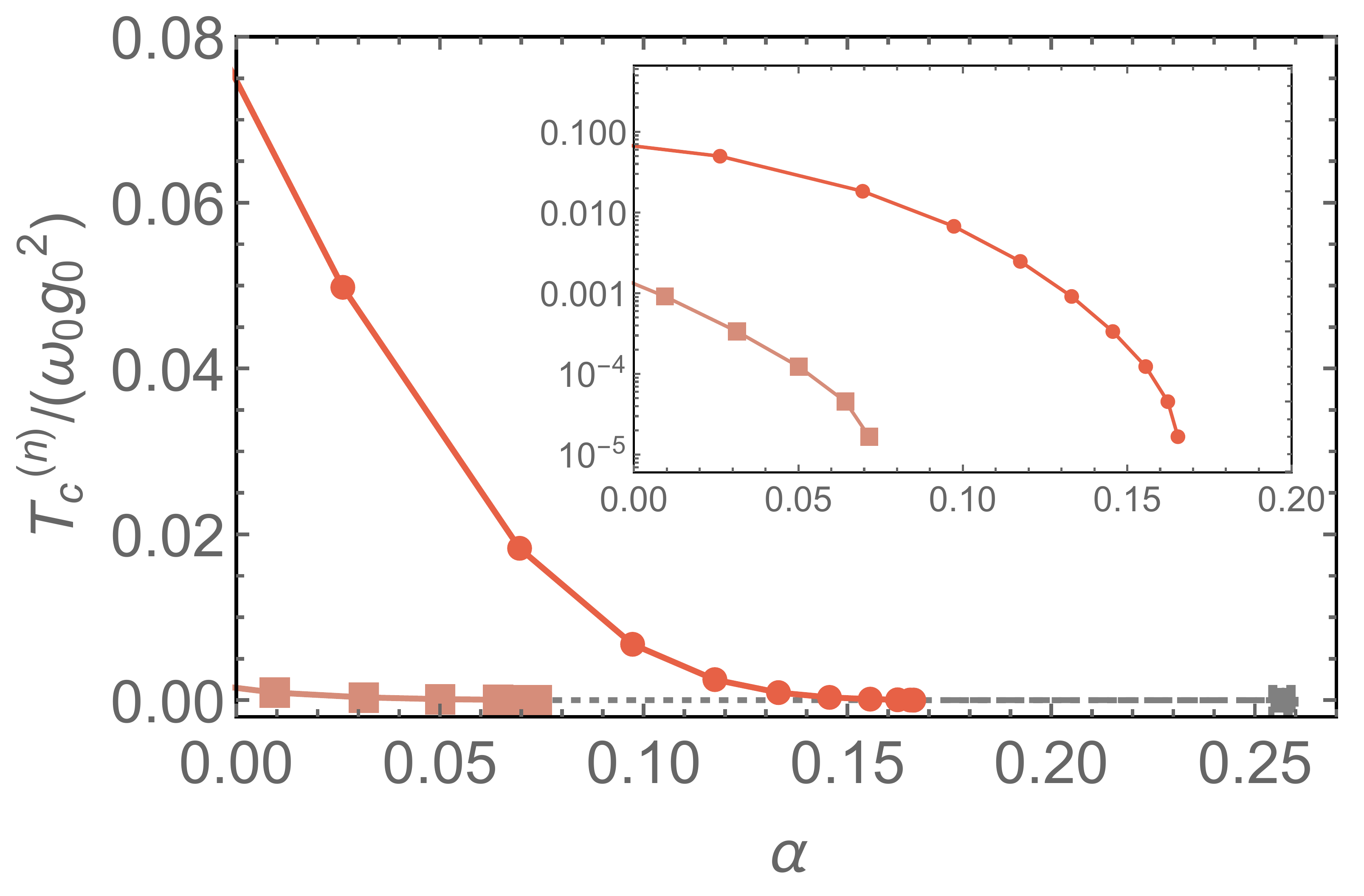}\qquad
\includegraphics[width=.29\columnwidth]{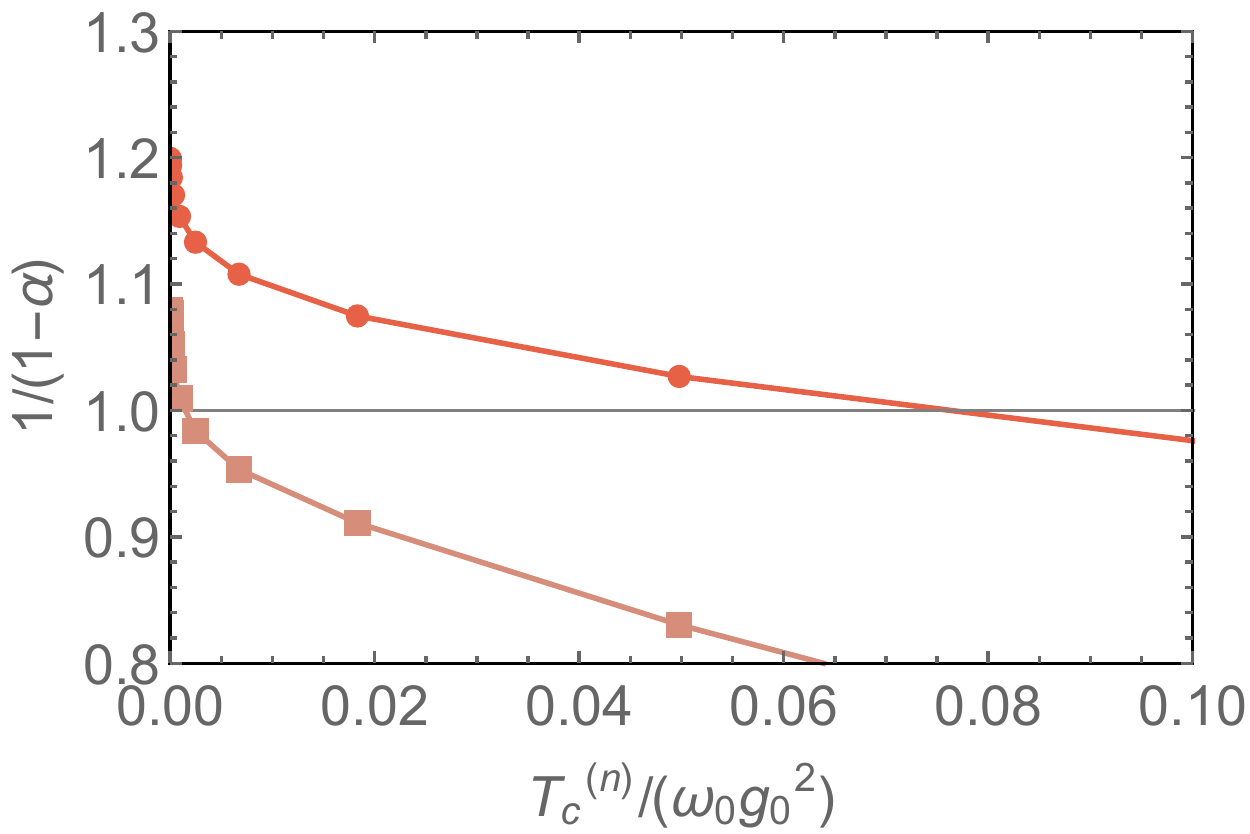}
\end{center}
\caption{Left: the largest two onset temperatures for varying $\alpha$ and $\eta=0.3$ (top) or $\eta=0.9$ (bottom). The gray symbol marks the critical $\alpha_c$, where we expect $T_c^{(n)}$ to vanish from our zero-temperature analysis. It is not reached completely due to numerical restrictions. Insets show the same data on a logarithmic scale.  Right: for comparison to the $\gamma$-model, we show the same data plotted as $1/(1-\alpha)$ for varying $T_c^{(n)}$. As we explained in Sec.~\ref{sec:4}, we can approximately identify $1/(1-\alpha)$ with $N$.
 Data points with $1/(1-\alpha)<1$ do not correspond to a physical solution, i.e. no superconductivity develops in this region.}
\label{fig:Tcn39}
\end{figure}

As described in the main text, we have determined the critical temperature for superconductivity based on the linearized gap equation Eq.~\eqref{eq:gap}.
If superconductivity develops out of the SYK or impurity regime, there is an infinite number of solutions of the gap equation with decreasing critical temperatures $T_c^{(0)}>T_c^{1}>\ldots$ that correspond to different eigenvalues of the gap equation. In the main text, we have shown the numerically determined $T_c^{(0)}$ and $T_c^{(1)}$ for $\eta=0.68$.
 Here, we also plot $T_c^{(n)}$ for $\eta=0.3$ and $\eta=0.9$ in Fig.~\ref{fig:Tcn39}.

\ew

\end{appendix}

\bibliography{bibSYK}

\end{document}